\documentclass{amsart}

\usepackage{latexsym,pifont}
\usepackage{epsfig}
\usepackage{rotating}

\usepackage{ifpdf}
\ifpdf
\usepackage{epstopdf}
\usepackage{hyperref}
\else
\usepackage[hypertex]{hyperref}
\fi

\usepackage{amsfonts,amsma th,amssymb,mathrsfs,extarrows,MnSymbol}
\usepackage[all]{xy}

\usepackage{tikz}
\usetikzlibrary{matrix,arrows}

\usepackage[makeroom]{cancel}

\newcommand{\alxydim}[2]{\begin{aligned}\xymatrix#1{#2}\end{aligned}}

\newcommand{\brem}{\begin{Rem}}
\newcommand{\erem}{\end{Rem}\medskip}
\newcommand{\beg}{\begin{Eg}}
\newcommand{\eeg}{\end{Eg}}
\newcommand{\bedef}{\begin{Def}}
\newcommand{\exdef}{\begin{flushright}$\diamond$\end{flushright}
\end{Def}\vskip0.1cm}
\newcommand{\berop}{\begin{Prop}}
\newcommand{\eerop}{\end{Prop}}
\newcommand{\belem}{\begin{Lem}}
\newcommand{\elem}{\end{Lem}}
\newcommand{\bethe}{\begin{Thm}}
\newcommand{\ethe}{\end{Thm}}
\newcommand{\becor}{\begin{Cor}}
\newcommand{\ecor}{\end{Cor}}
\newcommand{\beroof}{\noindent\begin{proof}}
\newcommand{\eroof}{\end{proof}}
\newcommand{\becon}{\begin{Conv}}
\newcommand{\econ}{\begin{flushright}$\checkmark$\end{flushright}\end{Conv}}
\newcommand{\befact}{\begin{Fact}}
\newcommand{\efact}{\begin{flushright}$\checkmark$\end{flushright}\end{Fact}}
\newcommand{\bequest}{\begin{Quest}}
\newcommand{\equest}{\end{Quest}}
\newcommand{\brob}{\begin{Prob}}
\newcommand{\erob}{\end{Prob}}

\newcommand{\barr}{\begin{array}}
\newcommand{\earr}{\end{array}}
\newcommand{\ben}{\begin{enumerate}}
\newcommand{\een}{\end{enumerate}}
\newcommand{\bit}{\begin{itemize}}
\newcommand{\eit}{\end{itemize}}

\newcommand{\qq}{\begin{eqnarray}}
\newcommand{\qqq}{\end{eqnarray}}

\newcommand{\nn}{\nonumber}

\newcommand{\ovl}[1]{\overline{#1}}
\newcommand{\unl}[1]{\underline{#1}}

\newcommand{\Reqref}[1]{Eq.\,\eqref{#1}}
\newcommand{\Rcite}[1]{Ref.\,\cite{#1}}
\newcommand{\Rxcite}[2]{Ref.\,\cite[#1]{#2}}

\newcommand\void[1]{}

\newcommand{\tx}[1]{\textrm{#1}} 
\newcommand{\ciut}[1]{\tiny$#1$}

\newcommand{\gt}[1]{\mathfrak{#1}}

\def\cA{\mathcal{A}}

\def\cD{\mathcal{D}}

\def\cF{\mathcal{F}}
\def\cG{\mathcal{G}}

\def\cI{\mathcal{I}}

\def\cK{\mathcal{K}}
\def\ceL{\mathcal{L}}
\def\cM{\mathcal{M}}
\def\cN{\mathcal{N}}
\def\cO{\mathcal{O}}

\def\cS{\mathcal{S}}
\def\cT{\mathcal{T}}
\def\cU{\mathcal{U}}
\def\cV{\mathcal{V}}
\def\cW{\mathcal{W}}

\def\cZ{\mathcal{Z}}

\def\xcE{\mathscr{E}}

\def\xcG{\mathscr{G}}

\def\xcL{\mathscr{L}}

\def\t{\mathbf{t}}

\def\bC{{\mathbb{C}}}

\def\bN{{\mathbb{N}}}

\def\bR{{\mathbb{R}}}
\def\bS{{\mathbb{S}}}

\def\bX{{\mathbb{X}}}

\def\bZ{{\mathbb{Z}}}

\def\a{\alpha}
\def\b{\beta}
\def\g{\gamma}
\def\G{\Gamma}
\def\d{\delta}
\def\D{\Delta}
\def\ep{\epsilon}
\def\vep{\varepsilon}

\def\k{\kappa}

\def\la{\lambda}

\def\om{\omega}
\def\Om{\Omega}

\def\si{\sigma}
\def\Si{\Sigma}

\def\t{\tau}

\def\z{\zeta}

\def\agt{\gt{a}}

\def\Bgt{\gt{B}}

\def\dgt{\gt{d}}

\def\egt{\gt{e}}

\def\ggt{\gt{g}}

\def\hgt{\gt{h}}

\def\kgt{\gt{k}}

\def\lgt{\gt{l}}

\def\mgt{\gt{m}}

\def\sgt{\gt{s}}

\def\tgt{\gt{t}}
\def\Tgt{\gt{T}}

\newcommand{\sfa}{{\mathsf a}}

\newcommand{\sfd}{{\mathsf d}}

\newcommand{\sfi}{{\mathsf i}}

\newcommand{\sfN}{{\mathsf N}}

\newcommand{\sfP}{{\mathsf P}}

\newcommand{\sfT}{{\mathsf T}}

\newcommand{\sfY}{{\mathsf Y}}

\newcommand{\txa}{{\rm a}}
\newcommand{\txA}{{\rm A}}

\newcommand{\txB}{{\rm B}}

\newcommand{\ee}{{\rm e}}
\newcommand{\txE}{{\rm E}}

\newcommand{\txF}{{\rm F}}
\newcommand{\txg}{{\rm g}}
\newcommand{\txG}{{\rm G}}

\newcommand{\txH}{{\rm H}}

\newcommand{\txK}{{\rm K}}

\newcommand{\txm}{{\rm m}}

\def\exp{{\rm exp}}
\def\id{{\rm id}}
\newcommand{\pr}{{\rm pr}}

\def\too{\longrightarrow}
\def\ev{{\rm ev}}

\def\mor{{\rm Hom}}

\def\1morf{1{\rm -Mor}}
\def\2morf{2{\rm -Mor}}
\def\dim{{\rm dim}}

\def\ker{{\rm ker}}

\def\End{{\rm End}}

\def\Inv{{\rm Inv}}

\newcommand{\sMan}{{\rm {\bf sMan}}}

\def\Vol{{\rm Vol}}

\newcommand{\pLie}[1]{\,{-\hspace{-8pt}\xcL}_{#1}}
\def\p{\partial}

\def\con{\righthalfcup}

\def\emb{\hookrightarrow}

\def\bd1{{\boldsymbol{1}}}
\def\brd0{{\boldsymbol{0}}}

\def\rk{{\rm rk}}
\def\det{{\rm det}}

\def\diag{\textrm{diag}}

\def\ad{{\rm ad}}
\def\Ad{{\rm Ad}}

\def\Cliff{{\rm Cliff}}

\newcommand{\uj}{{\rm U}(1)}

\def\x{\times}
\def\ox{\otimes}

\def\lx{{\hspace{-0.04cm}\ltimes\hspace{-0.05cm}}}
\def\rx{\rtimes}

\def\must{\stackrel{!}{=}}

\def\rstr{\mathord{\restriction}}

\newcommand{\corr}[1]{\left\langle #1 \right\rangle}

\newtheorem{Thm}{Theorem}
\newtheorem{Prop}[Thm]{Proposition}
\newtheorem{Lem}[Thm]{Lemma}
\newtheorem{Cor}[Thm]{Corollary}
\theoremstyle{definition}
\newtheorem{Rem}[Thm]{Remark}
\newtheorem{Def}[Thm]{Definition}
\newtheorem{Eg}[Thm]{Example}
\newtheorem{Conv}[Thm]{Convention}
\newtheorem{Fact}[Thm]{Fact}
\newtheorem{Quest}[Thm]{Question}
\newtheorem{Prob}[Thm]{Problem}

\numberwithin{equation}{section} 

\newcount\hour\newcount\minute
        \hour=\time \divide\hour by60 \minute=\time
        {\multiply\hour by60 \global\advance\minute by-\hour}
        \edef\militarytime{\number\hour:\ifnum\minute<10 0\fi\number\minute}

\begin{document}

\title{The higher-algebraic skeleton of the superstring\\ -- a case study}

\author{Rafa\l ~R.\ ~Suszek}
\address{R.R.S.:\ Katedra Metod Matematycznych Fizyki,\ Wydzia\l ~Fizyki
Uniwersytetu Warszawskiego,\ ul.\ Pasteura 5,\ PL-02-093 Warszawa,
Poland} \email{suszek@fuw.edu.pl}

\begin{abstract}
A novel Lie-superalgebraic description of the superstring in the super-Minkowskian background is extracted from the Cartan--Eilenberg super-1-gerbe geometrising the higher gauge field (the Green--Schwarz super-3-cocycle) that couples to the supercharge carried by the superstring.\ The description assumes the form of a hierarchy of Lie superalgebras integrable to a hierarchy of Lie supergroups and provides a manifestly supersymmetric model of a family of supermanifolds defining a trivialisation of the super-1-gerbe over the embedded superstring worldsheet.\ The trivialisation,\ obtained in a purely topological formulation of the superstring dynamics dual to the standard Nambu--Goto-type one,\ conforms with the gerbe-theoretic representation of extended sources of higher gauge fields known from previous studies of the $\si$-model of the bosonic string.
\end{abstract}

\maketitle

\tableofcontents

\section{Introduction}

The idea to probe spacetime geometry with the dynamics of distributions of charged matter has been around for a long time,\ {\it cp} Refs.\,\cite{Fock:1929gdq,Fock:1929qdp,Weyl:1929eg,Fock:1929gde,Fock:1929dgr,Dirac:1931fb},\ yielding an enhancement of the simple model of a metric spacetime $\,(M,\txg)\,$ accessible to a neutral pointlike particle.\ The enhancement incorporates the `higher' geometry of the gauge fields $\,\txH\in Z^{p+2}_{\rm dR}(M),\ p\in\bN\,$ (and their nonabelian counterparts) coupling to the respective charges,\ first neatly packaged by Lubkin \cite{Lubkin:1963},\ Trautman \cite{Trautman:1970fb} {\it et al.}\ in the structure of fibre bundles,\ and later generalised as bundle ($p$-)gerbes\footnote{For a gentle introduction to the general theory,\ {\it cp} \Rcite{Johnson:2003}.\ An overview was given by Murray in \Rcite{Murray:2007ps}.} and related objects by Murray {\it et al.} \cite{Murray:1994db,Murray:1999ew,Stevenson:2001grb2},\ {\it cp},\ in particular,\ \Rcite{Gajer:1996}.\ The latter form a descent hierarchy of geometrisations of integral classes in (a suitable refinement of) the de Rham cohomology of $\,M\,$ whose local sections provide us with the (suitably extended) Deligne--Beilinson cohomological data of the gauge fields,\ employed in the construction of simple models of charge dynamics and their geometric quantisation already by Alvarez \cite{Alvarez:1984es} and Gaw\c{e}dzki \cite{Gawedzki:1987ak}.

In the presence of Killing vector fields $\,\cK_A\in\G(\sfT M),\ A\in\ovl{1,K}\,$ of the background metric $\,\txg\,$ whose flows preserve the action functional defining the charge dynamics,\ the enhancement may take the form of a \emph{deformation} or an \emph{extension} of the Lie algebra of these vector fields.\ Notable instantiations of the former include the algebra $\,[P_\mu,P_\nu]=2qH_{\mu\nu}\,,\ \mu,\nu\in\ovl{0,3}\,$ of lifts of translations $\,\{P_\mu\}_{\mu\in\ovl{0,3}}\,$ to the space of states of a pointlike particle of charge $\,q\,$ in a constant electromagnetic field $\,\txH=H_{\mu\nu}\,\sfd x^\mu\wedge\sfd x^\nu\in\Om^2({\rm Mink}(3,1))$,\ and the Poisson germ of the Drinfeld--Jimbo quantum-group structure in the (chiral) Wess--Zumino--Witten model on $\,{\rm SU}(2)\,$ (with the Cartan 3-form as the gauge field) in \Rxcite{Sec.\,4}{Gawedzki:1990jc}.\ The latter are amply exemplified by the infinite sequence of extensions $\,{\rm Maxwell}_n,\ n\in\bN^\x\,$ of the Poincar\'e algebra presented in Refs.\,\cite{Bonanos:2008ez,Gomis:2017cmt} as algebraic structures encoding the rich dynamics of a (possibly backreacting) multipole distribution of charges in an external electromagnetic field (generalising the pioneering constructions:\ the kinematical algebras of \Rcite{Bacry:1970ye} and the Maxwell algebra of \Rcite{Schrader:1972zd}) and by the Free (super-)Differential-Algebra (FsDA) extensions of the super-Poincar\'e algebra considered in \Rcite{Chryssomalakos:2000xd} in the context of superstring theory in Minkowskian spacetime that build upon the earlier constructions of Refs.\,\cite{Green:1989nn,Siegel:1994xr,Bergshoeff:1995hm}.\ The supersymmetric extensions,\ of central relevance to us in what follows,\ seem to be encompassed by the structure of the Free Lie super-Algebra (FLsA) laid out in the recent study \cite{Gomis:2018xmo}.\ The common source of the enhancement in the examples listed is an interplay between the geometry (topology) of the distribution of charge consistent with its dynamics and the intrinsic cohomology of the gauge field $\,\txH\,$ (tied intimately with the aforementioned higher geometry),\ the latter being typically assumed to satisfy the strong invariance condition $\,\cK_A\con\txH\in B^\bullet_{\rm dR}(M)\,$ (ensuring quasi-invariance of the lagrangian density) in the case of charge distributions localised on closed submanifolds of $\,M$,\ {\it cp},\ in particular,\ Refs.\,\cite{deAzcarraga:1989mza},\ \cite[Cor.\,2.2]{Gawedzki:2010rn} and \cite[Sec.\,3]{Suszek:2018bvx}.\ The two are jointly\footnote{This is,\ arguably,\ most convincingly illustrated in the `top-down' treatment of charge dynamics in Refs.\,\cite{Bonanos:2008ez,Gomis:2017cmt}.} encoded by the (pre)symplectic form of the lagrangian model of dynamics,\ as given by the first-order formalism of Refs.\,\cite{Gawedzki:1972ms,Kijowski:1973gi,Kijowski:1974mp,Kijowski:1976ze,Szczyrba:1976,Kijowski:1979dj},\ and so a question arises how to isolate information on (the geometry of) a particular classical solution of the dynamics given an enhancement of a reference (neutral) Killing algebra (note that the enhancement captures symmetries of the entire \emph{space} of classical solutions).\ This is the general problem that we tackle in the present case study in the framework of supersymmetric dynamics of super-$p$-branes,\ to be investigated using methods of higher (super)geometry that we review systematically below.\\[-5pt]

The algebraic mechanisms from the previous paragraph have been encountered and employed extensively as model-building tools in the setting of (super)field theory with non-linearly realised symmetry \cite{Coleman:1969sm,Callan:1969sn,Salam:1969rq,Salam:1970qk,Isham:1971dv} and supersymmetry \cite{Volkov:1972jx,Volkov:1973ix,Ivanov:1978mx,Lindstrom:1979kq,Uematsu:1981rj,Ivanov:1982bpa,Samuel:1982uh,Ferrara:1983fi,Bagger:1983mv} (originally contemplated by Schwinger \cite{Schwinger:1967tc} and Wigner \cite{Weinberg:1968de} in the context of effective field theory with chiral symmetries) in which the fibre of the covariant configuration bundle (or the `field space') carries the structure of a homogeneous space $\,\txG/\txH\,$ of a (super)symmetry group $\,\txG\,$ relative to its distinguished closed subgroup $\,\txH\,$ with the tangent Lie algebra $\,\hgt\equiv{\rm Lie}(\txH)\,$ defining a \emph{reductive} decomposition $\,\tgt\oplus\hgt=\ggt\equiv{\rm Lie}(\txG)\,$ of the tangent Lie (super)algebra $\,\ggt\,$ of $\,\txG$,\ {\it i.e.},\ such that $\,[\hgt,\tgt]\subset\tgt$.\ Here,\ the dynamics is modelled in terms of $\txH$-basic tensors on $\,\txG\,$ taken from the tensor algebra of the linear space of ($\txG$-)left-invariant (LI) (super-)1-forms on the (super)manifold $\,\txG\,$ and pulled back to the field space $\,\txG/\txH\,$ along local sections of the principal $\txH$-bundle $\,\txG\xrightarrow{\ \pi_{\txG/\txH}\ }\txG/\txH\,$ ($\pi_{\txG/\txH}\,$ is the quotient map) whose potential nontriviality was accounted for in \cite[Sec.\,5]{Suszek:2019cum} and whose local sections particularly favoured by physical considerations were analysed at length in the $\bZ/2\bZ$-graded setting in \cite[Sec.\,2]{Suszek:2020xcu}.\ An in-depth study of the mechanism of spontaneous (super)symmetry breakdown by a \emph{classical solution} to the dynamics in this setting,\ in conjunction with a clever application of the so-called Inverse Higgs Effect originally discovered by Ivanov and Ogievetsky \cite{Ivanov:1975zq},\ have provided us with a reinterpretation of some standard action functionals modelling charge dynamics as Goldstone fields conjugate to distinguished central charges defining extensions of geometric ({\it i.e.},\ neutral) (super)symmetry algebras \cite{Gauntlett:1990nk},\ and -- crucially from the vantage point adopted in the present paper -- have led to a \emph{purely topological} reformulation of the Nambu--Goto-type `metric' components of action functionals for structureless (as in \Rcite{Dirac:1962iy},\ {\it cp} also \Rcite{deAzcarraga:1989vh}) extended distributions of charged matter in,\ {\it i.a.},\ Refs.\,\cite{West:2000hr,Gomis:2006xw,Gomis:2006wu,McArthur:2010zm},\ along the lines of the original idea of Hughes and Polchinski \cite{Hughes:1986dn} developed by Gauntlett,\ Itoh and Townsend in \Rcite{Gauntlett:1989qe}.

The specific choice,\ referred to in the last paragraph,\ of a constructive paradigm of study of charge dynamics with the help of the Cartan calculus on a Lie (super)group paves the way for a systematic application of the techniques of Free ($\bZ/2\bZ$-graded) Differential Algebras (FDA) of Refs.\,\cite{DAuria:1982uck,vanNieuwenhuizen:1982zf,Castellani:1982kd}.\ In the context of interest,\ these are specialised to an augmentation of the canonical FDA $\,\ceL\cI(\txG)\,$ of LI (super-)1-forms on the Lie (super)group $\,\txG\,$ by the (super-)$(p+1)$-form potential of the relevant gauge field.\ For these,\ a finite ladder of \emph{integrable} (super)central extensions $\,\brd0\too\agt_n\too\ggt_{n+1}\too\ggt_n\too\brd0\,$ (altogether combining into a generically \emph{non}-(super)central extension $\,\sfY\ggt\equiv\ggt_N\too\ggt_0\equiv\ggt\,$ of the original Lie (super)algebra $\,\ggt$) is sought that yields a resolution of the gauge field in the Cartan--Eilenberg (CaE) cohomology $\,{\rm CaE}^{p+2}(\sfY\txG)\equiv H^{p+2}_{\rm dR}(\sfY\txG)^{\sfY\txG}\,$ of the Lie (super)group $\,\sfY\txG\,$ which integrates $\,\sfY\ggt$.\ Each rung of the ladder is determined by a (super-)2-cocycle in the decomposition of the pullback of the gauge field in terms of elements of $\,\ceL\cI(\txG_n)\,$ (for $\,\txG_n\,$ the Lie (super)group of $\,\ggt_n$) in conformity with the standard classification of (super)central extensions of a Lie (super)algebra $\,\ggt_n\,$ by a (super)commutative Lie (super)algebra $\,\agt_n\,$ by elements of the group $\,H^2(\ggt_n,\agt_n)\,$ in the $\agt_n$-valued cohomology of $\,\ggt_n$,\ {\it cp} \Rcite{Leites:1975}.\ While it is \emph{not} clear {\it a priori} that a finite resolution $\,\sfY\ggt\,$ of this kind exists\footnote{It is certainly more natural to associate with the representative $\,\txH\,$ of a class in $\,{\rm CaE}^{p+2}(\txG)\,$ a slim Lie $(p+1)$-(super)algebra of Baez,\ Crans and Huerta \cite{Baez:2004hda6,Baez:2010ye,Huerta:2011ic},\ itself a special example of the more general structure (an $L_\infty$-(super)algebra) encountered in the study of string field theory \cite{Stasheff:1992slinf,Lada:1992wc}.\ However,\ to the best of the Author's knowledge,\ there do \emph{not} exist,\ to date,\ any \emph{explicit} constructions of the corresponding integrated structures (the so-called Lie $(p+1)$-supergroups) for the known super-$p$-branes with $\,p>0\,$ that would be amenable to direct analysis,\ which precludes the discussion of a number of concrete issues of physical relevance,\ {\it cp} below.} ({\it cp} the Theorem in \Rxcite{Sec.\,7}{vanNieuwenhuizen:1982zf}),\ an algorithm devised by de Azc\'arraga and collaborators in \Rcite{Chryssomalakos:2000xd},\ which essentially boils down to a systematic reconstruction of a finite quotient within the FLsA of \cite{Gomis:2018xmo},\ does produce the desired result in the very special setting of the Green--Schwarz-type (GS) super-$\si$-models of super-$p$-brane dynamics in super-Minkowskian (super)geometry\footnote{Exploration of curved supergeometries was pioneered in Refs.\,\cite{Bergshoeff:1985su,Bergshoeff:1987cm,Bandos:1997ui,deWit:1998yu,Claus:1998fh,Metsaev:1998it,Zhou:1999sm,Arutyunov:2008if,Gomis:2008jt,Fre:2008qc,DAuria:2008vov}.} $\,{\rm sMink}(d,1|ND_{d,1})\equiv\bR^{d,1|N}\,$ \cite{Casalbuoni:1975hx,Brink:1981nb,deAzcarraga:1982dhu,Green:1983wt,Green:1983sg,Achucarro:1987nc} (here,\ $\,N\in\bN^\x\,$ is the number of supercharges in a Majorana-spinor representation of $\,\Cliff(\bR^{d,1})\,$ of dimension $\,D_{d,1}\,$ that generate an $N$-extended supersymmetry) that we work with in the present paper and on which,\ consequently,\ we focus henceforth.\ In fact,\ it was recently argued by Grasso and McArthur in Refs.\,\cite{Grasso:2017dgx,Grasso:2017bhj} that these results are essentially  \label{foot:Sext} unique\footnote{Attaining a similar goal for physically relevant \emph{curved} supergeometries with a nontrivial topology,\ such as,\ {\it e.g.},\ the homogeneous spaces:\ $\,{\rm SU}(1,1|2)_2/({\rm SO}(1,1)\x{\rm SO}(2))\equiv{\rm s}({\rm AdS}_2\x\bS^2)\,$ (viewed as the supertarget of the Zhou superstring \cite{Zhou:1999sm}), $\,({\rm SU}(1,1|2)\x{\rm SU}(1,1|2))_2/({\rm SO}(2,1)\x{\rm SO}(3))\equiv{\rm s}({\rm AdS}_3\x\bS^3)\,$ (for the Park--Rey superstring \cite{Park:1998un}) and $\,{\rm SU}(2,2|4)/({\rm SO}(4,1)\x{\rm SO}(5))\equiv{\rm s}({\rm AdS}_5\x\bS^5)\,$ (for the Metsaev--Tseytlin superstring \cite{Metsaev:1998it}),\ seems to call for an augmentation of the original organising principle.\ One natural possibility,\ suggested by the prime r\^ole of the asymptotic correspondence between the curved dynamics (its supergeometric data and the lagrangian model) and its flat-superspace counterpart in the construction of the former,\ {\rm cp} \Rcite{Metsaev:1998it},\ is the requirement that the \.In\"on\"u--Wigner contraction underlying the asymptotic flattening should lift to the full-fledged geometrisation of the relevant gauge superfield.\ The principle was laid out in \Rcite{Suszek:2018bvx} (with several no-go results for the Metsaev--Tseytlin superstring in its current formulation),\ elaborated and successfully realised for the Zhou superparticle in $\,{\rm s}({\rm AdS}_2\x\bS^2)\,$ in \Rcite{Suszek:2018ugf},\ and formalised concisely in \Rcite{Suszek:2020rev}.\ It is expected to work out in conjunction with the $S$-expansion scheme put forward in \Rcite{Hatsuda:2001pp} and later generalised in Refs.\,\cite{deAzcarraga:2002xi,Izaurieta:2006zz}.\ This expectation is currently being investigated.} when viewed as solutions to a cohomological problem in $\,{\rm CaE}^{p+2}(\sfY{\rm sMink}(d,1|ND_{d,1}))\,$ (their argument exploits the assumed triviality of the de Rham cohomology of the extension).

In order to be able to interpret the extension $\,\sfY\txG\too\txG\,$ as a partial geometrisation of the GS super-$(p+2)$-cocycle $\,\txH\in Z^{p+2}_{\rm dR}(\txG)^\txG\,$ in the (standard) sense of Murray,\ one should put the CaE cohomology of $\,\txG\,$ on the same footing as the underlying de Rham cohomology.\ That this makes sense is suggested by an old argument due to Rabin and Crane \cite{Rabin:1984rm,Rabin:1985tv} that essentially explains the discrepancy between $\,{\rm CaE}^\bullet({\rm sMink}(d,1|ND_{d,1}))\,$ (for $\,N=1$) and $\,H^\bullet_{\rm dR}({\rm sMink}(d,1|ND_{d,1}))\equiv 0\,$ as coming,\ {\it via} the standard duality,\ from the homology of an orbifold $\,{\rm sMink}(d,1|ND_{d,1})/\G_{\rm KR}\,$ of $\,{\rm sMink}(d,1|ND_{d,1})\,$ relative to a discrete subgroup $\,\G_{\rm KR}\subset{\rm sMink}(d,1|ND_{d,1})\,$ that had been encountered previously by Kosteleck\'y and Rabin in their study of supersymmetric field theory on the lattice \cite{Kostelecky:1983qu}.\ The orbifold has the topological structure of a fibration over its body $\,{\rm Mink}(d,1)\,$ with compact Gra\ss mann-odd fibers \cite{Rabin:1984rm}.\ The argument led Rabin to postulate that the GS super-$\si$-model with the supertarget $\,{\rm sMink}(d,1|ND_{d,1})\,$ be interpreted as describing propagation of loop-like distributions of supercharge within $\,{\rm sMink}(d,1|ND_{d,1})/\G_{\rm KR}$.\ But then,\ by a standard argument ({\it cp},\ {\it e.g.},\ Refs.\,\cite{Dixon:1985jw,Dixon:1986jc} and,\ in particular,\ \cite[Sec.\,8.3]{Suszek:2012ddg} and \Rcite{Suszek:2013} in which the idea of a worldvolume orbifold was formalised with reference to the universal gauge principle derived in Refs.\,\cite{Gawedzki:2010rn,Gawedzki:2012fu}),\ one has to incorporate the $\G_{\rm KR}$-twisted sector in the superfield theory on the cover $\,{\rm sMink}(d,1|ND_{d,1})$,\ and,\ indeed,\ this yields,\ {\it e.g.},\ a Gra\ss mann-odd wrapping anomaly in the canonical picture of \cite[Sec.\,4.2]{Suszek:2018bvx} that reproduces the Green extension of the $\,{\rm sMink}(d,1|ND_{d,1})\,$ superalgebra resolving the GS super-3-cocycle.\ While vital for internal consistency of our treatment,\ the last result shows quite explicitly how an enhancement of a neutral Killing algebra in the presence of a distribution of charged matter \emph{combines} information on the cohomology of the gauge field and the topology of the distribution.\ We may now be more specific in defining our goal:\ We wish to extract a clearcut signature of the localisation of a classical superstring (and more generally super-$p$-brane) trajectory in the supertarget from the superalgebraic description of the gauge field that couples to it.\ To this end,\ we first need to complete the geometrisation of that field and recall from the extensive study of analogous geometrisations in the non-$\bZ/2\bZ$-graded setting the higher-geometric representation of extended objects to which the gauge field of the $\si$-model couples -- the D-branes \cite{Polchinski:1995mt},\ with the coupling encoded in the effective Dirac--Born--Infeld (DBI) lagrangian density \cite{Fradkin:1985qd,Abouelsaood:1986gd,Leigh:1989jq}.\\[-5pt]

The train of reasoning restated above has laid the foundation for the geometrisation programme,\ initiated by the Author in \Rcite{Suszek:2017xlw},\ elaborated in Refs.\,\cite{Suszek:2019cum,Suszek:2018bvx,Suszek:2018ugf,Suszek:2020xcu} and recently reviewed in \Rcite{Suszek:2020rev},\ which sets out to associate with the physically relevant CaE gauge-field super-$(p+2)$-cocycles that determine the known GS-type super-$\si$-models on homogeneous spaces $\,\txG/\txH\,$ of supersymmetry Lie supergroups $\,\txG$,\ as well as with the attendant supersymmetric defects \cite{Fuchs:2007fw,Runkel:2008gr,Suszek:2011hg,Suszek:2011},\ \emph{concrete} higher-geometric objects of the type conceived by Murray {\it et al.} in the non-$\bZ/2\bZ$-graded setting (and recently reconsidered in the $\bZ/2\bZ$-graded setting by Huerta \cite{Huerta:2020}),\ and to lift all essential geometric properties of the underlying superfield theories (such as,\ {\it e.g.},\ their $\k$-symmetry) and constitutive relations between them (such as,\ {\it e.g.},\ the fundamental asymptotic relation between the super-$p$-brane models with the super-${\rm AdS}_m\x\bS^m$ targets and their super-Minkowskian counterparts) to those higher (super)geometries and the associated higher categories.\ The rationale for the goal thus delineated is an early observation,\ due to Gaw\c{e}dzki \cite{Gawedzki:1987ak},\ that the higher-geometric objects canonically determine,\ through the so-called cohomological transgression,\ a geometric (pre)quantisation scheme for the simple charge geometrodynamics of the $\si$-model.\ Thus,\ the existence of the said lifts is to be viewed as a condition of quantum-mechanical consistency of the structures, properties and relations lifted.\ In the gerbe-theoretic picture,\ the D-branes of string theory are represented by trivialisations of the 1-gerbe of the gauge field of the $\si$-model over submanifolds of the target space $\,M\,$ \cite{Gawedzki:1999bq,Freed:1999vc,Carey:2002,Gawedzki:2002se,Gawedzki:2004tu} -- these are described by certain vector bundles whose connection acquires the interpretation of the (gerbe-twisted) gauge field of the DBI theory.

The basic geometric substrate of the principle of descent that lies at the core of Murray's geometrisation of the class $\,[\txH]\in H^{p+2}_{\rm dR}(M,\bZ)\,$ of a gauge field $\,\txH\,$ (assumed integral) \cite{Murray:1994db,Murray:2007ps} is a \emph{surjective submersion} $\,\sfY M\xrightarrow{\ \pi_{\sfY M}\ }M\,$ whose total space supports a smooth primitive $\,\txB\in\Om^{p+1}(\sfY M)\,$ for the pullback of $\,\txH$,\ {\it i.e.},\ such that $\,\pi_{\sfY M}^*\txH=\sfd\txB$.\ The $(p+1)$-form $\,\txB\,$ can then be viewed as the trivial $p$-gerbe $\,\cI^{(p)}_\txB\,$ of curvature $\,\sfd\txB\,$ and curving $\,\txB\,$ over $\,\sfY M$.\ The $p$-gerbe $\,\cG^{(p)}\,$ for $\,[\txH]\,$ is subsequently erected over the nerve of the small category $\,\alxydim{}{\sfY^{[2]}M\ar@<.5ex>[r]^{\quad s\equiv\pr_1} \ar@<-.5ex>[r]_{\quad t\equiv\pr_2} & \sfY M}$,\ defined by the ($\pi_{\sfY M}$-)fibred square $\,\sfY^{[2]}M\equiv\sfY M\x_M\sfY M\,$ of the surjective submersion ({\it cp} App.\,\ref{app:conv}),\ as a family $\,\xcG\equiv\{\cG^{(p-k)}\}_{k\in\ovl{1,p+1}}\,$ of $(p-k)$-gerbes (the $l$-gerbes with $\,l\in\{0,-1\}\,$ being identified with principal $\bC^\x$-bundles with a compatible connection ($l=0$) and connection-preserving isomorphisms between them ($l=-1$),\ respectively) over the respective fibred powers $\,\sfY^{[k+1]}M\equiv\sfY^{[k]}M\x_M\sfY M$,\ the members of $\,\xcG\,$ being subject to various coherence constraints.\ Accordingly,\ and in keeping with the underlying (super)field-theoretic paradigm in which the (super)field theory over $\,\txG/\txH\,$ is modelled over $\,\txG$,\ the point of departure of the programme advocated above is the \emph{epimorphism of Lie supergroups} $\,\sfY\txG\xrightarrow{\ \pi_{\sfY\txG}\ }\txG\,$ (alongside the LI primitive for the pullback of the LI gauge field $\,\txH\,$ along $\,\pi_{\sfY\txG}$) returned by the integrable-extension algorithm described in the previous paragraph.\ From this point onwards,\ one simply turns the crank of Murray's machine of descent and,\ recursively,\ that of de Azc\'arraga's extension procedure,\ insisting that all extensions are consistently $\ad_\hgt$-equivariant in the latter (a condition essentially built into the FDA techniques employed in the procedure in the guise of the so-called minimal subalgebra \cite{Sullivan:1977},\ {\it cp} \Rxcite{Sec.\,6}{vanNieuwenhuizen:1982zf}),\ and -- in the former -- that all secondary surjective submersions that arise in the process are Lie-supergroup epimorphisms,\ and that all (connection-preserving) isomorphisms of principal $\bC^\x$-bundles that mark the penultimate stage of the construction and of its sub-constructions are Lie-supergroup isomorphisms,\ so that,\ by the end of the long day,\ we obtain a `bundle $p$-gerbe object in the category of Lie supergroups'.\ The ensuing \textbf{Cartan--Eilenberg super-$p$-gerbe} $\,\cG^{(p)}\,$ still has to be descended to the relevant homogeneous space $\,\txG/\txH$.\ As demonstrated by Gaw\c{e}dzki,\ Waldorf and the Author in Refs.\,\cite{Gawedzki:2010rn,Gawedzki:2012fu,Suszek:2012ddg,Suszek:2011,Suszek:2013},\ this requires that $\,\cG^{(p)}\,$ carry a \emph{descendable} $\txH$-equivariant structure.\ One of the crucial features of the advocated geometrisation scheme is that such a structure is inscribed in the very definition of the super-$p$-gerbe,\ which lends weight to the claim to naturality of the scheme in the (super)field-theoretic context under consideration.\ To date,\ the results of the programme include an \emph{explicit} construction  \cite{Suszek:2017xlw} of the CaE super-$p$-gerbes over $\,{\rm sMink}(d,1|D_{d,1})\,$ for the GS super-$p$-branes with $\,p\in\{0,1,2\}$,\ an extensive study \cite{Suszek:2019cum} of their equivariance properties inspired by the analogy with the purely Gra\ss mann-even WZW $\si$-model \cite{Henneaux:1984mh},\ and an explicit construction \cite{Suszek:2018ugf} of a super-0-gerbe over $\,{\rm s}({\rm AdS}_2\x\bS^2)\,$ for the Zhou superparticle that provides a constructive application of the principle of \.In\"on\"u--Wigner contractibility proposed in \Rcite{Suszek:2018bvx} ({\it cp} the footnote on p.\,\pageref{foot:Sext}).\ The CaE super-$p$-gerbes (for a large class of known super-$p$-brane species) were also shown \cite{Suszek:2019cum,Suszek:2020xcu} to carry a canonical and canonically (linearised-)supersymmetric linearised $\k$-symmetry-equivariant structure,\ in conformity with an interpretation of $\k$-symmetry,\ worked out {\it ibid.},\ purely in terms of the supertarget geometry.\ In fact,\ it is from the latter interpretation that provides a solution to the problem posed in the present Introduction,\ and so we conclude the section with a recapitulation of the physical idea behind it.\\[-5pt]

The objective of the present study is to extract a supersymmetric \emph{target-space} higher-geometric description of the fundamental dynamical object of the GS super-$\si$-model with the supertarget $\,\txG/\txH$,\ {\it i.e.}, of the (closed) super-$p$-brane trajectory,\ from the CaE super-$p$-gerbe over $\,\txG\,$ associated with the GS super-$(p+2)$-cocycle that determines its Wess--Zumino term -- all that in the much tractable model setting:\ for the superstring ($p=1$) in the superspace $\,{\rm sMink}(d,1|D_{d,1})\equiv{\rm sISO}(d,1|D_{d,1})/{\rm Spin}(d,1)$.\ In the light of the hitherto discussion,\ the task boils down to identifying a (higher-)superalgebraic object related to the Lie superalgebra $\,\ggt\,$ of the supersymmetry supergroup $\,\txG\equiv{\rm sISO}(d,1|D_{d,1})\,$ that captures a classical ({\it i.e.},\ critical) embedding of the superstring worldsheet in the supertarget and,\ in particular,\ the supersymmetry that survives such localisation.\ That the well-posedness of this task is non-obvious is best illustrated by the discussion of a local (tangential) Gra\ss mann-odd supersymmetry of the GS super-$\si$-model in the standard NG formulation,\ aka $\k$-symmetry,\ discovered by de Azc\'arraga and Lukierski (for the superparticle) in \Rcite{deAzcarraga:1982njd} and subsequently rediscovered and elaborated by Siegel (for the superstring) in Refs.\,\cite{Siegel:1983hh,Siegel:1983ke},\ whose existence is tied with the mechanism of restitution of equibalance of the internal degreees of freedom of both Gra\ss mann parities in the vacuum of the GS superfield theory through a removal of fermionic Goldstone modes,\ consistent with the structure of the supersymmetry Lie superalgebra of the theory:\ The symmetry couples the metric and topological terms in the action functional ({\it i.e.},\ they are not invariant \emph{separately}),\ and that only for a finely tuned relative normalisation of the two.\ It also bracket-generates a (super)algebra whose on-shell closure requires incorporation of generators of diffeomorphisms of the worldsheet \cite{McArthur:1999dy}.\ A path to \emph{target-space} geometrisation of $\k$-symmetry and field equations of the GS super-$\si$-model on $\,\txG/\txH$,\ and so also towards a meaningful formulation of the problem of interest,\ was paved in Refs.\,\cite{Suszek:2019cum,Suszek:2020xcu} where a duality -- first noted in \Rcite{Hughes:1986dn},\ later elaborated substantially in \Rcite{Gauntlett:1989qe} and employed in a rederivation of a variety of (super-$\si$-)models of charge dynamics in Refs.\,\cite{McArthur:2010zm,West:2000hr,Gomis:2006xw,Gomis:2006wu} -- was formalised,\ geometrised and exploited that exists between the original NG formulation of the GS super-$\si$-model and a \emph{purely topological} (super)field theory,\ termed the Hughes--Polchinski (HP) formulation of the GS super-$\si$-model by the Author,\ with the superfield space $\,\txG/\txH_{\rm vac}\,$ associated to another reductive decomposition $\,\ggt=(\tgt\oplus\dgt)\oplus\hgt_{\rm vac},\ \dgt\oplus\hgt_{\rm vac}=\hgt\,$ with $\,\hgt_{\rm vac}={\rm Lie}(\txH_{\rm vac})\,$ that encodes the spontaneous breakdown $\,\txH\searrow\txH_{\rm vac}\,$ of the `invisible' gauge symmetry $\,\txH\,$ of the superfield theory.\ The field space of the new formulation contains additional degrees of freedom,\ to wit,\ the bosonic Goldstone fields transverse to (the body of) the vacuum of the GS super-$\si$-model and modelled on the vector space $\,\dgt\cong\hgt/\hgt_{\rm vac}$.\ Its topologicality is reflected by the replacement of the original metric term of the NG by the pullback of an $\txH_{\rm vac}$-basic LI super-$(p+1)$-form on $\,\txG$,\ itself (a distinguished scalar multiple of) a volume form $\,\Vol(\tgt^{(0)}_{\rm vac})\,$ on a fixed algebraic model $\,\tgt^{(0)}_{\rm vac}\subset\tgt^{(0)}\,$ of the body of the vacuum,\ to the worldvolume of the super-$p$-brane along suitably $\dgt$-augmented sections of the NG formulation.\ On the higher-geometric side,\ this means that the dual HP dynamics is entirely determined by a ($\txH_{\rm vac}$-equivariant) CaE super-$p$-gerbe -- the tensor product of the original super-$p$-gerbe for the GS super-$(p+2)$-cocycle with the trivial one with the curving given by the volume form on  $\,\tgt^{(0)}_{\rm vac}\,$ that we shall call,\ after \Rcite{Suszek:2019cum},\ the {\bf extended Hughes--Polchinski super-$p$-gerbe over $\,\txG\,$} and denote as ($\la_p^*\in\bR^\x\,$ is the scalar mentioned earlier)
\qq\label{eq:extHPspg}
\widehat\cG{}^{(p)}\equiv\cG{}^{(p)}\ox\cI^{(p)}_{\la_p^*\,\Vol(\tgt^{(0)}_{\rm vac})}\,.
\qqq 
The removal of the extra Goldstone modes through the Inverse Higgs Effect of \Rcite{Ivanov:1975zq} puts us back in the original NG formulation.\ It is realised by imposition of a subset of superfield equations of the HP formulation that can be interpreted as geometric constraints on the tangents of the fields of the model restricting the latter to a superdistribution in the tangent sheaf $\,\cT\txG\,$ of the target supermanifold $\,\txG$,\ dubbed the HP/NG correspondence superdistribution and denoted as $\,{\rm Corr}(\sgt\Bgt^{{\rm (HP)}}_{p,\la^*_p})$.\ That \emph{all} superfield equations geometrise in an analogous manner,\ as do the gauge-fixing conditions for the `invisible' local-symmetry group $\,\txH_{\rm vac}$,\ altogether giving rise to what was named the HP vacuum superdistribution and denoted as $\,{\rm Vac}(\sgt\Bgt^{{\rm (HP)}}_{p,\la^*_p})\,$ in \Rcite{Suszek:2020xcu},\ is a structural feature of the dual HP formulation that turns out to be instrumental in resolving the above-posed problem of extraction of super-$p$-brane data from $\,\cG^{(p)}\,$ in a manner that we outline below in the closing paragraph of the Introduction.

The (classical) vacuum of the GS super-$\si$-model in the HP formulation emerges as a sub-su\-per\-man\-i\-fold within $\,\txG\,$ defined as the intersection of the HP local sections of the principal $\txH_{\rm vac}$-bundle $\,\txG\too\txG/\txH_{\rm vac}\,$ used in the definition of the lagrangean superfield of the theory with an integral supermanifold of $\,{\rm Vac}(\sgt\Bgt^{{\rm (HP)}}_{p,\la^*_p})$.\ The existence of a foliation of the sections by such integral leaves calls for involutivity of $\,{\rm Vac}(\sgt\Bgt^{{\rm (HP)}}_{p,\la^*_p})\,$ that was examined in \Rcite{Suszek:2020xcu}.\ There is yet another superdistribution whose regular behaviour is of essence for the consistency of the entire framework,\ namely,\ the limit of the weak derived flag (in the sense of Tanaka\footnote{The concept is very closely related to that of the FLsA of \Rcite{Gomis:2018xmo}.} \cite{Tanaka:1970}) of the projection to $\,{\rm Vac}(\sgt\Bgt^{{\rm (HP)}}_{p,\la^*_p})\,$ of the linear span of the set of generators of an enhanced (right) gauge supersymmetry that arises upon restriction of the superfield of the super-$\si$-model to $\,{\rm Corr}(\sgt\Bgt^{{\rm (HP)}}_{p,\la_p^*})$.\ The projection removes the obvious Gra\ss mann-even component modelled on $\,\dgt\,$ (reflecting the enhancement $\,\hgt_{\rm vac}\nearrow\hgt\,$ of the `invisible' gauge-symmetry algebra that accompanies the transition between the two formulations) and leaves us with a superdistribution $\,\k(\sgt\Bgt^{{\rm (HP)}}_{p,\la_p^*})\,$ that contains a generic Gra\ss mann-odd component -- the latter \emph{is} the target space-geometric realisation of the $\k$-symmetry of the GS super-$\si$-model in the topological formulation,\ whence the name {\bf $\k$-symmetry superdistribution} given to it in \Rcite{Suszek:2020xcu}.\ The regularity alluded to above simply means that the limit should stay within $\,{\rm Vac}(\sgt\Bgt^{{\rm (HP)}}_{p,\la^*_p})$,\ so that it can be given the interpretation of a \emph{gauge supersymmetry of the vacuum},\ engendered by $\,\k(\sgt\Bgt^{{\rm (HP)}}_{p,\la_p^*})$.\ This happens iff the vacuum superdistribution is involutive,\ in which case $\,\k(\sgt\Bgt^{{\rm (HP)}}_{p,\la_p^*})\,$ is readily seen to bracket-generate $\,{\rm Vac}(\sgt\Bgt^{{\rm (HP)}}_{p,\la^*_p})\,$ -- the vacuum supermanifold becomes a single orbit of the gauge-symmetry supergroup obtained through integration of the Lie superalgebra $\,\gt{vac}\bigl(\sgt\Bgt^{{\rm (HP)}}_{p,\la^*_p}\bigr)\,$ modelling the limit.\ The last fact,\ taken in conjunction with the higher-geometric interpretation and implementation of gauge symmetries worked out in Refs.\,\cite{Gawedzki:2010rn,Gawedzki:2012fu,Suszek:2012ddg,Suszek:2011,Suszek:2013},\ leads us to the following hypothesis of Refs.\,\cite{Suszek:2020xcu,Suszek:2020rev}:

\begin{center}
{\it Upon restriction to the vacuum,\ the extended HP super-$p$-gerbe $\,\widehat\cG{}^{(p)}\,$ trivialises flatly as $\,\widehat\cG{}^{(p)}\cong\cI_0$.}
\end{center}

\noindent In the present paper,\ we prove the hypothesis for $\,\txG/\txH={\rm sISO}(d,1|D_{d,1})/{\rm Spin}(d,1)\equiv{\rm sMink}(d,1|D_{d,1})\,$ and $\,p=1\,$ with the relevant super-3-cocycle of the GS super-$\si$-model for the superstring (whose physical content and supersymmetry,\ both global and local ($\k$-symmetry),\ in the dual HP formulation is reviewed for later reference in Secs.\,\ref{sec:gdyn} and \ref{sec:susy},\ respectively) as in \Rcite{deAzcarraga:1989vh}.\ We do that by first lifting the CaE super-1-gerbe of \Rxcite{Sec.\,5.2}{Suszek:2017xlw},\ erected directly over the Lie supergroup $\,{\rm sMink}(d,1|D_{d,1})$,\ to the mother Lie supergroup $\,{\rm sISO}(d,1|D_{d,1})\,$ in Sect.\,\ref{sec:geometrise} ({\bf Theorem \ref{thm:liftGSs1g}}),\ and subsequently deriving the trivialisation 1-isomorphism in Sec.\,\ref{sub:hkappa} ({\bf Theorem \ref{thm:vactriv}}) upon demonstrating briefly the higher-geometric realisation of global supersymmetry in Sec.\,\ref{sub:hgsusy}.\ This yields the most basic solution to the extraction problem posed at the beginning of the Introduction,\ which is readily seen by rewriting the trivialisation (symbolically,\ and for the value $\,\la_1^*=2\,$ determined in Sec.\,\ref{sec:susy}) as
\qq\nn
\cG{}^{(1)}\rstr_{\rm vacuum}\cong\cI_{-2\Vol(\tgt^{(0)}_{\rm vac})}\rstr_{\rm vacuum}\,,
\qqq
{\it cp} \Reqref{eq:extHPspg}.\ The solution has an essential weakness:\ In consequence of the lack of an obvious Lie-supergroup structure on the vacuum,\ the 1-isomorphism can only be and is a non-supersymmetric one,\ and so it exists outside the framework systematically constructed in the first part of the paper (and introduced in the original papers).\ We strengthen the `raw' result stated in Theorem \ref{thm:vactriv} in the last part of Sec.\,\ref{sub:hkappa} by passing to the tangent sheaf of the higher-geometric object that represents the 1-isomorphism and extracting a hierarchy of Lie superalgebras \eqref{diag:sLieAlgshad} associated with the various supermanifold components of that object and interrelated analogously but by Lie-superalgebra homomorphisms ({\bf Theorem \ref{thm:salgskel}}) -- \emph{this} is the structure encoding the supersymmetric supergeometry of the vacuum that we have been after,\ and it seems appropriate to call it the {\bf ${\bf sLieAlg}$-skeleton of the vacuum}.\ We conclude our study with one further step in which we integrate the ${\bf sLieAlg}$-skeleton to a hierarchy of Lie supergroups ({\bf Theorem \ref{thm:sgpmod}}),\ whereby the {\bf ${\bf sLieGrp}$-model of the vacuum} \eqref{diag:sLieGrpmod} arises.

Theorems \ref{thm:vactriv}--\ref{thm:sgpmod} constitute the main results of the present study,\ consistent with the higher-geometric representations of physical objects charged under the gauge field geometrised,\ and form a solid basis for further investigation of geometrisations of gauge fields of the GS super-$\si$-models for super-$p$-branes on homogeneous spaces of Lie supergroups that we intend to take up in the future.

\section{The geometrodynamics of the HP superstring}\label{sec:gdyn}

In this opening section,\ we recall the definition of the supersymmetric field theory of interest,\ modelling the propagation in Minkowskian spacetime of a loop-like distribution of charges of both Gra\ss mann parities in equibalance.\ The definition calls for a supermanifold with the Minkowskian body and a supersymmetric de Rham 3-cocycle field that couples to the supercharge current engendered by the propagating loop.\ In our presentation,\ we emphasise,\ purposefully,\ the underlying Lie-supergroup structure and the associated tangential Lie-superalgebra structure.\medskip

The point of departure of our discussion is the $(d+1)$-dimensional Minkowski space (for some $\,d\in\bN^\x$)
\qq\nn
\bR^{d,1}\equiv\bigl(\bR^{\x d+1},\eta\bigr)\,,\qquad\qquad\qquad\qquad\eta=\eta_{ab}\,E^a\ox E^b\,,\qquad\qquad(\eta_{ab})=\diag\,(-1,\underbrace{1,1,\ldots,1}_{d\ \tx{times}})
\qqq
with its structure of an abelian Lie group determined by the binary operation (written in terms of the global cartesian coordinates $\,\{x^a\}^{a\in\ovl{0,d}}$)
\qq\nn
|\unl\txm|\ :\ \bR^{d,1}\x\bR^{d,1}\too\bR^{d,1}\ :\ \bigl(\bigl(x_1^a\bigr),\bigl(x_2^b\bigr)\bigr)\longmapsto\bigl(x_1^a+x_2^a\bigr)\,,
\qqq
with the commutative Lie algebra
\qq\nn
\gt{mink}(d,1)=\bigoplus_{a=0}^d\,\corr{P_a}\,,\qquad\qquad[P_a,P_b]=0
\qqq
and the associated left-invariant (LI) Maurer--Cartan form 
\qq\nn
E\equiv E^a\ox P_a\,,\qquad\qquad E^a(x)=\sfd x^a\,.
\qqq
To the above,\ there corresponds the Clifford algebra
\qq\nn
\Cliff\bigl(\bR^{d,1}\bigr)=\corr{\ \G_a\ \vert\ a\in\ovl{0,d}\ }\,,\qquad\qquad\{\G_a,\G_b\}=2\eta_{ab}\,\bd1
\qqq
that contains the spin group $\,{\rm Spin}(d,1)$,\ the universal cover of the connected component $\,{\rm SO}_0(d,1)\,$ of the identity element of the Lorentz group $\,{\rm SO}(d,1)\equiv{\rm SO}(\bR^{d,1})\,$ of $\,\bR^{d,1}$,
\qq\nn
\bd1\too\bZ/2\bZ\too{\rm Spin}(d,1)\xrightarrow{\ \pi_{\rm Spin}\ }{\rm SO}_0(d,1)\too\bd1\,.
\qqq
We pick up a vector space
\qq\nn
S_{d,1}\cong\bR^{\x D_{d,1}}
\qqq
that carries a Majorana-spinor realisation
\qq\nn
S\ :\ {\rm Spin}(d,1)\too{\rm End}(S_{d,1})
\qqq
of $\,{\rm Spin}(d,1)$,\ assuming the following identities to be satisfied in this realisation:\ the Fierz identities
\qq\label{eq:Fierz}\qquad\qquad
\G^a_{\a(\b}\,\G_{a\,\g\d)}=0\,,\qquad\qquad\qquad\a,\b,\g,\d\in\ovl{1,D_{d,1}}\,,\qquad\qquad\qquad\G^a\equiv\eta^{-1\,ab}\,\G_b\,,
\qqq
and -- for the corresponding charge-conjugation operator $\,C\in{\rm End}(S_{d,1})\,$ -- the symmetry relations
\qq\nn
C^{\rm T}=-C\,,\qquad\qquad\bigl(C\,\G_a\bigr)^{\rm T}=C\,\G_a\equiv\ovl\G{}_a
\qqq
which,\ in particular,\ rule out the possibilities $\,d\in\{5,6,7\}$.

Given these,\ we consider the associated super-Poincar\'e (super)group,\ that is the supermanifold
\qq\nn
{\rm sISO}(d,1|D_{d,1})=\bigl(\widetilde{{\rm ISO}}(d,1)\equiv\bR^{\x d+1}\rx_L{\rm Spin}(d,1),\cO_{{\rm sISO}(d,1|D_{d,1})}\equiv C^\infty(\cdot,\bR)\circ\pr_1\ox C^\infty(\cdot,\bR)\circ\pr_2\ox\bigwedge\hspace{-3pt}{}^\bullet\,S_{d,1}\bigr)
\qqq 
with the crossed product in the definition of its body $\,\widetilde{{\rm ISO}}(d,1)\,$ determined by the vector realisation 
\qq\nn
\alxydim{@C=1cm@R=1cm}{L & :\ & {\rm Spin}(d,1) \ar[rr] \ar[dr]_{\pi_{\rm Spin}} & & \End\bigl(\bR^{d,1}\bigr) \\ & & & {\rm SO}_0(d,1) \ar@{^{(}->}[ur] & }
\qqq
of the spin group,\ and the structure sheaf $\,\cO_{{\rm sISO}(d,1|D_{d,1})}\,$ written in terms of the structure sheaf $\,C^\infty(\cdot,\bR)\circ\pr_1\ox C^\infty(\cdot,\bR)\circ\pr_2\equiv\cO_{\widetilde{{\rm ISO}}(d,1)}\,$ of the body.\ The binary operation 
\qq\nn
\txm\ :\ {\rm sISO}(d,1|D_{d,1})\x{\rm sISO}(d,1|D_{d,1})\too{\rm sISO}(d,1|D_{d,1})
\qqq
of the Lie-supergroup structure on $\,{\rm sISO}(d,1|D_{d,1})\,$ is customarily described,\ in the $\cS$-point picture,\ in terms of the anticommuting generators $\,\theta^\a,\ \a\in\ovl{1,D_{d,1}}\,$ of $\,\bigwedge\hspace{-3pt}{}^\bullet\,S_{d,1}$,\ the global (cartesian-coordinate) generators $\,x^a,\ a\in\ovl{0,d}\,$ of the structure sheaf of $\,\bR^{\x d+1}$,\ and local (Lie-algebra) coordinates $\,\phi^{ab}\equiv\phi^{[ab]}\,$ on $\,{\rm Spin}(d,1)\,$ (their ensemble being identified with a point in the spin group by a mild abuse of the notation) as
\qq
\txm\bigl(\bigl(\theta_1^\a,x_1^a,\phi_1^{bc}\bigr),\bigl(\theta_2^\a,x_2^a,\phi_2^{bc}\bigr)\bigr)=\bigl(\theta_1^\a+S(\phi_1)^\a_{\ \b}\,\theta_2^\b,x_1^a+L(\phi_1)^a_{\ b}\,x_2^b-\tfrac{1}{2}\,\theta_1\,\ovl\G{}^a\,S(\phi_1)\,\theta_2,\bigl(\phi_1\star\phi_2\bigr)^{ab}\bigr)\,,\cr\label{eq:sISOgl}
\qqq
where 
\qq\nn
\theta_1\,\ovl\G{}^a\,S(\phi_1)\,\theta_2\equiv\theta_1^\a\,C_{\a\b}\,\G{}^{a\,\b}_{\ \ \ \g}\,S(\phi_1)^\g_{\ \d}\,\theta_2^\d
\qqq
and where $\,\star\,$ represents the standard binary operation on the spin group.\ In this picture,\ it is straightforward to write out coordinate expressions for the basis LI vector fields on the Lie supergroup $\,{\rm sISO}(d,1|D_{d,1})$,
\qq\nn
Q_\a(\theta,x,\phi)&=&S(\phi)^\b_{\ \a}\,\bigl(\tfrac{\vec\p\ }{\p\theta^\b}+\tfrac{1}{2}\,\theta^\g\,C_{\g\d}\,\G^{a\,\d}_{\ \ \ \b}\,\tfrac{\p\ }{\p x^a}\bigr)=:S(\phi)^\b_{\ \a}\,\unl Q{}_\b(\theta,x)\,,\cr\cr 
P_a(\theta,x,\phi)&=&L(\phi)^b_{\ a}\,\tfrac{\p\ }{\p x^b}=:L(\phi)^b_{\ a}\,\unl P{}_b(\theta,x)\,,\cr\cr 
J_{ab}(\theta,x,\phi)&=&\tfrac{\sfd\ }{\sfd t}\rstr_{t=0}\phi\star t\phi_{ab}\,,\qquad\qquad\bigl(\phi_{ab}\bigr)^{cd}=\d_a^{\ c}\,\d_b^{\ d}-\d_a^{\ d}\,\d_b^{\ c}\,,
\qqq
spanning the tangent sheaf $\,\cT{\rm sISO}(d,1|D_{d,1})\,$ of $\,{\rm sISO}(d,1|D_{d,1})$.\ These obey the superalgebra
\qq\nn
&\{Q_\a,Q_\b\}=\ovl\G{}_{\a\b}^a\,P_a\,,\qquad\qquad[P_a,P_b]=0\,,\qquad\qquad[Q_\a,P_a]=0\,,&\cr\cr
&[J_{ab},Q_\a]=\tfrac{1}{2}\,\bigl(Q\,\G_{ab}\bigr){}_\a=\tfrac{1}{2}\,\G_{ab}{}^\b_{\ \a}\,Q_\b\,,\qquad\qquad[J_{ab},P_c]=\eta_{bc}\,P_a-\eta_{ac}\,P_b\,,&\cr\cr
&[J_{ab},J_{cd}]=\eta_{ad}\,J_{bc}-\eta_{ac}\,J_{bd}+\eta_{bc}\,J_{ad}-\eta_{bd}\,J_{ac}\,,&
\qqq
expressed in terms of the antisymmetric products
\qq\nn
\G_{ab}=\tfrac{1}{2}\,[\G_a,\G_b]
\qqq
and called the super-Poincar\'e (super)algebra and denoted as ($D=\dim\,\gt{siso}(d,1|D_{d,1})-1$)
\qq\nn
\gt{siso}(d,1|D_{d,1})=\bigoplus_{\a=1}^{D_{d,1}}\,\corr{Q_\a}\oplus\bigoplus_{a=0}^d\,\corr{P_a}\oplus\bigoplus_{a<b=0}^d\,\corr{J_{ab}=-J_{ba}}\equiv\bigoplus_{A=0}^D\,\corr{t_A}\,.
\qqq
When referring symbolically to its (supercommutation) structure equations,\ we shall write (for homogeneous generators $\,t_A,\ A\in\ovl{0,D}\,$ of the respective Gra\ss mann parities $\,|t_A|\equiv|A|$)
\qq\nn
[t_A,t_B\}=f_{AB}^{\ \ \ C}\,t_C=(-1)^{|A|\,|B|+1}\,[t_B,t_A\}\,.
\qqq
The latter superalgebra is the key ingredient in the alternative (and equivalent) definition of the Lie supergroup $\,{\rm sISO}(d,1|D_{d,1})\,$ \`a la Kostant that identifies the supergroup with the super-Harish--Chandra pair
\qq\nn
{\rm sISO}(d,1|D_{d,1})\equiv\bigl(\widetilde{{\rm ISO}}(d,1),\gt{siso}(d,1|D_{d,1})\bigr)\,,
\qqq
with the body Lie group $\,\widetilde{{\rm ISO}}(d,1)\,$ realised on the Gra\ss mann-odd component 
\qq\nn
\gt{siso}(d,1|D_{d,1})^{(1)}\equiv\bigoplus_{\a=1}^{D_{d,1}}\,\corr{Q_\a}
\qqq 
of the Lie superalgebra $\,\gt{siso}(d,1|D_{d,1})\,$ as 
\qq\nn
\rho\ :\ \bR^{\x d+1}\rx_L{\rm Spin}(d,1)\too{\rm End}\bigl(\gt{siso}(d,1|D_{d,1})^{(1)}\bigr)\ :\ (x,\phi)\longmapsto S(\phi)^{\rm T}\equiv\rho(x,\phi)\,.
\qqq

The cotangent sheaf $\,\cT^*{\rm sISO}(d,1|D_{d,1})\,$ of $\,{\rm sISO}(d,1|D_{d,1})$,\ dual to $\,\cT{\rm sISO}(d,1|D_{d,1})\,$ (as a $\,\cO_{{\rm sISO}(d,1|D_{d,1})}$-module),\ is globally generated by the duals of the vector fields $\,Q_\a,\ P_a,\ J_{bc}$,\ {\it i.e.},\ the LI super-1-forms with the coordinate presentation
\qq\nn
q^\a(\theta,x,\phi)&=&S(\phi)^{-1\,\a}_{\ \ \ \ \b}\,\sfd\theta^\b=:S(\phi)^{-1\,\a}_{\ \ \ \ \b}\,\unl q{}^\b(\theta,x)\,,\cr\cr
p^a(\theta,x,\phi)&=&L(\phi)^{-1\,a}_{\ \ \ \ b}\,\bigl(\sfd x^b+\tfrac{1}{2}\,\theta\,\ovl\G{}^b\,\sfd\theta\bigr)=:L(\phi)^{-1\,a}_{\ \ \ \ b}\,\unl p{}^b(\theta,x)\,,\cr\cr
j^{ab}(\theta,x,\phi)&=&L(\phi)^{-1\,a}_{\ \ \ \ c}\,\sfd L(\phi)^c_{\ d}\,\eta^{-1\,db}\,.
\qqq
Through the ensuing super-Maurer--Cartan equations
\qq
\sfd q^\a=-\tfrac{1}{4}\,j^{ab}\wedge\bigl(\G_{ab}\,q\bigr)^\a\,,\qquad\qquad\sfd p^a=\tfrac{1}{2}\,q\wedge\ovl\G{}^a\,q-\eta_{bc}\,j^{ab}\wedge p^c\,,\qquad\qquad\sfd j^{ab}=-\eta_{cd}\,j^{ac}\wedge j^{bd}\,,\cr\label{eq:sMCeqs}
\qqq
they generate the Cartan--Eilenberg cochain complex of $\,{\rm sISO}(d,1|D_{d,1})$,
\qq\nn
\bigl(\Om^\bullet\bigl({\rm sISO}(d,1|D_{d,1})\bigr)^{{\rm sISO}(d,1|D_{d,1})}\equiv\corr{\ q^\a,p^a,j^{bc}\ \vert\ (\a,a)\in\ovl{1,D_{d,1}}\x\ovl{0,d}\,,\ b<c\in\ovl{0,d}\ },\sfd^\bullet\equiv\sfd\bigr)\,.
\qqq
Its cohomology,\ 
\qq\nn
H_{\rm dR}^\bullet\bigl({\rm sISO}(d,1|D_{d,1}),\bR\bigr)^{{\rm sISO}(d,1|D_{d,1})}\equiv{\rm CaE}^\bullet\bigl({\rm sISO}(d,1|D_{d,1})\bigr)\,,
\qqq
the supersymmetric refinement of the de Rham cohomology of $\,{\rm sISO}(d,1|D_{d,1})$,\ is termed the Cartan--Eilenberg cohomology of $\,{\rm sISO}(d,1|D_{d,1})$.\ By the $\bZ/2\bZ$-graded version of the classic Lie-algebraic result,\ it is isomorphic with the Chevalley--Eilenberg cohomology of the Lie superalgebra $\,\gt{siso}(d,1|D_{d,1})\,$ with values in the trivial $\gt{siso}(d,1|D_{d,1})$-module $\,\bR$,
\qq\nn
{\rm CaE}^\bullet\bigl({\rm sISO}(d,1|D_{d,1})\bigr)\cong H^\bullet\bigl(\gt{siso}(d,1|D_{d,1}),\bR\bigr)\,,
\qqq
a fact of prime significance for the geometrisation of physically relevant Cartan--Eilenberg (super-)cocyles discussed in Sec.\,\ref{sec:geometrise}.\ Among nontrivial classes in $\,{\rm CaE}^\bullet({\rm sISO}(d,1|D_{d,1}))$,\ we find that of the the Green--Schwarz super-3-cocycle
\qq\label{eq:GS3coc}
\underset{\tx{\ciut{(3)}}}{\chi}=q\wedge\ovl\G{}_a\,q\wedge p^a
\qqq
whose closedness follows directly from the Fierz identities \eqref{eq:Fierz}.

Consider,\ next,\ the two homogeneous spaces of the super-Poincar\'e group associated with the respective reductive decompositions of its tangent Lie superalgebra:
\qq\nn
&\gt{siso}(d,1|D_{d,1})=\gt{smink}(d,1|D_{d,1})\oplus\gt{spin}(d,1)\,,&\cr\cr
&[\gt{spin}(d,1),\gt{smink}(d,1|D_{d,1})]\subset\gt{smink}(d,1|D_{d,1})\,,&
\qqq
with
\qq\nn
\gt{smink}(d,1|D_{d,1})=\bigoplus_{\a=1}^{D_{d,1}}\,\corr{Q_\a}\oplus\bigoplus_{a=0}^d\,\corr{P_a}\,,\qquad\qquad\gt{spin}(d,1)=\bigoplus_{a<b=0}^d\,\corr{J_{ab}}\,,
\qqq
and
\qq\nn
&\gt{siso}(d,1|D_{d,1})=\bigl(\gt{smink}(d,1|D_{d,1})\oplus\dgt\bigr)\oplus\gt{spin}(d,1)_{\rm vac}\,,&\cr\cr
&[\gt{spin}(d,1)_{\rm vac},\gt{smink}(d,1|D_{d,1})\oplus\dgt]\subset\gt{smink}(d,1|D_{d,1})\oplus\dgt\,,&
\qqq
with
\qq\nn
\dgt=\bigoplus_{(\unl a,\widehat b)\in\{0,1\}\x\ovl{2,d}}\,\corr{J_{\unl a\widehat b}}\,,\qquad\qquad\gt{spin}(d,1)_{\rm vac}=\corr{J_{01}}\oplus\bigoplus_{\widehat a<\widehat b\in\ovl{2,d}}\,\corr{J_{\widehat a\widehat b}}\equiv\gt{spin}(1,1)\oplus\gt{spin}(d-1)\,,
\qqq
coming with the supervector-space projections
\qq\nn
p\equiv\pr_1\ &:&\ \gt{smink}(d,1|D_{d,1})\oplus\gt{spin}(d,1)\too\gt{smink}(d,1|D_{d,1})\,,\cr\cr
p_{\rm vac}\equiv\pr_1\ &:&\ \bigl(\gt{smink}(d,1|D_{d,1})\oplus\dgt\bigr)\oplus\gt{spin}(d,1)_{\rm vac}\too\gt{smink}(d,1|D_{d,1})\oplus\dgt\,.
\qqq
The former is the super-Minkowski space
\qq\nn
{\rm sMink}(d,1|D_{d,1})={\rm sISO}(d,1|D_{d,1})/{\rm Spin}(d,1)\,,
\qqq
and the latter is
\qq\nn
{\rm sISO}(d,1|D_{d,1})/{\rm Spin}(d,1)_{\rm vac}\equiv{\rm sMink}(d,1|D_{d,1})\x{\rm Spin}(d,1)/{\rm Spin}(d,1)_{\rm vac}\,,
\qqq
where
\qq\nn
{\rm Spin}(d,1)_{\rm vac}\equiv{\rm Spin}(1,1)\x{\rm Spin}(d-1)\,.
\qqq
They are bases of the respective principal (super-)bundles
\qq\label{diag:princsbndlsMink}
\alxydim{@C=1.5cm@R=1.5cm}{ {\rm Spin}(d,1) \ar[r] & {\rm sISO}(d,1|D_{d,1})\equiv{\rm sMink}(d,1|D_{d,1})\rx_{L,S}{\rm Spin}(d,1) \ar[d]^{\pi\equiv\pr_1} \\ & {\rm sMink}(d,1|D_{d,1})}
\qqq
and
\qq\label{diag:princsbndlvac}
\alxydim{@C=1.5cm@R=1.5cm}{ {\rm Spin}(d,1)_{\rm vac} \ar[r] & {\rm sISO}(d,1|D_{d,1}) \ar[d]^{\pi_{\rm vac}} \\ & {\rm sISO}(d,1|D_{d,1})/{\rm Spin}(d,1)_{\rm vac}}\,,
\qqq
on which the Lie supergroup acts (from the left) as
\qq\nn
[\ell]^\txK\ :\ {\rm sISO}(d,1|D_{d,1})\x{\rm sISO}(d,1|D_{d,1})/\txK\too{\rm sISO}(d,1|D_{d,1})/\txK
\qqq
in such a manner that
\qq\nn
[\ell]^\txK\circ\bigl(\id_{{\rm sISO}(d,1|D_{d,1})}\x\pi_\txK\bigr)=\pi_\txK\circ\ell\,,\qquad\qquad\ell\equiv\txm\,,
\qqq
where $\,\txK\in\{{\rm Spin}(d,1),{\rm Spin}(d,1)_{\rm vac}\}\,$ and $\,(\pi_{{\rm Spin}(d,1)},\pi_{{\rm Spin}(d,1)_{\rm vac}})=(\pi,\pi_{\rm vac})$.\ This is a simple example of the general situation described at length in \Rcite{Kostant:1975} and,\ more recently,\ in \Rcite{Fioresi:2007zz}.\ The Lie supergroup $\,{\rm sISO}(d,1|D_{d,1})\,$ acquires the interpretation of the {\bf supersymmetry group} in this context,\ and its tangent Lie superalgebra becomes the {\bf supersymmetry algebra}.\ It acts on itself also from the \emph{right},
\qq\nn
\wp\equiv\txm\ :\ {\rm sISO}(d,1|D_{d,1})\x{\rm sISO}(d,1|D_{d,1})\too{\rm sISO}(d,1|D_{d,1})
\qqq
but the latter action does not descend to the homogeneous space. 

The existence of the surjective submersions (in $\,\sMan$) $\,\pi_\txK\,$ can also be used to descend LI tensors from the mother Lie supergroup $\,{\rm sISO}(d,1|D_{d,1})\,$ to the homogeneous space $\,{\rm sISO}(d,1|D_{d,1})/\txK\,$ along local trivialising sections,\ an observation behind the long-established model-building technique of the so-called nonlinear realisations of (super)symmetry of Refs.\,\cite{Schwinger:1967tc,Weinberg:1968de,Coleman:1969sm,Callan:1969sn,Salam:1969rq,Salam:1970qk,Isham:1971dv,Volkov:1972jx,Volkov:1973ix,Ivanov:1978mx,Lindstrom:1979kq,Uematsu:1981rj,Ivanov:1982bpa,Samuel:1982uh,Ferrara:1983fi,Bagger:1983mv,West:2000hr,Gomis:2006wu,McArthur:1999dy,McArthur:2010zm}.\ In the case of a covariant tensor $\,T\,$ (of rank $n$),\ for the descent to yield a globally defined object on the homogeneous space,\ $\,T\,$ must be (right-)$\txK$-basic.\ Denote the tangent Lie algebra of the structure group $\,\txK\,$ as $\,\kgt\in\{\gt{spin}(d,1),\gt{spin}(d,1)_{\rm vac}\}$,\ with the understanding that $\,\kgt\subset\mgt\oplus\kgt=\gt{siso}(d,1|D_{d,1})\,$ (where $\,\mgt\in\{\gt{smink}(d,1|D_{d,1}),\gt{smink}(d,1|D_{d,1})\oplus\dgt\}\,$ is the formerly indicated direct-sum complement of $\,\kgt$,\ such that $\,[\kgt,\mgt]\subset\mgt$),\ and consider the induced \emph{element}wise realisation of the body Lie group $\,\widetilde{{\rm ISO}}(d,1)\,$ on $\,{\rm sISO}(d,1|D_{d,1})\,$ by automorphisms in the category $\,\sMan\,$ of supermanifolds,\ given by
\qq\nn
|\wp|_\cdot\ :\ \widetilde{{\rm ISO}}(d,1)\too{\rm Aut}_{\sMan}\bigl({\rm sISO}(d,1|D_{d,1})\bigr)\ :\ g\longmapsto\wp\circ\bigl(\id_{{\rm sISO}(d,1|D_{d,1})}\x\widehat g\bigr)\equiv|\wp|_g\,,
\qqq
where 
\qq\nn
\widehat g\in\mor_\sMan\bigl(\bR^{0|0},{\rm sISO}(d,1|D_{d,1})\bigr)
\qqq
is the topological point in $\,{\rm sISO}(d,1|D_{d,1})\,$ corresponding to $\,g\in\widetilde{{\rm ISO}}(d,1)$,\ whence
\qq\nn
|\wp|_g\ &:&\ {\rm sISO}(d,1|D_{d,1})\equiv{\rm sISO}(d,1|D_{d,1})\x\bR^{0|0}\xrightarrow{\ \id_{{\rm sISO}(d,1|D_{d,1})}\x\widehat g\ }{\rm sISO}(d,1|D_{d,1})\x{\rm sISO}(d,1|D_{d,1})\cr\cr
&&\xrightarrow{\ \txm\ }{\rm sISO}(d,1|D_{d,1})\,,
\qqq
as desired.\ With these in hand,\ we can make the concept of $\txK$-basicness precise.\ Thus,\ a covariant tensor $\,T\,$ on $\,{\rm sISO}(d,1|D_{d,1})\,$ is (right-)$\txK$-basic if it is $\kgt$-horizontal,
\qq\nn
\forall_{(X_1,X_2,\ldots,X_n)\in\gt{siso}(d,1|D_{d,1})^{\x n}}\ :\ \bigl(\ \exists_{i\in\ovl{1,n}}\ :\ X_i\in\kgt\quad \Longrightarrow\quad T(X_1,X_2,\ldots,X_n)=0\ \bigr)\,,
\qqq
and (right-)$\txK$-invariant,
\qq\nn
\forall_{(k,X)\in\txK\x\kgt}\ :\ \bigl(\ |\wp|_k^*T=T\quad\land\quad\pLie{X}T=0\ \bigr)\,.
\qqq
It is now easy to see that the distinguished super-3-cocycle \eqref{eq:GS3coc} is ${\rm Spin}(d,1)$-basic,\ and so also ${\rm Spin}(d,1)_{\rm vac}$-basic,\ as is the degenerate metric tensor
\qq\nn
\widehat\eta=\eta_{ab}\,p^a\ox p^b\,.
\qqq
Indeed,\ the LI super-1-forms $\,\theta_{\rm L}^\z,\ \z\in\ovl{0,\dim\,\mgt-1}\,$ dual to the basis LI vector fields from $\,\mgt\,$ are $\kgt$-horizontal by definition and transform linearly as 
\qq\nn
|\wp|_{k^{-1}}^*\theta_{\rm L}^\z=\rho(k)^\z_{\ \z'}\,\theta_{\rm L}^{\z'}
\qqq
in consequence of the assumed reductivity of the decomposition $\,\gt{siso}(d,1|D_{d,1})=\mgt\oplus\kgt$.\ Hence,\ it suffices to take a linear combination of their tensor products,\ 
\qq\nn
T=\la_{\z_1\z_2\ldots\z_n}\,\theta_{\rm L}^{\z_1}\ox\theta_{\rm L}^{\z_2}\ox\cdots\ox\theta_{\rm L}^{\z_n}\,,
\qqq
with components $\,\la_{\z_1\z_2\ldots\z_n}\,$ of a constant $\rho(\txK)$-invariant tensor as coefficients,
\qq\nn
\forall_{k\in\txK}\ :\ \la_{\z_1\z_2\ldots\z_n}\,\rho(k)^{\z_1}_{\ \z_1'}\,\rho(k)^{\z_2}_{\ \z_2'}\,\cdots\,\rho(k)^{\z_n}_{\ \z_n'}=\la_{\z_1'\z_2'\ldots\z_n'}\,,
\qqq 
to obtain a $\txK$-basic tensor $\,T$.\ The $\rho({\rm Spin}(d,1))$-invariance of the Minkowski metric is obvious,\ and that of $\,\ovl\G{}_{a\,\a\b}\,$ follows straightforwardly from the elementary properties of the generators of the Clifford algebra and of the charge-conjugation operator:
\qq\label{eq:covGam}
S(\phi)^{-1}\,\G^a\,S(\phi)=L(\phi)^a_{\ b}\,\G^b\,,\qquad\qquad C^{-1}\,S(\phi)^{\rm T}\,C=S(\phi)^{-1}\,.
\qqq
An example of a ${\rm Spin}(d,1)_{\rm vac}$-basic tensor that is \emph{not} ${\rm Spin}(d,1)$-basic is provided by the volume super-2-form on the subspace 
\qq\nn
\tgt^{(0)}_{\rm vac}:=\corr{P_0,P_1}\,,
\qqq
that is ($\ep_{\unl a\unl b}\,$ is the totally antisymmetric tensor with $\,\ep_{01}=1$)
\qq\nn
\Vol\bigl(\tgt^{(0)}_{\rm vac}\bigr)=p^0\wedge p^1\equiv\tfrac{1}{2}\,\ep_{\unl a\unl b}\,p^{\unl a}\wedge p^{\unl b}\,.
\qqq
Its (right-)${\rm Spin}(d,1)_{\rm vac}$-invariance is ensured by the fact that $\,\rho({\rm Spin}(d,1)_{\rm vac})\,$ restrict to $\,\tgt^{(0)}_{\rm vac}\,$ as unimodular automorphisms,
\qq\nn
\forall_{\phi\equiv(\phi_1,\phi_2)\in{\rm Spin}(1,1)\x{\rm Spin}(d-1)}\ :\ \det\,\bigl(\rho(\phi)\rstr_{\tgt^{(0)}_{\rm vac}}\bigr)\equiv\det\,\bigl(L(\phi)\rstr_{\tgt^{(0)}_{\rm vac}}\bigr)\equiv\det\,L(\phi_1)=1\,.
\qqq

Hence,\ in particular,\ there exist:\ a super-3-form $\,\underset{\tx{\ciut{(3)}}}{\txH}\,$ and a symmetric $(2,0)$-tensor $\,\widehat{\unl\eta}\,$ on $\,{\rm sMink}(d,1|D_{d,1})$,\ as well as a super-3-form $\,\widetilde{\underset{\tx{\ciut{(3)}}}{\txH}}$,\ a super-2-form $\,\underset{\tx{\ciut{(2)}}}{\upsilon}\,$ and a symmetric $(2,0)$-tensor $\,\widehat{\unl{\widetilde\eta}}\,$ on $\,{\rm sISO}(d,1|D_{d,1})/{\rm Spin}(d,1)_{\rm vac}\,$ such that
\qq\nn
&\underset{\tx{\ciut{(3)}}}{\chi}=\pi^*\underset{\tx{\ciut{(3)}}}{\txH}\,,\qquad\qquad\widehat\eta=\pi^*\widehat{\unl\eta}\,,&\cr\cr
&\underset{\tx{\ciut{(3)}}}{\chi}=\pi_{\rm vac}^*\widetilde{\underset{\tx{\ciut{(3)}}}{\txH}}\,,\qquad\qquad\widehat\eta=\pi_{\rm vac}^*\widehat{\unl{\widetilde\eta}}\,,\qquad\qquad\Vol\bigl(\tgt^{(0)}_{\rm vac}\bigr)=\pi_{\rm vac}^*\underset{\tx{\ciut{(2)}}}{\upsilon}\,.
\qqq
Upon putting together the explicit coordinate presentations of the various LI super-1-forms involved and identities \eqref{eq:covGam},\ we readily derive
\qq\nn
\underset{\tx{\ciut{(3)}}}{\txH}=\unl q\wedge\ovl\G{}_a\,\unl q\wedge\unl p^a\,,\qquad\qquad\widehat{\unl\eta}=\eta_{ab}\,\unl p^a\ox\unl p^b\,.
\qqq

Inspection of the group law \eqref{eq:sISOgl} reveals that the homogeneous space $\,{\rm sMink}(d,1|D_{d,1})\subset{\rm sISO}(d,1|D_{d,1})\,$ is,\ in fact,\ a Lie sub-supergroup with the binary operation
\qq\nn
\unl\txm\ :\ {\rm sMink}(d,1|D_{d,1})\x{\rm sMink}(d,1|D_{d,1})\too{\rm sMink}(d,1|D_{d,1})
\qqq
admitting the coordinate presentation 
\qq\nn
\unl\txm\bigl(\bigl(\theta_1^\a,x_1^a\bigr),\bigl(\theta_2^\a,x_2^a\bigr)\bigr)=\bigl(\theta_1^\a+\theta_2^\a,x_1^a+x_2^a-\tfrac{1}{2}\,\theta_1\,\ovl\G{}^a\,\theta_2\bigr)\,,
\qqq
with the corresponding basis LI vector fields
\qq\nn
\unl Q{}_\a(\theta,x)=\tfrac{\vec\p\ }{\p\theta^\a}+\tfrac{1}{2}\,\theta^\b\,C_{\b\g}\,\G^{a\,\g}_{\ \ \ \a}\,\tfrac{\p\ }{\p x^a}\,,\qquad\qquad\unl P{}_a(\theta,x)=\tfrac{\p\ }{\p x^a}
\qqq
spanning the super-minkowskian Lie superalgebra 
\qq\nn
\gt{smink}(d,1|D_{d,1})=\bigoplus_{\a=1}^{D_{d,1}}\,\corr{\unl Q{}_\a}\oplus\bigoplus_{a=0}^d\,\corr{\unl P{}_a}
\qqq
with the structure equations
\qq\nn
\{\unl Q{}_\a,\unl Q{}_\b\}=\ovl\G{}_{\a\b} ^a\,\unl P{}_a\,,\qquad\qquad[\unl P{}_a,\unl P{}_b]=0\,,\qquad\qquad[\unl Q{}_\a,\unl P{}_a]=0\,.
\qqq
Clearly,\ the super-1-forms $\,\unl q{}^\a\,$ and $\,\unl p{}^a\,$ are their (respective) duals,\ and the descended Green--Schwarz super-3-cocycle $\,\underset{\tx{\ciut{(3)}}}{\txH}\,$ defines a nontrivial class in 
\qq\nn
{\rm CaE}^\bullet\bigl({\rm sMink}(d,1|D_{d,1})\bigr)\cong H^\bullet\bigl(\gt{smink}(d,1|D_{d,1}),\bR\bigr)\,.
\qqq
As the de Rham cohomology of $\,{\rm sMink}(d,1|D_{d,1})\,$ is trivial,
\qq\nn
H^\bullet_{\rm dR}\bigl({\rm sMink}(d,1|D_{d,1})\bigr)=H^\bullet_{\rm dR}\bigl({\rm Mink}(d,1)\bigr)=\brd0\,,
\qqq
the Green--Schwarz super-3-cocycle admits a global primitive,\ albeit only a quasi-invariant one that can be chosen in the explicit form
\qq\nn
\underset{\tx{\ciut{(2)}}}{\txB}(\theta,x)=\theta\,\ovl\G{}_a\,\unl q(\theta,x)\wedge\unl p{}^a(\theta,x)\,.
\qqq
We shall write
\qq\nn
\underset{\tx{\ciut{(2)}}}{\b}:=\pi^*\underset{\tx{\ciut{(2)}}}{\txB}\,.
\qqq
~\medskip

The hitherto discussion provides us with all the ingredients of the two formulations of the {\bf Green--Schwarz (GS) super-$\si$-model of the superstring in $\,{\rm sMink}(d,1|D_{d,1})$}.\ The first of these is the {\bf Nambu--Goto} (NG) {\bf formulation} in which we have the theory of (inner-$\mor$) functorial embeddings 
\qq\nn
\xi\in[\Si,{\rm sMink}(d,1|D_{d,1})]\equiv\mor_\sMan\bigl(\Si\x-,{\rm sMink}(d,1|D_{d,1})\bigr)
\qqq
of a closed orientable two-dimensional manifold $\,\Si\,$ (the worldsheet) in $\,{\rm sMink}(d,1|D_{d,1})\,$ determined by the principle of least action applied to the Dirac--Feynman (DF) amplitude (with an obvious interpretation of the codomain)
\qq\nn
\cA_{\rm DF}^{\rm (NG)}\ :\ [\Si,{\rm sMink}(d,1|D_{d,1})]\too\uj\ :\ \xi\longmapsto\exp\left[\tfrac{\sfi}{\hbar}\,\left(\mu_1\,\int_\Si\,\sqrt{\det\,\bigl(\xi^*\widehat{\unl\eta}\bigr)}+\int_\Si\,\xi^*\underset{\tx{\ciut{(2)}}}{\txB}\right)\right]\,,
\qqq
in which $\,\mu_1\in\bR^\x\,$ is a parameter whose numerical value is fixed by the correspondence with the other formulation stated below,\ {\it cp} Refs.\,\cite{Green:1983sg,Green:1983wt}.\ The amplitude is to be evaluated on the Gra\ss mann-odd hyperplanes $\,\bR^{0|N},\ N\in\bN^\x$,\ whereby an $\bN^\x$-indexed family of supersymmetric two-dimensional field theories is obtained,\ {\it cp} \Rcite{Freed:1999}.\ The other one is the {\bf Hughes--Polchinski} (HP) {\bf formulation},\ first postulated in \cite{Hughes:1986dn},\ developed in Refs.\,\cite{Gauntlett:1989qe},\ applied in Refs.\,\cite{McArthur:1999dy,West:2000hr,Gomis:2006wu,Gomis:2006xw,McArthur:2010zm} and elaborated in \cite{Suszek:2019cum,Suszek:2020xcu},\ in which we deal with the theory of (inner-$\mor$) functorial embeddings 
\qq\nn
\widetilde\xi\in\bigl[\Si,{\rm sISO}(d,1|D_{d,1})/{\rm Spin}(d,1)_{\rm vac}\bigr]\equiv\mor_\sMan\bigl(\Si\x-,{\rm sISO}(d,1|D_{d,1})/{\rm Spin}(d,1)_{\rm vac}\bigr)
\qqq
of the same worldsheet $\,\Si\,$ in $\,{\rm sISO}(d,1|D_{d,1})/{\rm Spin}(d,1)_{\rm vac}\,$ determined by the principle of least action applied to the DF amplitude
\qq\nn
\cA_{\rm DF}^{\rm (HP)}\ :\ \bigl[\Si,{\rm sISO}(d,1|D_{d,1})/{\rm Spin}(d,1)_{\rm vac}\bigr]\too\uj\ :\ \widetilde\xi\longmapsto\exp\left[\tfrac{\sfi}{\hbar}\,\int_\Si\,\widetilde\xi{}^*\bigl(\la_1\,\underset{\tx{\ciut{(2)}}}{\upsilon}+p_1^*\underset{\tx{\ciut{(2)}}}{\txB}\bigr)\right]\,,
\qqq
in which 
\qq\nn
p_1\equiv\pr_1\ :\ {\rm sMink}(d,1|D_{d,1})\x\bigl({\rm Spin}(d,1)/{\rm Spin}(d,1)_{\rm vac}\bigr)\too{\rm sMink}(d,1|D_{d,1})
\qqq
and $\,\la_1\in\bR^\x\,$ is a parameter whose numerical value we establish through a symmetry analysis in Sec.\,\ref{sec:susy}.\ Here,\ it is presupposed that $\,\underset{\tx{\ciut{(2)}}}{\upsilon}\,$ is nondegenerate on classical field configurations (termed the vacua of the theory).\ In other words,\ we model the body of the vacuum on $\,\tgt^{(0)}_{\rm vac}$.\ In virtue of \Rxcite{Prop.\,5.3}{Suszek:2019cum} ({\it cp} also \Rxcite{Thm.\,3.4}{Suszek:2020xcu} for a more general result),\ the two formulations become equivalent upon partial reduction of the latter one through imposition of a subset of its Euler--Lagrange equations.\ This is quite nontrivial as the super-$\si$-model in the HP formulation is purely topological,\ unlike its NG counterpart.\ A careful analysis of the equivalence between the two formulations sets the stage for all our subsequent higher-geometric considerations,\ and so we shall now spend some time studying the relevant details in the spirit of \Rcite{Suszek:2020xcu}.\ In so doing,\ we shall refer to the two pairs:
\qq\nn
\gt{sB}^{{\rm (NG)}}_1=\bigl({\rm sMink}(d,1|D_{d,1}),\widehat\eta,\underset{\tx{\ciut{(3)}}}{\chi}\bigr)\,,
\qqq
and
\qq\nn 
\gt{sB}^{{\rm (HP)}}_{1,\la_1}=\bigl({\rm sISO}(d,1|D_{d,1})/{\rm Spin}(d,1)_{\rm vac},\la_1\,\sfd\Vol\bigl(\tgt^{(0)}_{\rm vac}\bigr)+\underset{\tx{\ciut{(3)}}}{\chi}\equiv\underset{\tx{\ciut{(3)}}}{\widehat\chi}{}^{(\la_1)}\bigr)
\qqq
as the {\bf NG superbackground} and {\bf HP superbackground},\ respectively.

The key to understanding the equivalence lies in a patchwise smooth realisation of the two homogeneous spaces:\ $\,{\rm sISO}(d,1|D_{d,1})/{\rm Spin}(d,1)_{\rm vac}\,$ and $\,{\rm sISO}(d,1|D_{d,1})/{\rm Spin}(d,1)\,$ within the mother Lie supergroup $\,{\rm sISO}(d,1|D_{d,1})\,$ by means of judiciously chosen local sections of the respective principal bundles \eqref{diag:princsbndlvac} and \eqref{diag:princsbndlsMink},\ along the lines of Refs.\,\cite{Fioresi:2007zz} and \cite{Suszek:2020xcu}.\ Let $\,O_0\ni e\,{\rm Spin}(d,1)_{\rm vac}\,$ be an open subset of $\,{\rm Spin}(d,1)/{\rm Spin}(d,1)_{\rm vac}\,$ that supports local (normal) coordinates $\,(\phi^{\unl a\widehat b})\ :\ O_0\xrightarrow{\ \cong\ }V_0\subset\dgt\,$ centred on the unital coset $\,e\,{\rm Spin}(d,1)_{\rm vac}\,$ ({\it i.e.},\ $\,\phi^{\unl a\widehat b}(e\,{\rm Spin}(d,1)_{\rm vac})=0$) and a local section of the principal ${\rm Spin}(d,1)_{\rm vac}$-bundle $\,{\rm Spin}(d,1)\too{\rm Spin}(d,1)/{\rm Spin}(d,1)_{\rm vac}\,$ (and so also its local trivialisation),\ and let $\,\{h_i\}_{i\in I}\subset{\rm Spin}(d,1),\ I\ni 0\,$ (with $\,h_0\equiv e$) be such that the translates $\,O_i=[|l|]_{h_i}(O_0)$,\ written in terms of the induced action 
\qq\nn
[l]\ :\ {\rm Spin}(d,1)\x{\rm Spin}(d,1)/{\rm Spin}(d,1)_{\rm vac}\too{\rm Spin}(d,1)/{\rm Spin}(d,1)_{\rm vac} 
\qqq
of the Lie group $\,{\rm Spin}(d,1)\,$ on the homogeneous space $\,{\rm Spin}(d,1)/{\rm Spin}(d,1)_{\rm vac}$,\ satisfying 
\qq\nn
[l]\circ\bigl(\id_{{\rm Spin}(d,1)}\x\pi_{{\rm Spin}(d,1)/{\rm Spin}(d,1)_{\rm vac}}\bigr)=\pi_{{\rm Spin}(d,1)/{\rm Spin}(d,1)_{\rm vac}}\circ l\,,\qquad\qquad l\equiv\star\,,
\qqq
cover $\,{\rm Spin}(d,1)/{\rm Spin}(d,1)_{\rm vac}$,\ 
\qq\nn
\bigcup_{i\in I}\,O_i={\rm Spin}(d,1)/{\rm Spin}(d,1)_{\rm vac}\,.
\qqq
The sets $\,U_i:={\rm Mink}(d,1)\x O_i,\ i\in I\,$ compose an open cover of the base $\,{\rm Mink}(d,1)\x{\rm Spin}(d,1)/{\rm Spin}(d,1)_{\rm vac}\,$ over which the principal ${\rm Spin}(d,1)_{\rm vac}$-bundle $\,\widetilde{{\rm ISO}}(d,1)\too{\rm Mink}(d,1)\x{\rm Spin}(d,1)/{\rm Spin}(d,1)_{\rm vac}\,$ trivialises.\ That base is the body of the base $\,{\rm sISO}(d,1|D_{d,1})/{\rm Spin}(d,1)_{\rm vac}\,$ of the principal ${\rm Spin}(d,1)_{\rm vac}$-bundle \eqref{diag:princsbndlvac},\ and it is over the $\,U_i\,$ that we define local sections of the latter.\ These we take in the (shifted-)exponential form\footnote{The sheaf-theoretic meaning of the $\,\si^{\rm vac}_i\,$ was given in \Rxcite{Sec.\,2}{Suszek:2020xcu}.} 
\qq
\si^{\rm vac}_i=|\ell|_{h_i}\circ\ee^{\theta^\a\circ\pr_1\ox Q_\a}\cdot\ee^{x^a\circ\pr_1\ox P_a}\cdot\ee^{\phi^{\unl a\widehat b}\circ\pr_2\ox J_{\unl a\widehat b}}\circ[|\ell|]_{h_i^{-1}}\ :\ \cU^{\rm vac}_i\equiv\bigl(U_i,\cO_{{\rm sISO}(d,1|D_{d,1})/{\rm Spin}(d,1)_{\rm vac}}\rstr_{U_i}\bigr)\cr\cr
\too{\rm sISO}(d,1|D_{d,1})\,, \label{eq:sivac}
\qqq
defined in terms of the left action
\qq\nn
|\ell|_\cdot\ :\ \widetilde{{\rm ISO}}(d,1)\too{\rm Aut}_{\sMan}\bigl({\rm sISO}(d,1|D_{d,1})\bigr)\ :\ g\longmapsto\txm\circ\bigl(\widehat g\x\id_{{\rm sISO}(d,1|D_{d,1})}\bigr)\equiv|\ell|_g
\qqq
and the corresponding induced action
\qq\nn
[|\ell|]_\cdot\ &:&\ \widetilde{{\rm ISO}}(d,1)\too{\rm Aut}_{\sMan}\bigl({\rm sISO}(d,1|D_{d,1})/{\rm Spin}(d,1)_{\rm vac}\bigr)\cr\cr 
&:&\ g\longmapsto[\ell]\circ\bigl(\widehat g\x\id_{{\rm sISO}(d,1|D_{d,1})/{\rm Spin}(d,1)_{\rm vac}}\bigr)\equiv[|\ell|]_g\,.
\qqq
They give rise to the {\bf Hughes--Polchinski section}
\qq\nn
\Si^{\rm HP}:=\bigsqcup_{i\in I}\,\cV_i\,,\qquad\qquad\cV_i:=\si^{\rm vac}_i\bigl(\cU_i^{\rm vac}\bigr)\,.
\qqq
Upon choosing a tessellation $\,\triangle_\Si\,$ of $\,\Si\,$ (composed of plaquettes that make up a set $\,\Tgt_2\subset\triangle_\Si$,\ edges and vertices) subordinate to $\,\{\cU^{\rm vac}_i\}_{i\in I}$,
\qq\nn
\forall_{\t\in\Tgt_2}\ \exists_{i_\t\in I}\ :\ |\widetilde\xi|(\t)\subset U_{i_\t}\,,
\qqq
we may,\ next,\ rewrite the DF amplitude of the above GS super-$\si$-model in the HP formulation as ($\widetilde\xi_\t\equiv\widetilde\xi\rstr_\t$)
\qq\nn
\cA_{\rm DF}^{\rm (HP)}[\widetilde\xi]=\exp\left[\tfrac{\sfi}{\hbar}\,\sum_{\t\in\Tgt_2}\,\int_\t\,\widetilde\xi{}_\t^*\si^{{\rm vac}*}_{i_\t}\bigl(\la_1\,\Vol\bigl(\tgt^{(0)}_{\rm vac}\bigr)+\underset{\tx{\ciut{(2)}}}{\b}\bigr)\right]\,.
\qqq
With the degrees of freedom of the super-$\si$-model (super)field thus separated into the super-minkowskian sector $\,(\theta^\a,x^a)\,$ and the spin-group sector $\,(\phi^{\unl a\widehat b}_i)\,$ (we use the subscript to mark the local coordinates on $\,O_i$),\ the equivalence is completely straightforward to state:\ Upon expressing the non-dynamical Goldstone spin-group fields $\,(\phi^{\unl a\widehat b}_i)\,$ in $\,\cA_{\rm DF}^{\rm (HP)}\,$ in terms of the remaining degrees of freedom $\,(\theta^\a,x^a)\,$ as dictated by the Euler--Lagrange equations of the GS super-$\si$-model in the HP formulation obtained by varying the DF amplitude for $\,\widetilde\xi\,$ in the direction of the spin-group fields $\,(\phi^{\unl a\widehat b}_i)$,\
\qq\label{eq:IHC}
\widetilde\xi{}_\t^*\si^{{\rm vac}*}_{i_\t}p^{\widehat a}=0\,,\qquad\widehat a\in\ovl{2,d}\,,
\qqq
we recover the DF amplitude of the NG formulation for $\,p_1\circ\widetilde\xi$,\ written in terms of the global coordinates $\,(\theta^\a,x^a)\,$ on $\,{\rm sMink}(d,1|D_{d,1})\,$ for a value $\,\mu_{1\,*}\equiv\mu_1(\la_1)\,$ of the parameter $\,\mu_1$ determined unquely by $\,\la_1\,$ -- this is the so-called {\bf Inverse Higgs Effect} of \Rcite{Ivanov:1975zq}.\ It permits us to restrict our subsequent discussion to the HP formulation,\ with the understanding that conclusions pertinent to the standard NG formulation of the super-$\si$-model can be drawn only upon restricting the tangents of the field configurations $\,\si^{\rm vac}_{i_\t}\circ\widetilde\xi\,$ to the {\bf NG/HP correspondence superdistribution}
\qq\nn
{\rm Corr}_{\rm HP}\bigl(\gt{sB}^{{\rm (HP)}}_{1,\la_1}\bigr)\subset\cT\Si^{\rm HP}\,,\qquad\qquad{\rm Corr}_{\rm HP}\bigl(\gt{sB}^{{\rm (HP)}}_{1,\la_1}\bigr)\rstr_{\cV_i}=\bigcap_{\widehat a=2}^d\,\ker\,p^{\widehat a}\cap\cT\cV_i\,.
\qqq
Advantages of this approach shall become clear along the way.

We conclude the present field-theoretic introduction by deriving the Euler--Lagrange equations of the super-$\si$-model in the HP formulation.\ To this end,\ we write the variation $\,\cV_\t\in[\t,\cT\si^{{\rm vac}}_{i_\t}(\cU^{\rm vac}_{i_\t})]\,$ of the composite embedding $\,\bX_\t\equiv\si^{{\rm vac}}_{i_\t}\circ\widetilde\xi{}_\t\,$ as
\qq\nn
\cV_\t=\d\theta^\a\,Q_\a(\bX_\t)+\d x^a\,P_a(\bX_\t)+\d\phi^{\unl a\widehat b}_{i_\t}\,J_{\unl a\widehat b}(\bX_\t)+\D^{\unl S}_{\t\,1}\,J_{\unl S}(\bX_\t)\,,
\qqq
in which the last term 
\qq\nn
\D^{\unl S}_{\t\,1}\,J_{\unl S}(\bX_\t)\equiv\D^{01}_{\t\,1}\,J_{01}(\bX_\t)+\D^{\widehat a\widehat b}_{\t\,2}\,J_{\widehat a\widehat b}(\bX_\t)
\qqq
represents $\gt{spin}(d,1)_{\rm vac}$-vertical corrections that render $\,\cV_\t\,$ tangent to the local section $\,\si^{{\rm vac}}_{i_\t}(\cU^{\rm vac}_{i_\t})$,\ {\it cp} \Rxcite{Prop.\,3.6}{Suszek:2020xcu},\ and calculate,\ with the help of the super-Maurer--Cartan equations \eqref{eq:sMCeqs},\ 
\qq\nn
&&-\sfi\,\hbar\,\cV_\cdot\con\d\log\,\cA_{\rm DF}^{\rm (HP)}[\widetilde\xi]=\sum_{\t\in\Tgt_2}\,\int_\t\,\cV_\t\con\widetilde\xi{}_\t^*\si^{{\rm vac}*}_{i_\t}\underset{\tx{\ciut{(3)}}}{\widehat\chi}{}^{(\la_1)}\cr\cr
&=&\sum_{\t\in\Tgt_2}\,\int_\t\,\cV_\t\con\widetilde\xi{}_\t^*\si^{{\rm vac}*}_{i_\t}\bigl(-\la_1\,\ep_{\unl a\unl b}\,\d_{\widehat c\widehat d}\,j^{\unl a\widehat c}\wedge p^{\widehat d}\wedge p^{\unl b}+q\wedge\ovl\G{}_{\widehat a}\,q\wedge p^{\widehat a}+2\eta_{\unl a\unl b}\,q\wedge\ovl\G{}^{\unl a}\,\bigl(\tfrac{\bd1_{D_{d,1}}-\tfrac{\la_1}{2}\,\G^0\,\G^1}{2}\bigr)\,q\wedge p^{\unl b}\bigr)\,,
\qqq
so that for $\,(\d\theta^\a,\d x^a)=0$,\ we obtain
\qq\nn
-\sfi\,\hbar\,\cV_\cdot\con\d\log\,\cA_{\rm DF}^{\rm (HP)}[\widetilde\xi]=-\sum_{\t\in\Tgt_2}\,\int_\t\,\la_1\,\ep_{\unl a\unl b}\,\d_{\widehat c\widehat d}\,\d\phi^{\unl a\widehat c}_{i_\t}\,\widetilde\xi{}_\t^*\si^{{\rm vac}*}_{i_\t}\bigl(p^{\widehat d}\wedge p^{\unl b}\bigr)\,.
\qqq
At this stage,\ we invoke the assumption of nondegeneracy of the volume form $\,\Vol(\tgt^{(0)}_{\rm vac})\,$ in the vacuum (which can be viewed as a condition of its partial localisation),\ implying that the tangent sheaf of the latter (in $\,{\rm sISO}(d,1|D_{d,1})$) is (locally) spanned on the vector fields 
\qq\nn
\cW_{\a\,i}=\D_{\a\,i}^\b\,Q_\b\,,\qquad\a\in\ovl{1,D_{d,1}}\,,\qquad\qquad\cW_{\unl a\,i}=P_{\unl a}\rstr_{\cV_i} +\D_{\unl a\,i}^{\widehat b}\,P_{\widehat b}+\D_{\unl a\,i}^{\unl b\widehat c}\,J_{\unl b\widehat c}+\D_{\unl a\,i}^{\unl S}\,J_{\unl S}\,,\qquad\unl a\in\{0,1\}\,,
\qqq
written in terms of certain (even) sections $\,\D_{\a\,i}^\b,\D_{\unl a\,i}^{\widehat b},\D_{\unl a\,i}^{\unl b\widehat c}\,$ and $\,\D_{\unl a\,i}^{\unl S}\,$ of the structure sheaf of $\,\cV_i\,$ to be determined below,\ with the $\gt{spin}(d,1)_{\rm vac}$-vertical component correcting the one along $\,\dgt\,$ in such a way that the sum is in $\,\cT\cV_i$,\ {\it cp} \Rxcite{Prop.\,3.6}{Suszek:2020xcu}.\ Taking the above into account,\ we obtain the formerly stated \Reqref{eq:IHC},\ or 
\qq\nn
\D_{\unl a\,i}^{\widehat b}=0\,,\qquad(\unl a,\widehat b)\in\{0,1\}\x\ovl{2,d}\,,
\qqq
which we write as
\qq\label{eq:IHCs}
p^{\widehat a}\approx 0\,,\qquad\widehat a\in\ovl{2,d}
\qqq
henceforth.\ Imposition of the latter leaves us with the reduced expression for the variation
\qq\nn
-\sfi\,\hbar\,\cV_\cdot\con\d\log\,\cA_{\rm DF}^{\rm (HP)}[\widetilde\xi]=2\eta_{\unl a\unl b}\,\sum_{\t\in\Tgt_2}\,\int_\t\,\cV_\t\con\widetilde\xi{}_\t^*\si^{{\rm vac}*}_{i_\t}\bigl(q\wedge\ovl\G{}^{\unl a}\,\bigl(\tfrac{\bd1_{D_{d,1}}-\tfrac{\la_1}{2}\,\G^0\,\G^1}{2}\bigr)\,q\wedge p^{\unl b}\bigr)\,,
\qqq
from which we readily extract the remaining Euler--Lagrange equations (written in the above notation)
\qq\nn
\sfP_{\la_1}\,q\approx 0\,,\qquad\qquad\sfP_{\la_1}\equiv\tfrac{\bd1_{D_{d,1}}-\tfrac{\la_1}{2}\,\G^0\,\G^1}{2}
\qqq
by setting $\,\d x^a=0$.\ If (and only if) $\,\la_1\in\{-2,2\}$,\ then the operator $\,\sfP_{\la_1}\,$ is a projector,
\qq\nn
\sfP_{\pm 2}^2=\sfP_{\pm 2}
\qqq
of rank
\qq\nn
\rk\,\sfP_{\pm 2}=\tfrac{D_{d,1}}{2}\,,
\qqq
with the additional properties
\qq\nn
\sfP_{\pm 2}\,\G^{\unl a}=\G^{\unl a}\,\bigl(\bd1_{D_{d,1}}-\sfP_{\pm 2}\bigr)\,,\qquad\qquad\sfP_{\pm 2}\,\G^{\widehat a}=\G^{\widehat a}\,\sfP_{\pm 2}\,,\qquad\qquad C^{-1}\,\sfP_{\pm 2}^{\rm T}\,C=\bd1_{D_{d,1}}-\sfP_{\pm 2}
\qqq
which leads to the emergence of a Lie sub-superalgebra
\qq\nn
&\bigl\{\bigl(Q\,\sfP_{\pm 2}\bigr)_\a,\bigl(Q\,\sfP_{\pm 2}\bigr)_\b\bigr\}=\bigl(\ovl\G{}^{\unl a}\,\sfP_{\pm 2}\bigr)_{\a\b}\,P_{\unl a}\,,\qquad\qquad[P_{\unl a},P_{\unl b}]=0\,,\qquad\qquad\bigl[\bigl(Q\,\sfP_{\pm 2}\bigr)_\a,P_{\unl a}\bigr]=0\,,&\cr\cr
&[J_{01},J_{01}]=0\,,\qquad\qquad[J_{01},J_{\widehat a\widehat b}]=0\,,\qquad\qquad[J_{\widehat a\widehat b},J_{\widehat c\widehat d}]=\d_{\widehat a\widehat d}\,J_{\widehat b\widehat c}-\d_{\widehat a\widehat c}\,J_{\widehat b\widehat d}+\d_{\widehat b\widehat c}\,J_{\widehat a\widehat d}-\d_{\widehat b\widehat d}\,J_{\widehat a\widehat c}\,,&\cr\cr
&\bigl[J_{01},\bigl(Q\,\sfP_{\pm 2}\bigr)_\a\bigr]=\tfrac{1}{2}\,\bigl(\bigl(Q\,\sfP_{\pm 2}\bigr)\,\G_{01}\bigr)_\a\,,\qquad\qquad\bigl[J_{\widehat a\widehat b},\bigl(Q\,\sfP_{\pm 2}\bigr)_\a\bigr]=\tfrac{1}{2}\,\bigl(\bigl(Q\,\sfP_{\pm 2}\bigr)\,\G_{\widehat a\widehat b}\bigr)_\a\,,&\cr\cr
&[J_{01},P_{\unl a}]=\eta_{\unl a 1}\,P_0-\eta_{\unl a 0}\,P_1\,,\qquad\qquad[J_{\widehat a\widehat b},P_{\unl a}]=0&
\qqq
within $\,\gt{siso}(d,1|D_{d,1})$.\ The above emphasises the indispensability of the projector $\,\sfP_{\pm 2}\,$ for the consistency of the field theory under consideration,\ and so fixes the absolute value of $\,\la_1$,\ leaving us only the immaterial choice of its sign,\ which we declare to be ``$+$",\ with 
\qq\nn
\sfP^{(1)}:=\sfP_{-2}
\qqq
and
\qq\nn
\underset{\tx{\ciut{(3)}}}{\widehat\chi}\equiv\underset{\tx{\ciut{(3)}}}{\widehat\chi}{}^{(2)}=2\eta_{\unl a\unl b}\,q\wedge\ovl\G{}^{\unl a}\,\bigl(\bd1_{D_{d,1}}-\sfP^{(1)}\bigr)\,q\wedge p^{\unl b}+2\ep_{\unl a\unl b}\,\d_{\widehat c\widehat d}\,p^{\widehat c}\wedge p^{\unl a}\wedge j^{\unl b\widehat d}+q\wedge\ovl\G{}_{\widehat a}\,q\wedge p^{\widehat a}\,.
\qqq
Indeed,\ the projector enforces a reduction of the Gra\ss mann-odd degrees of freedom necessary for the restoration of balance between them and their Gra\ss mann-even counterparts in the vacuum,\ the latter being subject to the (even) Inverse Higgs Constraints \eqref{eq:IHCs}.\ The Constraints are transmitted unto the Gra\ss mann-odd sector {\it via} the anticommutator of supercharges in the supersymmetry superalgebra $\,\gt{siso}(d,1|D_{d,1})$,\ and so for the sake of a residual supersymmetry in the vacuum,\ we need a subspace in the odd component $\,\gt{siso}(d,1|D_{d,1})^{(1)}\subset\gt{siso}(d,1|D_{d,1})\,$ which the superbracket maps to the surviving Gra\ss mann-even supersymmetries $\,P_0\,$ and $\,P_1$.\ The ratio
\qq\nn
{\rm BPS}\bigl(\gt{sB}^{{\rm (HP)}}_{1,2}\bigr)=\tfrac{\rk\,\sfP^{(1)}}{D_{d,1}}\equiv\tfrac{1}{2}
\qqq
goes under the name of the {\bf BPS fraction of the vacuum}.

Upon fixing a basis $\,\{\breve Q{}_{\unl\a}\}_{\unl\a\in\ovl{1,\frac{D_{d,1}}{2}}}\,$ in $\,{\rm im}\,\sfP^{(1)}{}^{\rm T}$,
\qq\nn
{\rm im}\,\sfP^{(1)}{}^{\rm T}=\corr{\ Q_\b\,\sfP^{(1)}{}^\b_{\ \a}\ \vert\ \a\in\ovl{1,D_{d,1}}\ }\equiv\bigoplus_{\unl\a=1}^{\frac{D_{d,1}}{2}}\,\corr{\breve Q{}_{\unl\a}}\,,
\qqq
we may write down (local) generators of the the {\bf vacuum superdistribution} 
\qq\nn
{\rm Vac}\bigl(\gt{sB}^{{\rm (HP)}}_{1,2}\bigr)\subset{\rm Corr}_{\rm HP}\bigl(\gt{sB}^{{\rm (HP)}}_{1,2}\bigr)\subset\cT\Si^{\rm HP}
\qqq
within the tangent sheaf $\,\cT\Si^{\rm HP}\,$ of the HP section $\,\Si^{\rm HP}\,$ that is determined by the Euler--Lagrange equations of the super-$\si$-model,
\qq\nn
\cW_{\unl\a\,i}=\breve Q{}_{\unl\a}\rstr_{\cV_i}\,,\qquad\a\in\ovl{1,\tfrac{D_{d,1}}{2}}\,,\qquad\qquad\cW_{\unl a\,i}=P_{\unl a}\rstr_{\cV_i} +\D_{\unl a\,i}^{\unl b\widehat c}\,J_{\unl b\widehat c}+\D_{\unl a\,i}^{\unl S}\,J_{\unl S}\,,\qquad\unl a\in\{0,1\}\,.
\qqq
In the light of the physical interpretation of the superdistribution,\ it is natural to demand involutivity of the latter,\ so that it defines -- in virtue of the Frobenius Theorem,\ \Rxcite{Thm.\,6.2.1}{Carmeli:2011} -- a foliation of the HP supertarget by embedded sub-supermanifolds,\ identified as the vacua of the supersymmetric field theory under consideration.\ Transform the matrices $\,\ovl\G{}^{\unl a}\,$ and $\,\G_{\unl S}$,\ commuting with $\,\sfP^{(1)}$,\ into a basis of the Majorana-spinor module adapted to the decomposition of the dual module $\,\gt{smink}(d,1|D_{d,1})^{(1)}\,$ into $\,{\rm im}\,\sfP^{(1)\,{\rm T}}\,$ and its direct-sum complement,
\qq\nn
{\rm im}\,\bigl(\bd1_{D_{d,1}}-\sfP^{(1)}\bigr)^{\rm T}\equiv\bigoplus_{\widehat\a=\frac{D_{d,1}}{2}+1}^{D_{d,1}}\,\corr{\widehat Q{}_{\widehat\a}}\,,
\qqq
whereupon they become block-diagonal,\ and denote the ensuing vacuum blocks as
\qq\nn
&\ovl\G{}^{\unl a}\rstr_{{\rm im}\,\sfP^{(1)}}=:\bigl(\ovl\g{}^{\unl a}_{\unl\a\unl\b}\bigr)_{\unl\a,\unl\b\in\ovl{1,\frac{D_{d,1}}{2}}}\equiv\ovl\g{}^{\unl a}\,,\qquad\qquad\det\,\ovl\g{}^{\unl a}\neq 0\,,\qquad\qquad\ovl\g{}^1=-\ovl\g{}^0\,,&\cr\cr
&\G_{01}\rstr_{{\rm im}\,\sfP^{(1)}}\equiv-\bd1_{\frac{D_{d,1}}{2}}\,,\qquad\qquad\G_{\widehat a\widehat b}\rstr_{{\rm im}\,\sfP^{(1)}}=:\bigl(\g_{\widehat a\widehat b}{}^{\unl\a}_{\ \unl\b}\bigr)_{\unl\a,\unl\b\in\ovl{1,\frac{D_{d,1}}{2}}}\equiv\g_{\widehat a\widehat b}\,.&
\qqq
Our discussion of involutivity of $\,{\rm Vac}(\gt{sB}^{{\rm (HP)}}_{1,2})\,$ begins with the inspection of the anticommutators
\qq\nn
\{\cW_{\unl\a\,i},\cW_{\unl\b\,i}\}=\ovl\g{}^0_{\unl\a\unl\b}\,(P_0-P_1)\rstr_{\cV_i}\equiv\ovl\g{}^0_{\unl\a\unl\b}\,\bigl(\cW_{0\,i}-\cW_{1\,i}\bigr)+\ovl\g{}^0_{\unl\a\unl\b}\,\bigl(\D_{1\,i}^{ab}-\D_{0\,i}^{ab}\bigr)\,J_{a<b}
\qqq
from which we read off the constraints
\qq\nn
\D_{1\,i}^{ab}=\D_{0\,i}^{ab}\equiv\D_i^{ab}\,,\qquad a<b\in\ovl{0,d}\,,
\qqq
implying
\qq\nn
\cW_{\unl a\,i}=P_{\unl a}\rstr_{\cV_i} +\D_i^{\unl b\widehat c}\,J_{\unl b\widehat c}+\D_i^{\unl S}\,J_{\unl S}\,.
\qqq
Next,\ we compute the commutators
\qq\nn
[\cW_{0\,i},\cW_{1\,i}]=\D_i^{01}\,\bigl(\cW_{0\,i}-\cW_{1\,i}\bigr) -\bigl(\D_i^{0\widehat a}+\D_i^{1\widehat a}\bigr)\,P_{\widehat a}+(P_0-P_1)\con\sfd\D_i^{ab}\,J_{a<b}\,,
\qqq
and infer the constraints
\qq\nn
\D_i^{0\widehat a}=-\D_i^{1\widehat a}\equiv\D_i^{\widehat a}\,,\qquad\widehat a\in\ovl{2,d}\,,\qquad\qquad(P_0-P_1)\con\sfd\D_i^{ab}=0\,,\qquad a<b\in\ovl{0,d}\,.
\qqq
Thus,
\qq\nn
\cW_{\unl a\,i}=P_{\unl a}\rstr_{\cV_i} +\D_i^{\widehat b}\,\bigl(J_{0\widehat b}-J_{1\widehat b}\bigr)+\D_i^{\unl S}\,J_{\unl S}\,,
\qqq
with
\qq\nn
(P_0-P_1)\con\sfd\D_i^{\widehat b}=0=(P_0-P_1)\con\sfd\D_i^{\unl S}\,.
\qqq
Finally,\ we examine the commutators
\qq\nn
[\cW_{\unl a\,i},\cW_{\unl\a\,i}]=-\tfrac{1}{2}\,\D_i^{01}\,\cW_{\unl\a}+\tfrac{1}{2}\,\D_i^{\widehat b\widehat c}\,\g_{\widehat b<\widehat c}{}^{\unl\b}_{\ \unl\a}\,\cW_{\unl\b}-\breve Q{}_{\unl\a}\con\sfd\D_i^{bc}\,J_{b<c}\,,
\qqq
whereby we arrive at the constraints
\qq\nn
\breve Q{}_{\unl\a}\con\sfd\D_i^{ab}=0\,,\qquad a<b\in\ovl{0,d}\,.
\qqq
We conclude that an \emph{involutive} vacuum superdistribution (of the type assumed) is spanned on fields 
\qq\nn
\cW_{\unl\a\,i}=\breve Q{}_{\unl\a}\rstr_{\cV_i}\,,\qquad\a\in\ovl{1,\tfrac{D_{d,1}}{2}}\,,\qquad\qquad\cW_{\unl a\,i}=P_{\unl a}\rstr_{\cV_i}+\D_i^{\widehat b}\,\bigl(J_{0\widehat b}-J_{1\widehat b}\bigr)+\D_i^{\unl S}\,J_{\unl S}\,,
\qqq
with
\qq\nn
(P_0-P_1)\con\sfd\D_i^{\widehat b}=0=(P_0-P_1)\con\sfd\D_i^{\unl S}\,,\qquad\qquad\breve Q{}_{\unl\a}\con\sfd\D_i^{\widehat b}=0=\breve Q{}_{\unl\a}\con\sfd\D_i^{\unl S}\,.
\qqq
Note that for the special choice
\qq\nn
\D_i^{\widehat b}\equiv 0\qquad\bigl(\qquad\Longrightarrow\qquad \D_i^{\unl S}\equiv 0\bigr)
\qqq
the vacuum superdistribution is modelled on the super-minkowskian component of the Lie sub-superalgebra
\qq\label{eq:vacbas}
\gt{vac}\bigl(\gt{sB}^{{\rm (HP)}}_{1,2}\bigr)=\bigoplus_{\unl\a=1}^{\frac{D_{d,1}}{2}}\,\corr{\breve Q{}_{\unl\a}}\oplus\corr{P_0,P_1}\oplus\gt{spin}(d,1)_{\rm vac}
\qqq
(the hidden gauge-symmetry algebra $\,\gt{spin}(d,1)_{\rm vac}\,$ is realised trivially) with the superbrackets
\qq\nn
&\bigl\{\breve Q{}_{\unl\a},\breve Q{}_{\unl\b}\bigr\}=\ovl\g{}^{\unl a}_{\unl\a\unl\b}\,P_{\unl a}\equiv\ovl\g{}^0_{\unl\a\unl\b}\,(P_0-P_1)\,,\qquad\qquad[P_{\unl a},P_{\unl b}]=0\,,\qquad\qquad\bigl[\breve Q{}_{\unl\a},P_{\unl a}\bigr]=0\,,&\cr\cr
&[J_{01},J_{01}]=0\,,\qquad\qquad[J_{01},J_{\widehat a\widehat b}]=0\,,\qquad\qquad[J_{\widehat a\widehat b},J_{\widehat c\widehat d}]=\d_{\widehat a\widehat d}\,J_{\widehat b\widehat c}-\d_{\widehat a\widehat c}\,J_{\widehat b\widehat d}+\d_{\widehat b\widehat c}\,J_{\widehat a\widehat d}-\d_{\widehat b\widehat d}\,J_{\widehat a\widehat c}\,,&\cr\cr
&\bigl[J_{01},\breve Q{}_{\unl\a}\bigr]=-\tfrac{1}{2}\,\breve Q{}_{\unl\a}\,,\qquad\qquad\bigl[J_{\widehat a\widehat b},\breve Q{}_{\unl\a}\bigr]=\tfrac{1}{2}\,\bigl(\breve Q\,\g_{\widehat a\widehat b}\bigr)_{\unl\a}\,,&\cr\cr
&[J_{01},P_{\unl a}]=\d_{\unl a 1}\,P_0+\d_{\unl a 0}\,P_1\,,\qquad\qquad[J_{\widehat a\widehat b},P_{\unl a}]=0&
\qqq
that we call the {\bf vacuum algebra of the superstring in $\,{\rm sMink}(d,1|D_{d,1})$}.\ In the next section,\ we shall interpret the departure from the simple model $\,\gt{vac}(\gt{sB}^{{\rm (HP)}}_{1,2})\,$ in the structure of the vacuum superdistribution in terms of gauge symmetries of the field theory.\ In the meantime,\ we mark the presence of the above sheaf parameters $\,\D^{\widehat b}\,$ and $\,\D^{\unl S}\,$ (with restrictions $\,\D^{\widehat b}\rstr_{\cV_i}=\D_i^{\widehat b}\,$ and $\,\D^{\unl S}\rstr_{\cV_i}=\D_i^{\unl S}$,\ respectively) in our notation as
\qq\nn
{\rm Vac}_{(\D)}\bigl(\gt{sB}^{{\rm (HP)}}_{1,2}\bigr)\equiv\bigoplus_{\unl\a=1}^{\frac{D_{d,1}}{2}}\,\corr{\cW_{\unl\a}\equiv\breve Q{}_{\unl\a}\rstr_{\Si^{\rm HP}}}\oplus\bigoplus_{\unl a\in\{0,1\}}\,\corr{\cW_{\unl a}\equiv P_{\unl a}\rstr_{\Si^{\rm HP}}+\D^{\widehat b}\,\bigl(J_{0\widehat b}-J_{1\widehat b}\bigr)+\D^{\unl S}\,J_{\unl S}}\,.
\qqq
The disjoint union of integral leaves $\,\cD_{i,\upsilon_i}\subset\cV_i\,$ (indices $\,\upsilon_i\,$ from an index set $\,\Upsilon_i\,$ enumerate the different leaves within $\,\cV_i$) of the involutive vacuum superdistribution shall be denoted as 
\qq\label{eq:vacfol}
\Si^{\rm HP}_{\rm vac}=\bigsqcup_{i\in I}\,\bigsqcup_{\upsilon_i\in\Upsilon_i}\,\cD_{i,\upsilon_i}\,,
\qqq
and termed the {\bf Hughes--Polchinski vacuum foliation}.\ It is embedded in the HP section,\ which we write as  
\qq\nn
\iota_{\rm vac}\ :\ \Si^{\rm HP}_{\rm vac}\emb\Si^{\rm HP}\,,
\qqq
and projects to the \textbf{physical vacuum foliation}
\qq\nn
\Si^{\rm HP}_{\rm phys\ vac}\equiv\bigcup_{i\in I}\,\bigcup_{\upsilon_i\in\Upsilon_i}\,\pi_{\rm vac}\bigl(\cD_{i,\upsilon_i}\bigr)\,.
\qqq
An alternative interpretation of the vacuum superdistribution and the vacuum algebra shall be given in the next section.

\section{Supersymmetries of the super-$\si$-model}\label{sec:susy}

The principle of supersymmetry lies at the core of the construction of the super-$\si$-model. Its field-theoretic implementation has its peculiarities that we review below upon identifying the various species of supersymmetry present.

\subsection{Global supersymmetry}

The GS super-$\si$-model in either formulation has a built-in global supersymmetry realised by the respective induced actions $\,[\ell]^\txK\,$ under which the integrand of the (metric) volume term is manifestly invariant (being defined in terms of the left invariant super-1-forms $\,p^a$),\ whereas that of the topological term is \emph{quasi}-invariant,\ {\it i.e.},\ invariant up to a total exterior derivative,\ as demonstrated explicitly (in the $\cS$-point picture) in
\qq\nn
[\ell]^*_{(\vep,y,\psi)}\underset{\tx{\ciut{(2)}}}{\txB}(\theta,x)=\underset{\tx{\ciut{(2)}}}{\txB}(\theta,x)+\sfd\bigl(S(\psi)^{-1}\,\vep\,\ovl\G{}_a\,\theta\,\bigl(\sfd x^a+\tfrac{1}{6}\,\theta\,\ovl\G{}^a\,\sfd\theta\bigr)\bigr)\,,
\qqq 
whence the said invariance of the DF amplitude for $\,\Si\,$ closed.\ The global supersymmetry of the field theory under consideration is reflected in the existence of a $\gt{siso}(d,1|D_{d,1})$-indexed family of Noether hamiltonians $\,\{h_X\}_{X\in\gt{siso}(d,1|D_{d,1})}\,$ on its space of states.\ These we derive in the NG formulation in which a state is represented by the Cauchy data $\,\Psi\equiv((\theta,x)\rstr_{\bS^1}\equiv\xi\rstr_{\bS^1},P)\,$ (a configuration $\,\xi\rstr_{\bS^1}\in[\bS^1,{\rm sMink}(d,1|D_{d,1})]\,$ and its LI \emph{kinetic} momentum $\,P$) of a vacuum localised on an equitemporal slice of the spacetime $\,\Si\,$ which we take to be (modelled on) $\,\bS^1\subset\Si$.\ The presymplectic form of the super-$\si$-model in this formulation reads
\qq\nn
\Om_\si[\Psi]=\d\vartheta[\Psi]+\int_{\bS^1}\,\ev^*\xi^*\underset{\tx{\ciut{(3)}}}{\txH}\,,
\qqq
where 
\qq\nn
\vartheta[\Psi]=\int_{\bS^1}\,\Vol(\bS^1)\,P_a\,\xi^*\unl p{}^a
\qqq
is the canonical (kinetic-)action 1-form on the space(s) of states $\,\cT^*[\bS^1,{\rm sMink}(d,1|D_{d,1})]$,\ and
\qq\nn
\ev\ :\ \bS^1\x\bigl[\bS^1,{\rm sMink}(d,1|D_{d,1})\bigr]\too{\rm sMink}(d,1|D_{d,1})
\qqq
is the evaluation mapping.\ The presymplectic form defines a Poisson superbracket on the space of hamiltonians on the space of states,\ {\it i.e.},\ on those sections $\,h\,$ of the structure sheaf of the latter that satisfy the relation 
\qq\nn
\d h=-\cV_h\con\Om_\si
\qqq
for some vector field $\,\cV_h$,\ termed hamiltonian -- the Poisson superbracket of two such sections:\ $\,h_1\,$ and $\,h_2\,$ with the respective hamiltonian vector fields $\,\cV_{h_1}\,$ and $\,\cV_{h_2}\,$ is given by 
\qq\nn
[h_1,h_2\}_{\Om_\si}=\cV_{h_2}\con\cV_{h_1}\con\Om_\si\,.
\qqq
In particular,\ upon contracting $\,\Om_\si\,$ with the \emph{covariant}\footnote{Here,\ covariance is determined by the super-1-form $\,\vartheta\,$ and expressed by the requirement:\ $\,\pLie{\widetilde\cK{}_X}\vartheta\must 0$.} lift
\qq\nn
\widetilde\cK{}_X[\Psi]\equiv\int_{\bS^1}\,\Vol(\bS^1)\,\cK_X(\xi)+\widetilde\D{}_X[\Psi]
\qqq
to $\,\cT^*[\bS^1,{\rm sMink}(d,1|D_{d,1})]\,$ of the suitably $\gt{spin}(d,1)$-vertically corrected right-invariant (RI) vector field $\,\cK_X\in\G(\cT{\rm sMink}(d,1|D_{d,1})),\ X\in\gt{siso}(d,1|D_{d,1})\,$ on $\,{\rm sMink}(d,1|D_{d,1})\equiv{\rm sMink}(d,1|D_{d,1})\x\{1\}\subset{\rm sISO}(d,1|D_{d,1})\,$ of the form
\qq\nn
\cK_X=R_X\rstr_{{\rm sMink}(d,1|D_{d,1})} +\Xi_X^{\ \ \ \unl S}\,J_{\unl S}\,,\qquad\qquad X\equiv X^A\,t_A\in\gt{siso}(d,1|D_{d,1})\,,
\qqq
expressed in terms of the sections $\,\Xi_X^{\ \ \ \unl S}\,$ of $\,\cO_{{\rm sISO}(d,1|D_{d,1})}({\rm sMink}(d,1|D_{d,1})\,$ of \Rxcite{Prop.\,5.1}{Suszek:2020xcu},\ we obtain the corresponding hamiltonian
\qq\nn
\widetilde\cK{}_X\con\Om_{\rm HP}=:-\d h_X\,.
\qqq
The relevant basis RI vector fields are
\qq\nn
R_{Q_\a}(\theta,x,0)&=&\tfrac{\vec\p\ }{\p\theta^\a}-\tfrac{1}{2}\,\theta^\b\,C_{\b\g}\,\G^{a\,\g}_{\ \ \ \a}\,\tfrac{\p\ }{\p x^a}\,,\cr\cr
R_{P_a}(\theta,x,0)&=&\tfrac{\p\ }{\p x^a}\,,\cr\cr
R_{J_{ab}}(\theta,x,0)&=&x^c\,\bigl(\eta_{cb}\,\tfrac{\p\ }{\p x^a}-\eta_{ca}\,\tfrac{\p\ }{\p x^b}\bigr)+\tfrac{1}{2}\,\G_{ab\,\b}^\a\,\theta^\b\,\tfrac{\vec\p\ }{\p\theta^\a}+\tfrac{\sfd\ }{\sfd t}\rstr_{t=0}t\phi_{ab}\star 0
\qqq
and give rise to the hamiltonians
\qq\nn
X^A\,h_A\equiv h_X=\int_{\bS^1}\,\bigl(\Vol(\bS^1)\,P_a\,\xi^*\bigl(R_X\con\theta^a_{\rm L}\bigr)+\ev^*\xi^*\k_X\bigr)\,,
\qqq
with
\qq\nn
\k_{Q_\a}(\theta,x)&=&-\ovl\G{}_{a\,\a\b}\,\theta^\b\,\bigl(2\sfd x^a-\tfrac{1}{3}\,\theta\,\ovl\G{}^a\,\sfd\theta\bigr)\,,\cr\cr
\k_{P_a}(\theta,x)&=&-\theta\,\ovl\G{}_a\,\sfd\theta\,,\cr\cr
\k_{J_{ab}}(\theta,x)&=&-x^c\,\theta\,\ovl\G{}_{ab}\,\G_c\,\sfd\theta\,.
\qqq
These furnish a realisation of a centrally extended supersymmetry Lie superalgebra $\,\gt{siso}(d,1|D_{d,1})\,$ within the Poisson superalgebra of observables on the space of states of the super-$\si$-model,
\qq\nn
[h_B,h_A\}_{\Om_{\rm HP}}=f_{AB}^{\ \ \ C}\,h_C+\cW_{AB}
\qqq
(here,\ the $\,f_{AB}^{\ \ \ C}\,$ are the structure constants of $\,{\rm sMink}(d,1|D_{d,1})$),\ with the components of the {\bf wrapping anomaly},\ 
\qq\nn
\cW_{AB}=-\int_{\bS^1}\,\ev^*\xi^*\bigl(R_{t_A}\con\sfd\k_{t_B}+f_{AB}^{\ \ \ C}\,\k_{t_C}\bigr)=:\int_{\bS^1}\,\ev^*w_{AB}\,,
\qqq
given by
\qq
&w_{\a\b}=\sfd\bigl(2\ovl\G{}_{a\,\a\b}\,x^a)\,,\qquad\qquad w_{ab}=0\,,\qquad\qquad w_{a\a}=\sfd\bigl(-2\ovl\G{}_{a\,\a\b}\,\theta^\b\bigr)\,,&\cr\cr
&w_{ab\,cd}=\sfd\bigl(-\tfrac{1}{2}\,x^e\,\theta\,\ovl\G{}_{ab}\,\G_e\,\G_{cd}\,\theta\bigr)\,,\qquad\qquad w_{ab\,c}=\sfd\bigl(\tfrac{1}{2}\,\theta\,\ovl\G{}_{abc}\,\theta\bigr)\,,& \label{eq:wrapansource}\\ \cr
&w_{ab\,\a}=\sfd\bigl(-2x^c\,\bigl(\eta_{ca}\,\ovl\G{}_b-\eta_{cb}\,\ovl\G{}_a\bigr)_{\a\b}\,\theta^\b-\tfrac{1}{6}\,\theta\,\ovl\G{}_{abc}\,\theta\,\ovl\G{}^c_{\a\b}\,\theta^\b\bigr)\,,&
\nn
\qqq
{\it cp} \Rcite{Suszek:2018bvx}.\ Clearly,\ in the trivial super-minkowskian topology,\ we obtain the result
\qq\nn
\cW_{AB}\equiv 0\,.
\qqq
Its refinement,\ to be considered in Sect.\,\ref{sec:geometrise},\ is the first step towards geometrisation of the cohomological content of the super-$\si$-model.

\subsection{Local supersymmetry of the vacuum}\label{sub:kappasym} The GS super-$\si$-model with the physical supertarget $\,{\rm sISO}(d,1|D_{d,1})/\txK\,$ \emph{realised within} $\,{\rm sISO}(d,1|D_{d,1})\,$ by means of the local sections has an implicit local symmetry modelled on the right action of $\,\txK$.\ In particular,\ in the NG formulation,\ we have the large hidden gauge group $\,{\rm Spin}(d,1)$.\ Therefore,\ we anticipate an enhancement of the (tangential) local symmetry in the HP formulation according to the scheme 
\qq\nn
\gt{spin}(d,1)_{\rm vac}\nearrow\gt{spin}(d,1)
\qqq
upon restriction to the correspondence superdistribution $\,{\rm Corr}_{\rm HP}(\gt{sB}^{{\rm (HP)}}_{1,2})$.\ Inspection of the expression
\qq\nn
\underset{\tx{\ciut{(3)}}}{\widehat\chi}\rstr_{{\rm Corr}_{\rm HP}(\gt{sB}^{{\rm (HP)}}_{1,2})}=2\eta_{\unl a\unl b}\,q\wedge\ovl\G{}^{\unl a}\,\bigl(\bd1_{D_{d,1}}-\sfP^{(1)}\bigr)\,q\wedge p^{\unl b}
\qqq
immediately corroborates our expectation:\ The vector fields
\qq\nn
\cT_{\unl a\widehat b}\in\G\bigl(\cT\Si^{\rm HP}\bigr)\,,\qquad\cT_{\unl a\widehat b}\rstr_{\cV_i}=J_{\unl a\widehat b}\rstr_{\cV_i}+T_{\unl a\widehat b\,i}^{\unl S}\,J_{\unl S}\equiv\cT_{\unl a\widehat b\,i}\,,\qquad\qquad(\unl a,\widehat b)\in\{0,1\}\x\ovl{2,d}\,,
\qqq
with correcting sections $\,T_{\unl a\widehat b\,i}^{\unl S}\in\cO_{\Si^{\rm HP}}(\cV_i)$ uniquely determined by the condition $\,\cT_{\unl a\widehat b\,i}\in\G(\cT\cV_i)$,\ {\it cp} \Rxcite{Prop.\,3.6}{Suszek:2020xcu},\ satisfy
\qq\nn
\forall_{(\unl a,\widehat b)\in\{0,1\}\x\ovl{2,d}}\ :\ \cT_{\unl a\widehat b}\con\underset{\tx{\ciut{(3)}}}{\widehat\chi}\rstr_{{\rm Corr}_{\rm HP}(\gt{sB}^{{\rm (HP)}}_{1,2})}=0\,.
\qqq
This enhancement does not carry any physically nontrivial information as it merely reflects the residual redundancy of our \emph{realisation} of the physical supertarget within the mother Lie supergroup.\ Accordingly,\ we are inclined to fix the hidden gauge by augmenting the set of Euler--Lagrange equations derived formerly with 
\qq\nn
j^{\unl a\widehat b}\approx 0\,,\qquad(\unl a,\widehat b)\in\{0,1\}\x\ovl{2,d}\,,
\qqq
so that altogether we arrive at the conjunction of constraints
\qq\nn
\bigl(\bd1_{D_{d,1}}-\sfP^{(1)}\bigr)\,q\approx 0\,,\qquad\qquad p^{\widehat a}\approx 0\,,\qquad\widehat a\in\ovl{2,d}\,,\qquad\qquad j^{\unl b\widehat c}\approx 0\,,\qquad(\unl b,\widehat c)\in\{0,1\}\x\ovl{2,d}
\qqq
as the definition of the ({\bf hidden}) {\bf gauge-fixed vacuum superdistribution}
\qq\nn
{\rm Vac}_{\rm hgf}\bigl(\gt{sB}^{{\rm (HP)}}_{1,2}\bigr)\equiv {\rm Vac}_{(\D=0)}\bigl(\gt{sB}^{{\rm (HP)}}_{1,2}\bigr)=\bigoplus_{\unl\a=1}^{\frac{D_{d,1}}{2}}\,\corr{\cW_{\unl\a}\equiv\breve Q{}_{\unl\a}\rstr_{\Si^{\rm HP}}}\oplus\bigoplus_{\unl a\in\{0,1\}}\,\corr{\cW_{\unl a}\equiv P_{\unl a}\rstr_{\Si^{\rm HP}}}\,.
\qqq
Clearly,\ the hidden gauge-symmetry (sub)distribution $\,\gt{spin}(d,1)_{\rm vac}\,$ is a symmetry of the above vacuum superdistribution,
\qq\nn
\bigl[\gt{spin}(d,1)_{\rm vac},{\rm Vac}_{\rm hgf}\bigl(\gt{sB}^{{\rm (HP)}}_{1,2}\bigr)\bigr]\subset{\rm Vac}_{\rm hgf}\bigl(\gt{sB}^{{\rm (HP)}}_{1,2}\bigr)\,,
\qqq
and so the vacuum foliation descends to the supertarget $\,{\rm sISO}(d,1|D_{d,1})/{\rm Spin}(d,1)_{\rm vac}$.\ 

But that is not all.\ Indeed,\ the very mechanism responsible for the restitution of supersymmetry in the vacuum gives rise to an extra and physically nontrivial local supersymmetry on restriction to $\,{\rm Corr}_{\rm HP}(\gt{sB}^{{\rm (HP)}}_{1,2})$,\ to wit,\ tangential Gra\ss mann-odd translations along $\,{\rm im}\,\sfP^{(1)}{}^{\rm T}$,
\qq\nn
\forall_{\unl\a\in\ovl{1,\frac{D_{d,1}}{2}}}\ :\ \cW_{\unl\a}\con\underset{\tx{\ciut{(3)}}}{\widehat\chi}\rstr_{{\rm Corr}_{\rm HP}(\gt{sB}^{{\rm (HP)}}_{1,2})}=0\,.
\qqq
Furthermore,\ we readily establish
\qq\nn
(\cW_0+\cW_1)\con\underset{\tx{\ciut{(3)}}}{\widehat\chi}\rstr_{{\rm Corr}_{\rm HP}(\gt{sB}^{{\rm (HP)}}_{1,2})}=0\,,
\qqq
and so,\ altogether,\ we obtain the {\bf enhanced gauge-symmetry superdistribution}
\qq\nn
\cG\cS_1\equiv\cG\cS\bigl(\gt{sB}^{{\rm (HP)}}_{1,2}\bigr)=\bigoplus_{\unl\a=1}^{\frac{D_{d,1}}{2}}\,\corr{\cW_{\unl\a}}\oplus\corr{\cW_0+\cW_1}\oplus\bigoplus_{(\unl a,\widehat b)\in\{0,1\}\x\ovl{2,d}}\,\corr{\cT_{\unl a\widehat b}}\subset{\rm Corr}_{\rm HP}\bigl(\gt{sB}^{{\rm (HP)}}_{1,2}\bigr)\,,
\qqq
modelled on the sub-space
\qq\nn
\gt{gs}_1\equiv\gt{gs}\bigl(\gt{sB}^{{\rm (HP)}}_{1,2}\bigr)={\rm im}\,\sfP^{(1)}{}^{\rm T}\oplus\corr{P_0+P_1}\oplus\gt{spin}(d,1)\,,
\qqq
with the component along $\,\gt{spin}(d,1)_{\rm vac}\,$ realised trivially.\ Actually,\ in order to be able to interpret $\,\cG\cS_1\,$ as a proper local-symmetry structure of the theory,\ we should demand that the limit of its weak derived flag,\ as introduced in \Rxcite{Def.\,4.9}{Suszek:2020xcu},\ stays in the correspondence superdistribution.\ This is,\ clearly,\ \emph{not} the case for $\,\cG\cS_1$,\ and so we are led to extract from it a sub-superdistribution that satisfies this extra condition -- in this manner,\ we arrive at the {\bf $\k$-symmetry superdistribution}
\qq\nn
\k\bigl(\gt{sB}^{{\rm (HP)}}_{1,2}\bigr)=\bigoplus_{\unl\a=1}^{\frac{D_{d,1}}{2}}\,\corr{\cW_{\unl\a}}\oplus\corr{\cW_0+\cW_1}\,.
\qqq
The latter immediately reveals its peculiarity:\ The limit of its weak derived flag,
\qq\nn
\k\bigl(\gt{sB}^{{\rm (HP)}}_{1,2}\bigr)^{-\infty}=\bigoplus_{\unl\a=1}^{\frac{D_{d,1}}{2}}\,\corr{\cW_{\unl\a}}\oplus\corr{\cW_0}\oplus\corr{\cW_1}
\qqq
is contained not only in $\,{\rm Corr}_{\rm HP}(\gt{sB}^{{\rm (HP)}}_{1,2})$,\ but in the vacuum superdistribution,\ or,\ more accurately,
\qq\nn
\k\bigl(\gt{sB}^{{\rm (HP)}}_{1,2}\bigr)^{-\infty}\equiv{\rm Vac}_{\rm hgf}\bigl(\gt{sB}^{{\rm (HP)}}_{1,2}\bigr)\,
\qqq
{\it i.e.},\ the $\k$-symmetry superdistribution is superbracket-generating for the tangent sheaf of the gauged-fixed HP vacuum foliation $\,\Si^{\rm HP}_{\rm vac}$,\ its flows enveloping the integral leaves of the latter.\ Hence,\ it makes sense to think of $\,\k(\gt{sB}^{{\rm (HP)}}_{1,2})^{-\infty}\,$ as the gauge (super)symmetry of the vacuum.\ The generating nature of the Gra\ss mann-odd component 
\qq\nn
\k\bigl(\gt{sB}^{{\rm (HP)}}_{1,2}\bigr)^{(1)}\equiv\bigoplus_{\unl\a=1}^{\frac{D_{d,1}}{2}}\,\corr{\cW_{\unl\a}}\subset\k\bigl(\gt{sB}^{{\rm (HP)}}_{1,2}\bigr)
\qqq
won it its name -- the square root of (the chiral half of) the vacuum -- in \Rcite{Suszek:2020xcu}.\ It is the HP counterpart of the odd gauge symmetry of the super-$\si$-model in the NG formulation,\ known under the name of $\k$-symmetry,\ that was originally found and studied by de Azc\'arraga and Lukierski in Refs.\,\cite{deAzcarraga:1982dhu,deAzcarraga:1982njd} in the setting of the super-$\si$-model of superparticle dynamics,\ and subsequently rediscovered and elaborated by Siegel in \Rcite{Siegel:1983hh} and,\ in the two-dimensional setting, in \Rcite{Siegel:1983ke}.\ In the present context,\ the Lie superalgebra $\,\gt{vac}(\gt{sB}^{{\rm (HP)}}_{1,2})\,$ acquires the interpretation of the gauge-symmetry algebra of the vacuum,\ confirmed trivially by its inclusion in the kernel of the suitably restricted presymplectic form 
\qq\nn
\Om_{\rm HP}=\sum_{\t\in\Tgt_2}\,\int_{\bS^1\cap\t}\,\ev^*\si_{i_\t}^{{\rm vac}\,*}\underset{\tx{\ciut{(3)}}}{\widehat\chi}
\qqq
of the GS super-$\si$-model in the HP formulation (in which the space of states is parametrised by configurations $\,\widetilde\xi\rstr_{\bS^1}$).\ The crucial feature of the gauge supersymmetry modelled by $\,\gt{vac}(\gt{sB}^{{\rm (HP)}}_{1,2})\,$ is its target space-geometric nature,\ to be contrasted with the mixed target-space/worldsheet and hence somewhat obscure nature of its NG predecessor,\ {\it cp} Refs.\,\cite{McArthur:1999dy,Gomis:2006wu} -- this,\ in conjunction with the geometrisation of the superfield equations,\ makes the HP formulation perfectly suited for a fully fledged higher-geometric analysis that we carry out in the remainder of this paper.\ We readily convince ourselves that it integrates to a Lie sub-supergroup of the supersymmetry group $\,{\rm sISO}(d,1|D_{d,1})\,$ represented by the super-Harish--Chandra pair
\qq\nn
{\rm sISO}(d,1|D_{d,1})_{\rm vac}=\bigl({\rm Mink}(d,1)_{\rm vac}\rx{\rm Spin}(d,1)_{\rm vac},\gt{vac}\bigl(\gt{sB}^{{\rm (HP)}}_{1,2}\bigr)\bigr)\subset{\rm sISO}(d,1|D_{d,1})
\qqq
with 
\qq\nn
{\rm Mink}(d,1)_{\rm vac}=\{\ \bigl(x^a\bigr)\in{\rm Mink}(d,1) \quad\vert\quad \forall_{\widehat a\in\ovl{2,d}}\ :\ x^{\widehat a}=0\ \}
\qqq
and with the action of the body Lie group on the Lie superalgebra $\,\gt{vac}(\gt{sB}^{{\rm (HP)}}_{1,2})\,$ inherited from $\,{\rm sISO}(d,1|D_{d,1})$.\ We shall,\ henceforth,\ refer to this Lie supergroup by the name of the \textbf{$\k$-symmetry group of the superstring in $\,{\rm sMink}(d,1|D_{d,1})$}.\ Its action on $\,{\rm sISO}(d,1|D_{d,1})\,$ (by right translations) splits the latter into orbits -- the integral leaves of the (integrable) superdistribution $\,\gt{vac}(\gt{sB}^{{\rm (HP)}}_{1,2})\subset\cT{\rm sISO}(d,1|D_{d,1})$,\ the very ones whose `intersections' with the $\,\cV_i,\ i\in I\,$ model the vacua within the mother Lie supergroup.

\section{A geometrisation of the superbackground -- the super-1-gerbe(s)}\label{sec:geometrise}

Behind every two-dimensional (super-)$\si$-model,\ there is a 1-gerbe.\ A 1-gerbe is a geometrisation,\ proposed by Murray in \Rcite{Murray:1994db} and recalled exhaustively in \Rxcite{Sec.\,2}{Suszek:2017xlw} ({\it cp} also \Rcite{Huerta:2020} for a recent rendering in the $\bZ/2\bZ$-graded setting),\ of (the cohomology class of) a de Rham 3-cocycle that couples to the charge current defined by the embedding of the worldsheet $\,\Si\,$ in the (super)target.\ This higher-geometric object gives a rigorous meaning to the topological term in the DF amplitude ({\it cp} \Rcite{Gawedzki:1987ak}),\ determines a prequantisation of the field theory through cohomological transgression ({\it ibid.}) and naturally encodes information on its (pre)quantisable global symmetries and their gauging ({\it cp} Refs.\,\cite{Gawedzki:2003pm,Gawedzki:2007uz,Gawedzki:2008um,Gawedzki:2010rn,Suszek:2012ddg,Suszek:2013,Gawedzki:2012fu}).\ More generally,\ the bicategory of 1-gerbes provides us with geometric and cohomological data necessary for a quantum-mechanically consistent definition of a poly-phase $\si$-model in which various phases are separated by edges of a so-called defect localised on a graph in $\,\Si$,\ {\it cp} Refs.\,\cite{Fuchs:2007fw,Runkel:2008gr,Runkel:2009sp},\ and distinguished 1-cells of the bicategory model dualities between equivalent $\si$-models (with, sometimes, different targets),\ {\it cp} Refs.\,\cite{Suszek:2011hg,Suszek:2011}.\ The Reader is urged to consult Refs.\,\cite{Suszek:2017xlw,Suszek:2019cum,Suszek:2020xcu} for a review of applications of gerbe theory in the field-theoretic setting of interest.\ Below,\ we merely recapitulate the logic behind the construction of the (super-)1-gerbe for the GS super-$\si$-model and present details of the construction itself.\medskip

A prerequisite of a meaningful geometrisation of the 3-form component of the (super)background of the loop's propagation is the identification of the cohomology relevant to the field theory in hand.\ The discussion carried out in the preceding sections leaves no room for doubt -- the cohomology to be considered in the setting of the super-$\si$-model is the supersymmetric refinement of the standard de Rham cohomology of the supertarget.\ Owing to the inherent noncompactness of the supersymmetry group $\,{\rm sMink}(d,1|D_{d,1})\,$ (and $\,{\rm sISO}(d,1|D_{d,1})$,\ for that matter),\ the latter cohomology differs from the standard de Rham cohomology (trivial for $\,{\rm sMink}(d,1|D_{d,1})$),\ and the GS super-3-cocycle of interest happens to define a nontrivial class in $\,{\rm CaE}^3({\rm sMink}(d,1|D_{d,1}))\equiv H^3_{\rm dR}({\rm sMink}(d,1|D_{d,1}))^{{\rm sMink}(d,1|D_{d,1})}\,$ (lifting to a nontrivial class in $\,{\rm CaE}^3({\rm sISO}(d,1|D_{d,1}))$).\ A non-pragmatic rationale for a geometrisation of the Cartan--Eilenberg cohomology of the supertarget $\,{\rm sMink}(d,1|D_{d,1})$,\ as stated -- after Rabin and Crane,\ {\it cp} Refs.\,\cite{Rabin:1984rm,Rabin:1985tv} -- in \Rxcite{Sec.\,3}{Suszek:2017xlw},\ is a topologisation of the said cohomology as the dual of the singular homology of an orbifold $\,{\rm sMink}(d,1|D_{d,1})/\G_{\rm KR}\,$ of the super-Minkowski space by the natural action of the Kosteleck\'y--Rabin discrete supersymmetry group $\,\G_{\rm KR}\,$ of \Rcite{Kostelecky:1983qu},\ generated by integer Gra\ss mann-odd translations (in the $\cS$-point picture,\ and for a suitable choice of the Majorana representation of $\,\Cliff(\bR^{d,1})$).\ Accordingly,\ the GS super-$\si$-model ought to be interpreted as an effective description of standard loop dynamics in $\,{\rm sMink}(d,1|D_{d,1})/\G_{\rm KR}$.\ Prior to investigating the consequences of the latter idea,\ we pause to decode its meaning and present a concrete realisation,\ in a semi-heuristic approach in which we gloss over any ({\it e.g.},\ topological) subtleties encountered along the way.\ Thus,\ we change the hitherto (Kostant's) perspective on supermanifolds and present $\,{\rm sMink}(d,1|D_{d,1})\,$ -- upon invoking \cite[Def.-Cor.]{Batchelor:1980} ({\it cp} also Refs.\,\cite{Gawedzki:1977pb,Batchelor:1979a}) -- as (a direct limit $\,N\to\infty\,$ of) a nested family,\ indexed by $\,\bN^\x\ni N$,\ of DeWitt's skeletons given essentially by (`soul') vector bundles 
\qq\nn
{\rm Skel}_N\bigl({\rm sMink}(d,1|D_{d,1})\bigr)\equiv\bigoplus_{a=0}^d\,\bigr(\bR\oplus\bigoplus_{k=1}^{E(\frac{N}{2})}\,\bigwedge\hspace{-3pt}{}^{2k}\,\bR^{\x N}\bigl)\oplus\bigoplus_{\a=1}^{D_{d,1}}\,\bigoplus_{l=0}^{E(\frac{N-1}{2})}\,\bigwedge\hspace{-3pt}{}^{2l+1}\,\bR^{\x N}\too\bigoplus_{a=0}^d\,\bR\equiv\bR^{\x d+1}
\qqq 
of rank $\,2^{N-1}\,(d+1+D_{d,1})-d-1\,$ over the body $\,\bR^{\x d+1}\equiv{\rm Mink}(d,1)$,\ {\it cp} \Rxcite{Sec.\,2.1}{DeWitt:1984}.\ Practically speaking,\ this amounts to realising the global coordinate generators $\,(\theta^\a,x^a)\,$ of the structure sheaf $\,\cO_{{\rm sMink}(d,1|D_{d,1})}\,$ in the $N{}^{\rm th}$ skeleton as (functional) linear combinations of elements of a basis $\,\{\bd1\}\cup\{e^{i_1}\wedge e^{i_2}\wedge\cdots\wedge e^{i_m}\}_{i_1<i_2<\ldots<i_m\in\ovl{1,N},\ m\in\ovl{1,N}}\,$ of the corresponding  exterior algebra $\,\bigwedge\hspace{-3pt}{}^\bullet\,\bR^{\x N}\,$ as
\qq\nn
\theta^\a_{(N)}=\sum_{l=0}^{E(\frac{N-1}{2})}\,\theta^\a_{i_1 i_2\ldots i_{2l+1}}\,e^{i_1}\wedge e^{i_2}\wedge\cdots\wedge e^{i_{2l+1}}\,,\qquad\qquad x^a_{(N)}=x^a_0+\sum_{k=1}^{E(\frac{N}{2})}\,x^a_{i_1 i_2\ldots i_{2k}}\,e^{i_1}\wedge e^{i_2}\wedge\cdots\wedge e^{i_{2k}}
\qqq
This presentation seems naturally compatible with Freed's identification of the super-$\si$-model mapping space $\,[\Si,{\rm sMink}(d,1|D_{d,1})]\,$ as the appropriate inner-$\mor$ functor -- indeed,\ we may think of the morphisms from $\,[\Si,{\rm sMink}(d,1|D_{d,1})](\bR^{0|N})\,$ as probing the $N{}^{\rm th}$ skeleton.\ Now, the Rabin--Crane argument \emph{at level} $N$ refers to the subgroup $\,\G_{\rm KR}^{(N)}\subset{\rm Skel}_N({\rm sMink}(d,1|D_{d,1}))\,$ of the $N{}^{\rm th}$ skeleton (with respect to super-minkowskian multiplication,\ realised in terms of the exterior product) generated\footnote{In a suitable Majorana representation of the Clifford algebra with integer-valued matrices of the generators,\ {\it cp} \Rcite{Kostelecky:1983qu}.} by odd vectors 
\qq\nn
\nu^\a_{(N)}=\sum_{l=0}^{E(\frac{N-1}{2})}\,n^\a_{i_1 i_2\ldots i_{2l+1}}\,e^{i_1}\wedge e^{i_2}\wedge\cdots\wedge e^{i_{2l+1}}\,,\qquad\a\in\ovl{1,D_{d,1}}
\qqq
with all (pure-soul) coefficients in
\qq\nn
\bZ\ni n^\a_{i_1 i_2\ldots i_{2l+1}}\,.
\qqq
An example of the orbifold 
\qq\nn
{\rm Skel}_N\bigl({\rm sMink}(d,1|D_{d,1})\bigr)/\G_{\rm KR}^{(N)}
\qqq
whose homology is readily seen to encode the CaE cohomology of $\,{\rm sMink}(d,1|D_{d,1})\,$ (owing to the polynomial character of the binary operation of this Lie supergroup in the $\cS$-point picture) was explicitly constructed in \Rxcite{App.}{Rabin:1984rm},\ and the crucial observation of a \emph{generic} nature is that it has compact \emph{odd} (and even) fibre directions.\ The nested character of the of DeWitt's skeletal presentation implies that the latter observation carries over to the direct limit,\ and so in view of our earlier remark on the interpretation of Freed's prescription in the present context,\ it becomes clear that we should allow for monodromies of the embedding fields of both parities ({\it i.e.},\ the so-called twisted sector) in the super-$\si$-model with the supertarget $\,{\rm sMink}(d,1|D_{d,1})\,$ when modelling the field theory with the Rabin--Crane orbifold $\,{\rm sMink}(d,1|D_{d,1})/\G_{\rm KR}$ as the supertarget.\ Taking into account the top line in \Reqref{eq:wrapansource},\ we are led to consider a supercentral wrapping-charge extension 
\qq\nn
\brd0\too\bR^{d|D_{d,1}}\too\widetilde{\gt{smink}}(d,1|D_{d,1})\too\gt{smink}(d,1|D_{d,1})\too\brd0
\qqq
of the original supersymmetry algebra $\,\gt{smink}(d,1|D_{d,1})\,$ with the supervector space structure 
\qq\nn
\widetilde{\gt{smink}}(d,1|D_{d,1})=\bigl(\bigoplus_{\a=1}^{D_{d,1}}\,\corr{\widetilde{\unl Q}{}_\a}\oplus\bigoplus_{a=0}^d\,\corr{\widetilde{\unl P}{}_a}\bigr)\oplus\bigl(\bigoplus_{\b=1}^{D_{d,1}}\,\corr{\unl Z{}^\b}\oplus\bigoplus_{b=1}^d\,\corr{\unl R{}^b}\bigr)\equiv\gt{smink}(d,1|D_{d,1})\oplus\bR^{d|D_{d,1}}
\qqq
and with the Lie-superalgebra structure determined by the relations
\qq\nn
&\{\widetilde{\unl Q}{}_\a,\widetilde{\unl Q}{}_\b\}=\ovl\G{}^a_{\a\b}\,\bigl(\widetilde{\unl P}{}_a+\eta_{ab}\,\unl R{}^b\bigr)\,,\qquad\qquad[\widetilde{\unl P}{}_a,\widetilde{\unl P}{}_b]=0\,,\qquad\qquad[\widetilde{\unl Q}{}_\a,\widetilde{\unl P}{}_a]=\ovl\G_{a\,\a\b}\,\unl Z{}^\b\,,&\cr\cr
&[\widetilde{\unl Q}{}_\a,\unl R{}^a]=0\,,\qquad\qquad[\widetilde{\unl P}{}_a,\unl R{}^b]=0\,,\qquad\qquad\{\widetilde{\unl Q}{}_\a,\unl Z{}^\b\}=0\,,\qquad\qquad[\widetilde{\unl P}{}_a,\unl Z{}^\a]=0\,,&\cr\cr
&[\unl R{}^a,\unl R{}^b]=0\,,\qquad\qquad\{\unl Z{}^\a,\unl Z{}^\b\}=0\,,\qquad\qquad[R_a,\unl Z{}^\a]=0\,.&
\qqq
We note,\ parenthetically,\ that considerations similar to ours were employed in \Rcite{deAzcarraga:1989mza} in a derivation of central topological charges associated with the (even) wrapping states of the super-$p$-brane.\ It is to be emphasised,\ though,\ that neither the Rabin--Crane argument,\ nor the monodromy in the Gra\ss mann-odd directions and the attendant subtlety of the twisted sector were considered in that work.

The Gra\ss mann-even ($d$-vector) component of the extension is trivial\footnote{This need \emph{not} be so on the level of the associated Lie supergroup,\ {\it cp} \Rxcite{Sec.\,2.3.1}{Chryssomalakos:2000xd},\ but we shall not pursue this point in what follows.} -- it can be removed by the simple redefinition 
\qq\nn
\widetilde{\unl P}{}_a\longmapsto\widetilde{\unl P}{}_a+\eta_{ab}\,\unl R{}^b
\qqq
that leaves us with the irreducible Gra\ss mann-odd extension 
\qq\label{eq:oddGSext}
\brd0\too\bR^{0|D_{d,1}}\too\sfY\gt{smink}(d,1|D_{d,1})\xrightarrow{\ \pi_{\sfY\gt{smink}(d,1|D_{d,1})}\ }\gt{smink}(d,1|D_{d,1})\too\brd0
\qqq
with the supervector-space structure 
\qq\nn
\sfY\gt{smink}(d,1|D_{d,1})=\bigl(\bigoplus_{\a=1}^{D_{d,1}}\,\corr{\sfY\unl Q{}_\a}\oplus\bigoplus_{a=0}^d\,\corr{\sfY\unl P{}_a}\bigr)\oplus\bigoplus_{\b=1}^{D_{d,1}}\,\corr{\unl Z{}^\b}\equiv\gt{smink}(d,1|D_{d,1})\oplus\bR^{0|D_{d,1}}
\qqq
and the Lie superbracket
\qq\nn
&\{\sfY\unl Q{}_\a,\sfY\unl Q{}_\b\}=\ovl\G{}^a_{\a\b}\,\sfY\unl P{}_a\,,\qquad\qquad[\sfY\unl P{}_a,\sfY\unl P{}_b]=0\,,\qquad\qquad[\sfY\unl Q{}_\a,\sfY\unl P{}_a]=\ovl\G_{a\,\a\b}\,\unl Z{}^\b\,,&\cr\cr
&\{\sfY\unl Q{}_\a,\unl Z{}^\b\}=0\,,\qquad\qquad[\sfY\unl P{}_a,\unl Z{}^\a]=0\,,\qquad\qquad\{\unl Z{}^\a,\unl Z{}^\b\}=0\,,&
\qqq
alongside a decoupled abelian algebra $\,\bR^{\x d}$,\ 
\qq\nn
\widetilde{\gt{smink}}(d,1|D_{d,1})\cong\sfY\gt{smink}(d,1|D_{d,1})\oplus\bR^{\x d}\,.
\qqq
The odd extension,\ which is none other than the Green superalgebra of \Rcite{Green:1989nn},\ has an attractive cohomological feature,\ to wit,\ the pullback of the nontrivial 3-cocycle $\,\underset{\tx{\ciut{(3)}}}{\txH}\,$ to $\,\sfY\gt{smink}(d,1|D_{d,1})\,$ trivialises.\ Indeed,\ denote the super-1-forms dual to the new generators $\,\unl Z{}^\a\,$ as $\,\unl z{}_\a\,$ to obtain\footnote{In the present paper,\ the super-forms appear in a (seemingly) double r\^ole:\ as sections of the sheaf of superdifferential forms on the Lie supergroup (regarded as a supermanifold) and as (super-)forms on its tangent Lie superalgebra.\ We use the same symbol(s) for both r\^oles,\ which,\ however,\ should not lead to confusion as it is always clear from the context which r\^ole is currently under consideration (in particular,\ we reserve the symbol $\,\widehat\d\,$ for the coboundary operator of the Lie-superalgebra cohomology).} 
\qq\nn
\pi_{\sfY\gt{smink}(d,1|D_{d,1})}^*\underset{\tx{\ciut{(3)}}}{\txH}=\widehat\d\unl z{}_\a\wedge\pi_{\sfY\gt{smink}(d,1|D_{d,1})}^*\unl q^\a=\widehat\d\bigl(\unl z{}_\a\wedge\pi_{\sfY\gt{smink}(d,1|D_{d,1})}^*\unl q^\a\bigr)\,.
\qqq
This is a manifestation of a $\bZ/2\bZ$-graded variant of the classic Lie-algebraic phenomenon:\ For any Lie (super)algebra $\,\ggt$,\ classes in the second group $\,H^2(\ggt,\agt)\,$ of the Chevalley--Eilenberg cohomology of $\,\ggt\,$ with values in a trivial (super)module $\,\agt\,$ are in a one-to-one correspondence with equivalence classes of (super)central extensions
\qq\nn
\brd0\too\agt\too\widetilde\ggt\too\ggt\too\brd0
\qqq
of $\,\ggt\,$ by $\,\agt$,\ {\it cp} \Rcite{Suszek:2017xlw}.\ Now,\ the supervector space 
\qq\nn
\Om^1\bigl({\rm sMink}(d,1|D_{d,1})\bigr)^{{\rm sMink}(d,1|D_{d,1})}=\bigoplus_{\a=1}^{D_{d,1}}\,\corr{\unl q{}^\a}\oplus\bigoplus_{a=0}^d\,\corr{\unl p{}^a}
\qqq
carries a natural structure of a $\gt{smink}(d,1|D_{d,1})$-module determined by the action
\qq\nn
\pLie{\cdot}\ &:&\ \gt{smink}(d,1|D_{d,1})\x\Om^1\bigl({\rm sMink}(d,1|D_{d,1})\bigr)^{{\rm sMink}(d,1|D_{d,1})}\too\Om^1\bigl({\rm sMink}(d,1|D_{d,1})\bigr)^{{\rm sMink}(d,1|D_{d,1})}\cr\cr
&:&\ \bigl(X,\om\bigr)\longmapsto\pLie{X}\om\,,
\qqq
and its Gra\ss mann-odd subspace
\qq\nn
\bigl(\Om^1\bigl({\rm sMink}(d,1|D_{d,1})\bigr)^{{\rm sMink}(d,1|D_{d,1})}\bigr)^{(1)}\equiv\bigoplus_{\a=1}^{D_{d,1}}\,\corr{\unl q{}^\a}
\qqq
is a trivial submodule.\ Accordingly,\ the GS super-3-cocycle
\qq\label{eq:GSs3cdec}
\underset{\tx{\ciut{(3)}}}{\txH}\equiv\bigl(\unl p{}^a\wedge\ovl\G{}_{a\,\a\b}\,\unl q{}^\b\bigr)\wedge\unl q{}^\a
\qqq
acquires the interpretation of a nontrivial $\bigl(\Om^1\bigl({\rm sMink}(d,1|D_{d,1})\bigr)^{{\rm sMink}(d,1|D_{d,1})}\bigr)^{(1)}$-valued super-2-cocycle on $\,\gt{smink}(d,1|D_{d,1})$,\ and as such it gives rise to the supercentral extension \eqref{eq:oddGSext}.\ The idea of trivialising the CaE super-$(p+2)$-cocycles that codefine the Green--Schwarz-type super-$\si$-models of the (half-BPS) super-minkowskian super-$p$-branes through the above Lie-superalgebraic mechanism (necessarily stepwise for $\,p>1$) was invented by de Azc\'arraga {\it et al.}\ in \Rcite{Chryssomalakos:2000xd}.\ Its adaptation to, interpretation and elaboration in the higher-geometric context under consideration constitutes the foundation of the geometrisation programme advanced by the Author in a series of papers \cite{Suszek:2017xlw,Suszek:2019cum,Suszek:2018bvx,Suszek:2018ugf,Suszek:2020xcu,Suszek:2020rev} that we turn to next.

The Lie-superalgebra extension \eqref{eq:oddGSext} integrates to a supercentral Lie-supergroup extension 
\qq\nn
\bd1\too\bR^{0|D_{d,1}}\too\sfY{\rm sMink}(d,1|D_{d,1})\too{\rm sMink}(d,1|D_{d,1})\too\bd1
\qqq
with the supermanifold structure
\qq\nn
\sfY{\rm sMink}(d,1|D_{d,1})={\rm sMink}(d,1|D_{d,1})\x\bR^{0|D_{d,1}}\ni\bigl(\theta^\a,x^a,\xi_\b\bigr)
\qqq
and the Lie-supergroup structure determined by the binary operation
\qq\nn
\sfY\unl\txm\ :\ \sfY{\rm sMink}(d,1|D_{d,1})\x\sfY{\rm sMink}(d,1|D_{d,1})\too\sfY{\rm sMink}(d,1|D_{d,1})
\qqq
with the coordinate presentation
\qq\nn
&&\sfY\unl\txm\bigl(\bigl(\theta_1^\a,x_1^a,\xi_{1\,\b}\bigr),\bigl(\theta_2^\a,x_2^a,\xi_{2\,\b}\bigr)\bigr)\cr\cr
&=&\bigl(\theta_1^\a+\theta_2^\a,x_1^a+x_2^a-\tfrac{1}{2}\,\theta_1\,\ovl\G{}^a\,\theta_2,\xi_{1\,\b}+\xi_{2\,\b}+\ovl\G{}_{b\,\b\g}\,\theta_1^\g\,x_2^b-\tfrac{1}{6}\,\bigl(\theta_1\,\ovl\G{}_b\,\theta_2\bigr)\,\ovl\G{}^b_{\b\g}\,\bigl(2\theta_1^\g+\theta_2^\g\bigr)\bigr)\,,
\qqq
ensuring the desired left-invariance of the super-1-form
\qq\nn
\unl z{}_\a(\theta,x,\xi)=\sfd\xi_\a-\ovl\G{}_{a\,\a\b}\,\theta^\b\,\bigl(\sfd x^a+\tfrac{1}{6}\,\theta\,\ovl\G{}^a\,\sfd\theta\bigr)\,.
\qqq
The corresponding basis LI vector fields are
\qq\nn
\sfY\unl Q{}_\a(\theta,x,\xi)&=&\tfrac{\vec\p\ }{\p\theta^\a}+\tfrac{1}{2}\,\ovl\G{}^a_{\a\b}\,\theta^\b\,\tfrac{\p\ }{\p x^a}+\tfrac{1}{3}\,\ovl\G{}_{a\,\a\b}\,\theta^\b\,\ovl\G{}^a_{\g\d}\,\theta^\g\,\tfrac{\vec\p\ }{\p\xi_\d}\,,\cr\cr
\sfY\unl P{}_a(\theta,x,\xi)&=&\tfrac{\p\ }{\p x^a}+\ovl\G{}_{a\,\a\b}\,\theta^\b\,\tfrac{\vec\p\ }{\p\xi_\a}\,,\cr\cr
\unl Z{}^\a(\theta,x,\xi)&=&\tfrac{\vec\p\ }{\p\xi_\a}\,.
\qqq
The idea of \Rcite{Suszek:2017xlw} was to take the epimorphism\footnote{Note that the cartesian product is \emph{not} that in $\,{\bf sLieGrp}$.}
\qq\nn
\pi_{\sfY{\rm sMink}(d,1|D_{d,1})}\equiv\pr_1\ :\ \sfY{\rm sMink}(d,1|D_{d,1})\equiv{\rm sMink}(d,1|D_{d,1})\x\bR^{0|D_{d,1}}\too{\rm sMink}(d,1|D_{d,1})
\qqq
\emph{in the category} $\,{\bf sLieGrp}\,$ of Lie supergroups,\ together with the LI primitive 
\qq\nn
\underset{\tx{\ciut{(2)}}}{\sfY\txB}:=\unl z{}_\a\wedge\pi_{\sfY{\rm sMink}(d,1|D_{d,1})}^*\unl q^\a
\qqq
of the CaE super-3-cocycle $\,\underset{\tx{\ciut{(3)}}}{\txH}\,$ on its total space $\,\sfY{\rm sMink}(d,1|D_{d,1})$, 
\qq\nn
\sfd\underset{\tx{\ciut{(2)}}}{\sfY\txB}=\pi_{\sfY{\rm sMink}(d,1|D_{d,1})}^*\underset{\tx{\ciut{(3)}}}{\txH}\,,
\qqq
as the point of departure ({\it i.e.},\ the surjective submersion and the curving,\ respectively) of the standard geometrisation procedure due to Murray,\ and subsequently bring the procedure to completion within $\,{\bf sLieGrp}$.

Thus, as the next step,\ we consider the fibred-product\footnote{Our convention on fibred products in the category $\,\sMan\,$ is given in App.\,\ref{app:conv}.} Lie supergroup
\qq\nn
\sfY^{[2]}{\rm sMink}(d,1|D_{d,1})&\equiv &\sfY{\rm sMink}(d,1|D_{d,1})\x_{{\rm sMink}(d,1|D_{d,1})}\sfY{\rm sMink}(d,1|D_{d,1})\cr\cr
&:=&\sfY{\rm sMink}(d,1|D_{d,1})\hspace{2pt}{}_{\pi_{{\rm sMink}(d,1|D_{d,1})}}\hspace{-2pt}\x_{\pi_{{\rm sMink}(d,1|D_{d,1})}}\sfY{\rm sMink}(d,1|D_{d,1})
\qqq
with the binary operation inherited from the cartesian product $\,\sfY{\rm sMink}(d,1|D_{d,1})\x\sfY{\rm sMink}(d,1|D_{d,1})\,$ of Lie supergroups through restriction.\ It admits global coordinates
\qq\nn
\sfY^{[2]}{\rm sMink}(d,1|D_{d,1})\ni\bigl((\theta,x,\xi_1),(\theta,x,\xi_2)\bigr)\,.
\qqq
On its tangent Lie superalgebra
\qq\nn
&&\sfY^{[2]}\gt{smink}(d,1|D_{d,1})\equiv\sfY\gt{smink}(d,1|D_{d,1})\oplus_{\gt{smink}(d,1|D_{d,1})}\sfY\gt{smink}(d,1|D_{d,1})\cr\cr
&=&\bigoplus_{\a=1}^{D_{d,1}}\,\corr{\bigl(\sfY\unl Q{}_\a,\sfY\unl Q{}_\a\bigr)}\oplus\bigoplus_{a=0}^d\,\corr{\bigl(\sfY\unl P{}_a,\sfY\unl P{}_a\bigr)}\oplus\bigoplus_{\b=1}^{D_{d,1}}\,\corr{\bigl(\unl Z{}^\b,0\bigr)}\oplus\bigoplus_{\g=1}^{D_{d,1}}\,\corr{\bigl(0,\unl Z{}^\g\bigr)}\,,
\qqq
endowed with (the restriction of) the standard direct-sum superbracket,\ to be denoted as $\,[\cdot,\cdot\}_\oplus$,\ we find the nontrivial super-2-cocycle
\qq\nn
\underset{\tx{\ciut{(2)}}}{\txF}=\bigl(\pr_2^*-\pr_1^*\bigr)\underset{\tx{\ciut{(2)}}}{\sfY\txB}\,,
\qqq
with the coordinate presentation
\qq\nn
\underset{\tx{\ciut{(2)}}}{\txF}\bigl((\theta,x,\xi_1),(\theta,x,\xi_2)\bigr)=\sfd\theta^\a\wedge\sfd\bigl(\xi_{2\,\a}-\xi_{1\,\a}\bigr)\,.
\qqq
In virtue of the formerly invoked correspondence,\ the super-2-cocycle determines a central extension 
\qq\nn
\brd0\too\bR\too\lgt\xrightarrow{\ \pi_\lgt\ }\sfY^{[2]}\gt{smink}(d,1|D_{d,1})\too\brd0
\qqq
with the supervector-space structure
\qq\nn
\lgt\cong\sfY^{[2]}\gt{smink}(d,1|D_{d,1})\oplus\bR\ni(X,r)
\qqq
with respect to which
\qq\nn
\pi_\lgt\equiv\pr_1\ :\ \sfY^{[2]}\gt{smink}(d,1|D_{d,1})\oplus\bR\too\sfY^{[2]}\gt{smink}(d,1|D_{d,1})\,,
\qqq
and with the Lie superbracket 
\qq\nn
[(X_1,r_1),(X,r_2)\}_{\underset{\tx{\ciut{(2)}}}{\txF}}=\bigl([X_1,X_2\}_\oplus,X_2\con X_1\con\underset{\tx{\ciut{(2)}}}{\txF}\bigr)\,,
\qqq
{\it cp} \Rxcite{App.\,C}{Suszek:2017xlw}.\ Thus,\ we have
\qq\nn
\lgt=\bigoplus_{\a=1}^{D_{d,1}}\,\corr{\xcL\unl Q{}_\a}\oplus\bigoplus_{a=0}^d\,\corr{\xcL\unl P{}_a}\oplus\bigoplus_{\b=1}^{D_{d,1}}\,\corr{\xcL\unl Z{}_{(1)}^\b}\oplus\bigoplus_{\g=1}^{D_{d,1}}\,\corr{\xcL\unl Z{}_{(2)}^\g}\oplus\corr{\unl\cZ}
\qqq
with the structure equations (in which we drop the subscript $\,\underset{\tx{\ciut{(2)}}}{\txF}\,$ on the superbrackets)
\qq\nn
&\{\xcL\unl Q{}_\a,\xcL\unl Q{}_\b\}=\ovl\G{}^a_{\a\b}\,\xcL\unl P{}_a\,,\qquad\qquad[\xcL\unl P{}_a,\xcL\unl P{}_b]=0\,,\qquad\qquad[\xcL\unl Q{}_\a,\xcL\unl P{}_a]=\ovl\G_{a\,\a\b}\,\bigl(\xcL\unl Z{}_{(1)}^\b+\xcL\unl Z{}_{(2)}^\b\bigr)\,,&\cr\cr
&-\{\xcL\unl Q{}_\a,\xcL\unl Z{}_{(1)}^\b\}=\d_\a^{\ \b}\,\unl\cZ=\{\xcL\unl Q{}_\a,\xcL\unl Z{}_{(2)}^\b\}\,,\qquad\qquad[\xcL\unl P{}_a,\xcL\unl Z{}_{(n)}^\b]=0\,,\qquad\qquad\{\xcL\unl Z{}_{(m)}^\a,\xcL\unl Z{}_{(n)}^\b\}=0\,,&\cr\cr
&[\xcL\unl Q{}_\a,\unl\cZ]=0\,,\qquad\qquad[\xcL\unl P{}_a,\unl\cZ]=0\,,\qquad\qquad[\xcL\unl Z{}_{(n)}^\a,\unl\cZ]=0\,,\qquad\qquad[\unl\cZ,\unl\cZ]=0\,.&
\qqq
Upon denoting the super-1-form dual to the central generator $\,\unl\cZ\,$ as $\,\unl\z$,\ we readily establish the identity
\qq\nn
\widehat\d\unl\z=\pi_\lgt^*\underset{\tx{\ciut{(2)}}}{\txF}\,.
\qqq
As before,\ the Lie-superalgebra extension integrates to a central Lie-supergroup extension -- this time,\ we obtain
\qq\nn
\bd1\too\bC^\x\too\xcL\too\sfY^{[2]}{\rm sMink}(d,1|D_{d,1})\too\bd1
\qqq
with the supermanifold structure
\qq\nn
\xcL=\sfY^{[2]}{\rm sMink}(d,1|D_{d,1})\x\bC^\x\ni\bigl(\bigl(\theta^\a,x^a,\xi_{1\,\b}\bigr), \bigl(\theta^\a,x^a,\xi_{2\,\b}\bigr),z\bigr)
\qqq
and the Lie-supergroup structure determined by the binary operation
\qq\nn
\xcL\unl\txm\ :\ \xcL\x\xcL\too\xcL
\qqq
with the coordinate presentation
\qq\nn
&&\xcL\unl\txm\bigl(\bigl((\theta_1,x_1,\xi_{1,1}),(\theta_1,x_1,\xi_{1,2}),z_1\bigr),\bigl((\theta_2,x_2,\xi_{2,1}),(\theta_2,x_2,\xi_{2,2}),z_2\bigr)\bigr)\cr\cr
&=&\bigl(\sfY\unl\txm\bigl((\theta_1,x_1,\xi_{1,1}),(\theta_2,x_2,\xi_{2,1})\bigr),\sfY\unl\txm\bigl((\theta_1,x_1,\xi_{1,2}),(\theta_2,x_2,\xi_{2,2})\bigr),\ee^{\sfi\,\theta_1^\a\,(\xi_{2,2}-\xi_{2,1})_\a}\cdot z_1\cdot z_2\bigr)
\qqq
such that the super-1-form
\qq\nn
\unl\z\bigl((\theta,x,\xi_1),(\theta,x,\xi_2),z\bigr)=\tfrac{\sfi\,\sfd z}{z}+\theta^\a\,\sfd\bigl(\xi_{2\,\a}-\xi_{1\,\a}\bigr)=:\tfrac{\sfi\,\sfd z}{z}+\txa\bigl((\theta,x,\xi_1),(\theta,x,\xi_2)\bigr)
\qqq
is LI.\ The extension has the structure of a (trivial) principal $\bC^\x$-bundle 
\qq\nn
\pi_\xcL\equiv\pr_1\ :\ \xcL\equiv\sfY^{[2]}{\rm sMink}(d,1|D_{d,1})\x\bC^\x\too\sfY^{[2]}{\rm sMink}(d,1|D_{d,1})
\qqq
with an obvious fibrewise action of the structure group $\,\bC^\x\,$ and with the LI principal connection super-1-form
\qq\nn
\underset{\tx{\ciut{(1)}}}{\txA}{}_\xcL\equiv\unl\z
\qqq
of curvature $\,\underset{\tx{\ciut{(2)}}}{\txF}$,
\qq\nn
\sfd\underset{\tx{\ciut{(1)}}}{\txA}{}_\xcL=\pi_\xcL^*\underset{\tx{\ciut{(2)}}}{\txF}\,.
\qqq

Finally,\ we consider the fibred-product Lie supergroup
\qq\nn
\sfY^{[3]}{\rm sMink}(d,1|D_{d,1})\equiv\sfY{\rm sMink}(d,1|D_{d,1})\x_{{\rm sMink}(d,1|D_{d,1})}\sfY{\rm sMink}(d,1|D_{d,1})\x_{{\rm sMink}(d,1|D_{d,1})}\sfY{\rm sMink}(d,1|D_{d,1})
\qqq
(defined analogously to the fibred square $\,\sfY^{[2]}{\rm sMink}(d,1|D_{d,1})$) and,\ over it,\ the pullback bundles
\qq\nn
\pi_{\pr_{i,j}^*\xcL}\equiv\pr_1\ :\ \pr_{i,j}^*\xcL\equiv\sfY^{[3]}{\rm sMink}(d,1|D_{d,1})\hspace{2pt}{}_{\pr_{i,j}}\hspace{-2pt}\x_{\pi_\xcL}\xcL\too \sfY^{[3]}{\rm sMink}(d,1|D_{d,1})
\qqq
endowed with the Lie-supergroup structure obtained,\ through restriction,\ from the product one on $\,\sfY^{[3]}{\rm sMink}(d,1|D_{d,1})\x\xcL\supset\sfY^{[3]}{\rm sMink}(d,1|D_{d,1})\x_{\pr_{i,j}}\xcL$.\ On the fibred cube $\,\sfY^{[3]}{\rm sMink}(d,1|D_{d,1})$,\ we have coordinates
\qq\nn
\bigl((\theta,x,\xi_1),(\theta,x,\xi_2),(\theta,x,\xi_3)\bigr)\equiv(y_1,y_2,y_3)\,,
\qqq
and so for the pullback bundles,\ we obtain coordinates
\qq\nn
\pr_{i,j}^*\xcL\ni\bigl((y_1,y_2,y_3),(y_i,y_j,z)\bigr)\equiv\bigl(y_{1,2,3},(y_{i,j},z)\bigr)\,.
\qqq
In these,\ the induced binary operation 
\qq\nn
\xcL_{i,j}\unl\txm\ :\ \pr_{i,j}^*\xcL\x\pr_{i,j}^*\xcL\too\pr_{i,j}^*\xcL
\qqq
reads
\qq\nn
&&\xcL_{i,j}\unl\txm\bigl(\bigl(y^1_{1,2,3},\bigl(y^1_{i,j},z_1\bigr)\bigr),\bigl(y^2_{1,2,3},\bigl(y^2_{i,j},z_2\bigr)\bigr)\bigr)\cr\cr
&=&\bigl(\bigl(\sfY\unl\txm(y^1_1,y^2_1),\sfY\unl\txm(y^1_2,y^2_2),\sfY\unl\txm(y^1_3,y^2_3)\bigr),\xcL\unl\txm\bigl((y^1_{i,j},z_1),(y^2_{i,j},z_2)\bigr)\bigr)\,.
\qqq
Out of the first two pullback bundles,\ $\,\pr_{1,2}^*\xcL\,$ and $\,\pr_{2,3}^*\xcL$,\ we form the tensor-product principal $\bC^\x$-bundle \label{pref:tensLie}
\qq\nn
[\pi_\xcL\circ\pr_1]\ :\ \pr_{1,2}^*\xcL\ox\pr_{2,3}^*\xcL\too\sfY^{[3]}{\rm sMink}(d,1|D_{d,1})\,,
\qqq
defined as the associated bundle
\qq\nn
\pr_{1,2}^*\xcL\ox\pr_{2,3}^*\xcL=\bigl(\pr_{1,2}^*\xcL\hspace{2pt}{}_{\pr_1}\hspace{-2pt}\x_{\pr_1}\pr_{2,3}^*\xcL\bigr)/\bC^\x\,,
\qqq
with the projection to the base (written out in coordinates)
\qq\nn
[\pi_\xcL\circ\pr_1]\bigl(\bigl(y_{1,2,3},(y_{1,2},1)\bigr)\ox\bigl(y_{1,2,3},(y_{2,3},z)\bigr)\bigr)=y_{1,2,3}\,.
\qqq
Here,\ we are quotienting out\footnote{Formally,\ we perform the quotienting in the body and subsequently take the sub-sheaf composed of $\la$-invariant sections in the structure sheaf of $\,\pr_{1,2}^*\xcL\x_{\sfY^{[3]}{\rm sMink}(d,1|D_{d,1})}\pr_{2,3}^*\xcL\,$ as the structure sheaf of the quotient supermanifold.} the `diagonal' action 
\qq\nn
\la\ :\ \bC^\x\x\bigl(\pr_{1,2}^*\xcL\hspace{2pt}{}_{\pr_1}\hspace{-2pt}\x_{\pr_1}\pr_{2,3}^*\xcL\bigr)\too \pr_{1,2}^*\xcL\hspace{2pt}{}_{\pr_1}\hspace{-2pt}\x_{\pr_1}\pr_{2,3}^*\xcL
\qqq
with the coordinate presentation
\qq\nn
\la\bigl(z,\bigl(y_{1,2,3},(y_{1,2},z_1)\bigr),\bigl(y_{1,2,3},(y_{2,3},z_2)\bigr)\bigr)=\bigl(\bigl(y_{1,2,3},(y_{1,2},z_1\cdot z)\bigr),\bigl(y_{1,2,3},\bigl(y_{2,3},z_2\cdot z^{-1}\bigr)\bigr)\bigr)\,.
\qqq
The tensor-product bundle inherits a natural Lie-supergroup structure from (the restricted product one on) $\,\pr_{1,2}^*\xcL\hspace{2pt}{}_{\pr_1}\hspace{-2pt}\x_{\pr_1}\pr_{2,3}^*\xcL$,
\qq\nn
[\xcL_{1,2;2,3}\unl\txm]\ :\ \bigl(\pr_{1,2}^*\xcL\ox\pr_{2,3}^*\xcL\bigr)\x\bigl(\pr_{1,2}^*\xcL\ox\pr_{2,3}^*\xcL\bigr)\too\pr_{1,2}^*\xcL\ox\pr_{2,3}^*\xcL\,,
\qqq
with the coordinate presentation
\qq\nn
&&[\xcL_{1,2;2,3}\unl\txm]\bigl(\bigl(y^1_{1,2,3},\bigl(y^1_{1,2},1\bigr)\bigr)\ox\bigl(y^1_{1,2,3},\bigl(y^1_{2,3},z_1\bigr)\bigr),\bigl(y^2_{1,2,3},\bigl(y^2_{1,2},1\bigr)\bigr)\ox\bigl(y^2_{1,2,3},\bigl(y^2_{2,3},z_2\bigr)\bigr)\bigr)\cr\cr
&=&\xcL_{1,2}\unl\txm\bigl(\bigl(y^1_{1,2,3},\bigl(y^1_{1,2},1\bigr)\bigr),\bigl(y^2_{1,2,3},\bigl(y^2_{1,2},1\bigr)\bigr)\bigr)\ox\xcL_{2,3}\unl\txm\bigl(\bigl(y^1_{1,2,3},\bigl(y^1_{2,3},z_1\bigr)\bigr),\bigl(y^2_{1,2,3},\bigl(y^2_{2,3},z_2\bigr)\bigr)\bigr)\,.
\qqq
At this stage,\ it suffices to compare the base components of the principal connection super-1-forms on $\,\pr_{1,2}^*\xcL\ox\pr_{2,3}^*\xcL\,$ and $\,\pr_{1,3}^*\xcL$,
\qq\nn
\bigl(\pr_{1,2}^*\txa+\pr_{2,3}^*\txa\bigr)(y_{1,2,3})=\pr_{1,3}^*\txa(y_{1,2,3})\,,
\qqq
to infer the existence of a connection-preserving isomorphism of principal $\bC^\x$-bundles
\qq\nn
\mu_\xcL\ :\ \pr_{1,2}^*\xcL\ox\pr_{2,3}^*\xcL\xrightarrow{\ \cong\ }\pr_{1,3}^*\xcL
\qqq
with the coordinate presentation
\qq\nn
\mu_\xcL\bigl(\bigl(y_{1,2,3},(y_{1,2},1)\bigr)\ox\bigl(y_{1,2,3},(y_{2,3},z)\bigr)\bigr)=\bigl(y_{1,2,3},(y_{1,3},z)\bigr)\,.
\qqq
We reserve the suggestive (symbolic) notation
\qq\nn
\mu_\xcL\equiv\bd1
\qqq
for an isomorphism of the above trivial form.\ The isomorphism satisfies the coherence (groupoid) identity
\qq\nn
\pr_{1,3,4}^*\mu_\xcL\circ\bigl(\pr_{1,2,3}^*\mu_\xcL\ox\id_{\pr_{3,4}^*\xcL}\bigr)=\pr_{1,2,4}^*\mu_\xcL\circ\bigl(\id_{\pr_{1,2}^*\xcL}\ox\pr_{2,3,4}^*\mu_\xcL\bigr)
\qqq
over the quadruple fibred product $\,\sfY^{[4]}{\rm sMink}(d,1|D_{d,1})$,\ the latter being equipped with the canonical projections $\,\pr_{i,j,k}\ :\ \sfY^{[4]}{\rm sMink}(d,1|D_{d,1})\too\sfY^{[3]}{\rm sMink}(d,1|D_{d,1}),\ (i,j,k)\in\{(1,2,3),(1,3,4),(2,3,4),$ $(1,2,4)\}\,$ and $\,\pr_{m,n}\ :\ \sfY^{[4]}{\rm sMink}(d,1|D_{d,1})\too\sfY^{[2]}{\rm sMink}(d,1|D_{d,1}),\ (m,n)\in\{(1,2),(3,4)\}\,$ (defined in an obvious manner).\ Clearly,\ $\,\mu_\xcL\,$ is also a Lie-supergroup isomorphism,
\qq\nn
\mu_\xcL\circ[\xcL_{1,2;2,3}\unl\txm]=\xcL_{1,3}\unl\txm\circ\bigl(\mu_\xcL\x\mu_\xcL\bigr)\,.
\qqq

The 1-gerbe
\qq\nn
\cG^{(1)}_{\rm GS}:=\bigl(\sfY{\rm sMink}(d,1|D_{d,1}),\pi_{\sfY{\rm sMink}(d,1|D_{d,1})},\underset{\tx{\ciut{(2)}}}{\sfY\txB},\xcL,\pi_\xcL,\underset{\tx{\ciut{(1)}}}{\txA}{}_\xcL,\mu_\xcL\bigr)
\qqq
was named the {\bf Green--Schwarz super-1-gerbe over $\,{\rm sMink}(d,1|D_{d,1})\,$} in \Rxcite{Def.\,5.9}{Suszek:2017xlw}.\ It is an example of a {\bf Cartan--Eilenberg super-1-gerbe} (over the Lie supergroup $\,{\rm sMink}(d,1|D_{d,1})$),\ that is a distinguished 1-gerbe in the category of Lie supergroups.\ Its existence and equivariance properties,\ the latter to be discussed at length in Sec.\,\ref{sec:susyequiv},\ are markers of a quantum-mechanical consistency of the GS super-$\si$-model of the superstring in $\,{\rm sMink}(d,1|D_{d,1})$.\ As it stands, the super-1-gerbe is naturally associated with the NG formulation that can be phrased in terms of the differential-geometric data of the homogeneous space $\,{\rm sMink}(d,1|D_{d,1})\,$ of the mother supersymmetry group $\,{\rm sISO}(d,1|D_{d,1})\,$ exclusively.\ Instrumental in its construction is the `accidental' Lie-supergroup structure on this particular homogeneous space.\ In the case of a generic homogeneous space $\,\txG/\txH\,$ of a supersymmetry Lie supergroup $\,\txG\,$ associated with a Lie subgroup $\,\txH\subset|\txG|\,$ of its body $\,|\txG|$,\ the geometrisation scheme exemplified above,\ making use of the relations between the Cartan--Eilenberg cohomology of the Lie supergroup and the Chevalley--Eilenberg cohomology of its tangent Lie superalgebra and the interpretation of the distinguished second cohomology group of the latter,\ has to start on $\,\txG\,$ and only in the end \emph{descend} to $\,\txG/\txH$.\ The cohomology to be geometrised under such circumstances is $\,{\rm CaE}^\bullet(\txG)\,$ further restricted to $\txH$-basic super-forms,\ and -- in the light of the findings of Refs.\,\cite{Gawedzki:2010rn,Suszek:2011,Suszek:2012ddg,Gawedzki:2012fu,Suszek:2013} -- the CaE super-1-gerbe that we seek to erect over $\,\txG\,$ has to carry a \emph{descendable} $\txH$-equivariant structure\footnote{This notion shall be recalled and illustrated in Sec.\,\ref{sec:susyequiv}}.\ The advantage of carrying out the geometrisation over $\,\txG\,$ is that it yields a higher-geometric object that can be restricted directly to the HP section $\,\Si^{\rm HP}\,$ and used there in a rigorous study of the dual HP formulation upon incorporating a suitable correction coming from the LI volume super-2-form $\,\Vol(\tgt^{(0)}_{\rm vac})$.\ In order to attain the same goal in the super-minkowskian setting under consideration,\ we need to walk the path laid out above in the reverse direction,\ that is we must lift the GS super-1-gerbe to $\,{\rm sISO}(d,1|D_{d,1})$.\ An obvious thing to do would be to pull back $\,\cG^{(1)}_{\rm GS}\,$ along the canonical projection $\,\pi\,$ of \Reqref{diag:princsbndlsMink} -- this operation was actually employed in the original studies of supersymmetry-equivariance of the GS super-1-gerbe of the superstring in $\,{\rm sMink}(d,1|D_{d,1})\,$ that was reported in \Rcite{Suszek:2019cum}.\ The obvious drawback of this idea is that it depends on the supermanifold morphism $\,\pi\,$ that is \emph{not} a Lie-supergroup homomorphism,\ and so the lift takes us out of the category of CaE super-1-gerbes that seems to be the most adequate one for our physically motivated purposes.\ Below, we propose an alternative construction that effectively circumnavigates the obstacle encoutered on our way towards `gerbification' of the super-minkowskian GS super-$\si$-model in the supergeometrically largely tractable HP formulation.\bigskip

The idea that we wish to pursue now consists in lifting the supercentral extension $\,\sfY\gt{smink}(d,1|D_{d,1})\,$ of the super-minkowskian Lie superalgebra \emph{equivariantly} to an extension of the full supersymmetry Lie superalgebra $\,\gt{siso}(d,1|D_{d,1})$.\ Taking a closer look at the lift \eqref{eq:GS3coc} of the GS super-3-cocycle \eqref{eq:GSs3cdec},
\qq\nn
\underset{\tx{\ciut{(3)}}}{\chi}\equiv\bigl(p^a\wedge\ovl\G{}_{a\,\a\b}\,q^\b\bigr)\wedge q^\a\,,
\qqq 
and the precise relation between the lifted LI super-1-forms $\,q^\a\,$ in it and the $\,\unl q{}^\a\,$ that we previously identified as the basis of the (trivial) $\gt{smink}(d,1|D_{d,1})$-module defining the extension, 
\qq\label{eq:liftZs}
q^\a(\theta,x,\phi)=S(\phi)^{-1\,\a}_{\ \ \ \ \b}\,\unl q{}^\b(\theta,x)\,,
\qqq
we are readily led to postulate the lift 
\qq\nn
\sfY\gt{siso}(d,1|D_{d,1})=\bigl(\bigoplus_{\a=1}^{D_{d,1}}\,\corr{\sfY Q_\a}\oplus\bigoplus_{a=0}^d\,\corr{\sfY P_a}\oplus\bigoplus_{\b=1}^{D_{d,1}}\,\corr{Z^\b}\bigr)\oplus\bigoplus_{b<c=0}^d\,\corr{\sfY J_{bc}}\equiv\sfY\gt{smink}(d,1|D_{d,1})\oplus\gt{spin}(d,1)
\qqq
in the form of a $\gt{spin}(d,1)$-module Lie superalgebra with superbrackets
\qq\nn
&\{\sfY Q_\a,\sfY Q_\b\}=\ovl\G{}^a_{\a\b}\,\sfY P_a\,,\qquad\qquad[\sfY P_a,\sfY P_b]=0\,,\qquad\qquad[\sfY Q_\a,\sfY P_a]=\ovl\G_{a\,\a\b}\,Z^\b\,,&\cr\cr
&\{\sfY Q_\a,Z^\b\}=0\,,\qquad\qquad[\sfY P_a,Z^\a]=0\,,\qquad\qquad\{Z^\a,Z^\b\}=0\,,&\cr\cr
&[\sfY J_{ab},\sfY Q_\a]=\tfrac{1}{2}\,\G_{ab}{}^\b_{\ \a}\,\sfY Q_\b\,,\qquad\qquad[\sfY J_{ab},\sfY P_c]=\eta_{bc}\,\sfY P_a-\eta_{ac}\,\sfY P_b\,,\qquad\qquad[\sfY J_{ab},Z^\a]=-\tfrac{1}{2}\,\G_{ab}{}^\a_{\ \b}\,Z^\b\,,&\cr\cr
&[\sfY J_{ab},\sfY J_{cd}]=\eta_{ad}\,\sfY J_{bc}-\eta_{ac}\,\sfY J_{bd}+\eta_{bc}\,\sfY J_{ad}-\eta_{bd}\,\sfY J_{ac}\,,&
\qqq
that extends the analogous structure on the equivariant lift $\,\gt{siso}(d,1|D_{d,1})\,$ of $\,\gt{smink}(d,1|D_{d,1})$.\ Thus,\ the extra generators $\,Z^\a\,$ transform under $\,\gt{spin}(d,1)\,$ as spinors,\ in conformity with \Reqref{eq:liftZs}.\ The only components of the ensuing super-Jacobiator that are not trivially null ({\it e.g.}, because of being identical with their un-extended counterparts) read
\qq\nn
{\rm sJac}\bigl(\sfY Q_\a,\sfY Q_\b,\sfY Q_\g\bigr)&=&3!\,\ovl\G{}_{a\,(\a\b}\,\ovl\G{}^a_{\g)\d}\,Z^\d\,,\cr\cr
{\rm sJac}\bigl(\sfY Q_\a,\sfY P_a,\sfY J_{bc}\bigr)&=&\bigl(\bigl(\ovl\G{}_a\,\G_{bc}\bigr)_{(\a\b)}-\eta_{ab}\,\ovl\G{}_{c\,\a\b}+\eta_{ac}\,\ovl\G{}_{b\,\a\b}\bigr)\,Z^\b\,,\cr\cr
{\rm sJac}\bigl(Z^\a,\sfY J_{ab},\sfY J_{cd}\bigr)&=&\tfrac{1}{4}\,\bigl(\G_{ab}\,\G_{cd}-\G_{cd}\,\G_{ab}-2\eta_{ad}\,\G_{bc}+2\eta_{ac}\,\G_{bd}-2\eta_{bc}\,\G_{ad}+2\eta_{bd}\,\G_{ac}\bigr)^\a_{\ \b}\,Z^\b\,.
\qqq
The first of these vanishes in virtue of the Fierz identities \eqref{eq:Fierz}.\ Next,\ we compute
\qq\nn
\bigl(\ovl\G{}_a\,\G_{bc}\bigr)^{\rm T}=\ovl\G{}_{cb}\,\G_a\,,
\qqq
and use it to perform the reduction 
\qq\nn
\ovl\G{}_a\,\G_{bc}+\bigl(\ovl\G{}_a\,\G_{bc}\bigr)^{\rm T}-2\eta_{ab}\,\ovl\G_c+2\eta_{ac}\,\ovl\G_b=0\,,
\qqq
which implies that the second component of the super-Jacobiator is zero.\ Finally,\ the third component can be rewritten as
\qq\nn
{\rm sJac}\bigl(Z^\a,\sfY J_{ab},\sfY J_{cd}\bigr)\equiv-q^\a\bigl({\rm sJac}\bigl(\sfY Q_\b,\sfY J_{ab},\sfY J_{cd}\bigr)\bigr)\,Z^\b=0\,,
\qqq
which concludes the proof of existence of a Lie-superalgebra structure on $\,\sfY\gt{siso}(d,1|D_{d,1})\,$ with the superbracket postulated above.\ Denote as
\qq\nn
\pi_{\sfY\gt{siso}(d,1|D_{d,1})}\ :\ \sfY\gt{siso}(d,1|D_{d,1})\too\gt{siso}(d,1|D_{d,1})
\qqq
the Lie-superalgebra epimorphism obtained by linearly extending the assignment
\qq\nn
\pi_{\sfY\gt{siso}(d,1|D_{d,1})}\ :\ \bigl(\sfY Q_\a,\sfY P_a,Z^\b,\sfY J_{bc}\bigr)\longmapsto\bigl(Q_\a,P_a,0,J_{bc}\bigr)\,,
\qqq
and let the duals of the generators $\,Z^\a\,$ be $\,z_\a$.\ We then obtain the desired identity
\qq\nn
\widehat\d\bigl(z_\a\wedge\pi_{\sfY\gt{siso}(d,1|D_{d,1})}^*q^\a\bigr)=\pi_{\sfY\gt{siso}(d,1|D_{d,1})}^*\underset{\tx{\ciut{(3)}}}{\chi}\,.
\qqq
Next,\ we readily enhance the Lie superalgebra $\,\sfY\gt{siso}(d,1|D_{d,1})\,$ to a super-Harish--Chandra pair ({\it i.e.},\ to a Lie supergroup)
\qq\nn
\sfY{\rm sISO}(d,1|D_{d,1})=\bigl(\widetilde{\rm ISO}(d,1),\sfY\gt{siso}(d,1|D_{d,1})\bigr)
\qqq
with the body Lie group $\,\widetilde{\rm ISO}(d,1)\,$ realised on the Gra\ss mann-odd component 
\qq\nn
\sfY\gt{siso}(d,1|D_{d,1})^{(1)}\equiv\bigoplus_{\a=1}^{D_{d,1}}\,\corr{\sfY Q_\a}\oplus\bigoplus_{\b=1}^{D_{d,1}}\,\corr{Z^\b}
\qqq
of $\,\sfY\gt{siso}(d,1|D_{d,1})\,$ as
\qq\nn
\sfY\rho\ :\ \bR^{\x d+1}\rx_L{\rm Spin}(d,1)\too\End\bigl(\sfY\gt{siso}(d,1|D_{d,1})^{(1)}\bigr)\ :\ (x,\phi)\longmapsto S(\phi)^{\rm T}\oplus S(\phi)^{-1}\equiv\sfY\rho(x,\phi)\,.
\qqq
Thus,\ $\,\pi_{\sfY\gt{siso}(d,1|D_{d,1})}\,$ integrates to a Lie-supergroup epimorphism 
\qq\nn
\pi_{\sfY{\rm sISO}(d,1|D_{d,1})}=\pr_1\x\id_{{\rm Spin}(d,1)}\ :\ \sfY{\rm sISO}(d,1|D_{d,1})\equiv\sfY{\rm sMink}(d,1|D_{d,1})\rx_{L,S,S^{-{\rm T}}}{\rm Spin}(d,1)\too{\rm sISO}(d,1|D_{d,1})\,,
\qqq
and we have a coordinate description of the Lie supergroup
\qq\nn
\sfY{\rm sMink}(d,1|D_{d,1})\rx_{L,S,S^{-{\rm T}}}{\rm Spin}(d,1)\ni\bigl(\theta^\a,x^a,\xi_\b,\phi^{bc}\bigr)
\qqq
in which the binary operation 
\qq\nn
\sfY\txm\ :\ \sfY{\rm sISO}(d,1|D_{d,1})\x\sfY{\rm sISO}(d,1|D_{d,1})\too\sfY{\rm sISO}(d,1|D_{d,1})
\qqq
takes the form
\qq\nn
\sfY\txm\bigl(\bigl(\theta_1^\a,x_1^a,\xi_{1\,\b},\phi_1^{bc}\bigr),\bigl(\theta_2^\a,x_2^a,\xi_{2\,\b},\phi_2^{bc}\bigr)\bigr)=\bigl(\theta_1^\a+S(\phi_1)^\a_{\ \g}\,\theta_2^\g,x_1^a+L(\phi_1)^a_{\ d}\,x_2^d-\tfrac{1}{2}\,\theta_1\,\ovl\G{}^a\,S(\phi_1)\,\theta_2,\cr\cr
\xi_{1\,\b}+\xi_{2\,\g}\,S(\phi_1)^{-1\,\g}_{\ \ \ \ \b}+\ovl\G{}_{d\,\b\g}\,\theta_1^\g\,L(\phi_1)^d_{\ e}\,x_2^e-\tfrac{1}{6}\,\bigl(\theta_1\,\ovl\G{}_d\,S(\phi_1)\,\theta_2\bigr)\,\ovl\G{}^d_{\b\g}\,\bigl(2\theta_1^\g+S(\phi_1)^\g_{\ \d}\,\theta_2^\d\bigr),\bigl(\phi_1\star\phi_2\bigr)^{bc}\bigr)\,.
\qqq
The LI super-1-forms $\,z_\a\,$ admit the explicit coordinate presentation
\qq\nn
z_\a(\theta,x,\xi,\phi)=\unl z{}_\b(\theta,x,\xi)\,S(\phi)^\b_{\ \a}
\qqq
and we arrive at the anticipated identity
\qq\nn
\pi_{\sfY{\rm sISO}(d,1|D_{d,1})}^*\underset{\tx{\ciut{(3)}}}{\chi}=\sfd\bigl(z_\a\wedge\pi_{\sfY{\rm sISO}(d,1|D_{d,1})}^*q^\a\bigr)\,,
\qqq
whence also the choice of the curving
\qq\nn
\underset{\tx{\ciut{(2)}}}{\sfY\b}:=z_\a\wedge\pi_{\sfY{\rm sISO}(d,1|D_{d,1})}^*q^\a
\qqq
of the CaE super-1-gerbe over $\,{\rm sISO}(d,1|D_{d,1})\,$ under reconstruction.\ The canonical projection $\,\pi\,$ of Diag.\,\eqref{diag:princsbndlsMink} lifts to the extensions as the supermanifold morphism
\qq\nn
\sfY\pi\equiv\pr_1\ :\ \sfY{\rm sISO}(d,1|D_{d,1})\equiv\sfY{\rm sMink}(d,1|D_{d,1})\x{\rm Spin}(d,1)\too\sfY{\rm sMink}(d,1|D_{d,1})
\qqq
with the property
\qq\nn
\pi\circ \pi_{\sfY{\rm sISO}(d,1|D_{d,1})}=\pi_{\sfY{\rm sMink}(d,1|D_{d,1})}\circ\sfY\pi\,,
\qqq
and we establish the descent relation
\qq\nn
\underset{\tx{\ciut{(2)}}}{\sfY\b}\equiv\sfY\pi^*\underset{\tx{\ciut{(2)}}}{\sfY\txB}\,.
\qqq
On the level of the underlying supervector spaces,\ we have the corresponding linear maps
\qq\nn
\sfY p\equiv\pr_1\ :\ \sfY\gt{smink}(d,1|D_{d,1})\oplus\gt{spin}(d,1)\too\sfY\gt{smink}(d,1|D_{d,1})
\qqq
trivially satisfying the identity
\qq\nn
\pi_{\sfY\gt{smink}(d,1|D_{d,1})}\circ\sfY p=p\circ\pi_{\sfY\gt{siso}(d,1|D_{d,1})}\,,
\qqq
and the corresponding relation
\qq\nn
\underset{\tx{\ciut{(2)}}}{\sfY\b}\equiv\sfY p^*\underset{\tx{\ciut{(2)}}}{\sfY\txB}
\qqq
between super-2-forms on the respective Lie superalgebras.

From this point onwards,\ the construction proceeds along the same lines as for $\,{\rm sMink}(d,1|D_{d,1})$.\ Thus,\ we take the fibred-square Lie super group
\qq\nn
\sfY^{[2]}{\rm sISO}(d,1|D_{d,1})&\equiv&\sfY{\rm sISO}(d,1|D_{d,1})\x_{{\rm sISO}(d,1|D_{d,1})}\sfY{\rm sISO}(d,1|D_{d,1})\cr\cr
&\cong&\sfY^{[2]}{\rm sMink}(d,1|D_{d,1})\rx_{L,S,S^{-{\rm T}}\x S^{-{\rm T}}}{\rm Spin}(d,1)\,,
\qqq
endowed with the canonical projection
\qq\nn
\sfY^{[2]}\pi\equiv\sfY\pi\x\sfY\pi\ :\ \sfY{\rm sISO}(d,1|D_{d,1})\x_{{\rm sISO}(d,1|D_{d,1})}\sfY{\rm sISO}(d,1|D_{d,1})\cr\cr
\too\sfY{\rm sMink}(d,1|D_{d,1})\x_{{\rm sMink}(d,1|D_{d,1})}\sfY{\rm sMink}(d,1|D_{d,1})\,,
\qqq
and its tangent Lie superalgebra
\qq\nn
&&\sfY^{[2]}\gt{siso}(d,1|D_{d,1})\equiv\sfY\gt{siso}(d,1|D_{d,1})\oplus_{\gt{siso}(d,1|D_{d,1})}\sfY\gt{siso}(d,1|D_{d,1})\cr\cr
&=&\bigoplus_{\a=1}^{D_{d,1}}\,\corr{\bigl(\sfY Q_\a,\sfY Q_\a\bigr)}\oplus\bigoplus_{a=0}^d\,\corr{\bigl(\sfY P_a,\sfY P_a\bigr)}\oplus\bigoplus_{\b=1}^{D_{d,1}}\,\corr{\bigl(Z^\b,0\bigr)}\oplus\bigoplus_{\g=1}^{D_{d,1}}\,\corr{\bigl(0,Z^\g\bigr)}\oplus\bigoplus_{b<c=0}^d\,\corr{\bigl(\sfY J_{bc},\sfY J_{bc}\bigr)}\cr\cr
&\cong &\sfY^{[2]}\gt{smink}(d,1|D_{d,1})\oplus\gt{spin}(d,1)\,,
\qqq
coming with the supervector-space projection
\qq\nn
\sfY^{[2]}p\equiv\sfY p\oplus\sfY p\equiv\pr_1\ :\ \sfY^{[2]}\gt{smink}(d,1|D_{d,1})\oplus\gt{spin}(d,1)\too\sfY^{[2]}\gt{smink}(d,1|D_{d,1})\,.
\qqq
The nontrivial super-2-cocycle 
\qq\nn
\underset{\tx{\ciut{(2)}}}{\cF}=\bigl(\pr_2^*-\pr_1^*\bigr)\underset{\tx{\ciut{(2)}}}{\sfY\b}\equiv\sfY^{[2]}p^*\underset{\tx{\ciut{(2)}}}{\txF}
\qqq
on $\,\sfY^{[2]}\gt{siso}(d,1|D_{d,1})\,$ engenders a central extension
\qq\label{eq:extbunext}
\brd0\too\bR\too\widetilde\lgt\xrightarrow{\ \pi_{\widetilde\lgt}\ }\sfY^{[2]}\gt{siso}(d,1|D_{d,1})\too\brd0
\qqq
with the supervector-space structure
\qq\nn
\widetilde\lgt=\bigl(\bigoplus_{\a=1}^{D_{d,1}}\,\corr{\widetilde\xcL Q_\a}\oplus\bigoplus_{a=0}^d\,\corr{\widetilde\xcL P_a}\oplus\bigoplus_{\b=1}^{D_{d,1}}\,\corr{\widetilde\xcL Z_{(1)}^\b}\oplus\bigoplus_{\g=1}^{D_{d,1}}\,\corr{\widetilde\xcL Z_{(2)}^\g}\bigr)\oplus\corr{\cZ}\oplus\bigoplus_{b<c=0}^d\,\corr{\widetilde\xcL J_{bc}}\equiv\lgt\oplus\gt{spin}(d,1)
\qqq
with respect to which
\qq\nn
\pi_{\widetilde\lgt}\equiv\pi_\lgt\oplus\id_{\gt{spin}(d,1)}\ :\ \lgt\oplus\gt{spin}(d,1)\too\sfY^{[2]}\gt{siso}(d,1|D_{d,1})\,,
\qqq
and with the Lie superbracket
\qq\nn
&\{\widetilde\xcL Q_\a,\widetilde\xcL Q_\b\}=\ovl\G{}^a_{\a\b}\,\widetilde\xcL P_a\,,\qquad\qquad[\widetilde\xcL P_a,\widetilde\xcL P_b]=0\,,\qquad\qquad[\widetilde\xcL Q_\a,\widetilde\xcL P_a]=\ovl\G_{a\,\a\b}\,\bigl(\widetilde\xcL Z_{(1)}^\b+\widetilde\xcL Z_{(2)}^\b\bigr)\,,&\cr\cr
&-\{\widetilde\xcL Q_\a,\widetilde\xcL Z_{(1)}^\b\}=\d_\a^{\ \b}\,\cZ=\{\widetilde\xcL Q_\a,\widetilde\xcL Z_{(2)}^\b\}\,,\qquad\qquad[\widetilde\xcL P_a,\widetilde\xcL Z_{(m)}^\a]=0\,,\qquad\qquad\{\widetilde\xcL Z_{(m)}^\a,\widetilde\xcL Z_{(n)}^\b\}=0\,,&\cr\cr
&[\widetilde\xcL Q_\a,\cZ]=0\,,\qquad\qquad[\widetilde\xcL P_a,\cZ]=0\,,\qquad\qquad[\widetilde\xcL Z_{(m)}^\a,\cZ]=0\,,\qquad\qquad[\cZ,\cZ]=0\,,&\cr\cr
&[\widetilde\xcL J_{ab},\widetilde\xcL Q_\a]=\tfrac{1}{2}\,\G_{ab}{}^\b_{\ \a}\,\widetilde\xcL Q_\b\,,\qquad\qquad[\widetilde\xcL J_{ab},\widetilde\xcL P_c]=\eta_{bc}\,\widetilde\xcL P_a-\eta_{ac}\,\widetilde\xcL P_b\,,&\cr\cr
&[\widetilde\xcL J_{ab},\widetilde\xcL Z_{(m)}^\a]=-\tfrac{1}{2}\,\G_{ab}{}^\a_{\ \b}\,\widetilde\xcL Z_{(m)}^\b\,,\qquad\qquad[\widetilde\xcL J_{ab},\cZ]=0\,,&\cr\cr
&[\widetilde\xcL J_{ab},\widetilde\xcL J_{cd}]=\eta_{ad}\,\widetilde\xcL J_{bc}-\eta_{ac}\,\widetilde\xcL J_{bd}+\eta_{bc}\,\widetilde\xcL J_{ad}-\eta_{bd}\,\widetilde\xcL J_{ac}\,.&
\qqq
With $\,\z\,$ denoting the super-1-form dual to $\,\cZ\,$ and 
\qq\nn
\widetilde\xcL p\equiv\pr_1\ :\ \lgt\oplus\gt{spin}(d,1)\too\lgt\,,
\qqq
we obtain,\ similarly as before,
\qq\nn
\widehat\d\z=\pi_{\widetilde\lgt}^*\underset{\tx{\ciut{(2)}}}{\cF}\equiv\widetilde\xcL p^*\pi_\lgt^*\underset{\tx{\ciut{(2)}}}{\txF}\,.
\qqq
The above Lie-superalgebra extension integrates to a central Lie-supergroup extension
\qq\nn
\bd1\too\bC^\x\too\widetilde\xcL\xrightarrow{\ \pi_{\widetilde\xcL}\ }\sfY^{[2]}{\rm sISO}(d,1|D_{d,1})\too\bd1
\qqq
with the supermanifold structure 
\qq\nn
\widetilde\xcL=\sfY^{[2]}{\rm sISO}(d,1|D_{d,1})\x\bC^\x\cong\xcL\x{\rm Spin}(d,1)
\qqq
for which
\qq\nn
\pi_{\widetilde\xcL}\equiv\pi_\xcL\x\id_{{\rm Spin}(d,1)}\ :\ \widetilde\xcL\too\sfY^{[2]}{\rm sMink}(d,1|D_{d,1})\x{\rm Spin}(d,1)\equiv\sfY^{[2]}{\rm sISO}(d,1|D_{d,1})\,,
\qqq
and with the Lie-supergroup structure determined by the binary operation
\qq\nn
\widetilde\xcL\txm\ :\ \widetilde\xcL\x\widetilde\xcL\too\widetilde\xcL
\qqq
with the coordinate presentation
\qq\nn
&&\widetilde\xcL\txm\bigl(\bigl((\theta_1,x_1,\xi_{1,1},\phi_1),(\theta_1,x_1,\xi_{1,2},\phi_1),z_1\bigr),\bigl((\theta_2,x_2,\xi_{2,1},\phi_2),(\theta_2,x_2,\xi_{2,2},\phi_2),z_2\bigr)\bigr)\cr\cr
&=&\bigl(\sfY\txm\bigl((\theta_1,x_1,\xi_{1,1},\phi_1),(\theta_2,x_2,\xi_{2,1},\phi_2)\bigr),\sfY\txm\bigl((\theta_1,x_1,\xi_{1,2},\phi_1),(\theta_2,x_2,\xi_{2,2},\phi_2)\bigr),\cr\cr
&&\ee^{\sfi\,\theta_1^\a\,(\xi_{2,2}-\xi_{2,1})_\b\,S(\phi_1)^{-1\,\b}_{\ \ \ \ \a}}\cdot z_1\cdot z_2\bigr)\,.
\qqq
Its form ensures left-invariance of the super-1-form
\qq\nn
\z\bigl((\theta,x,\xi_1,\phi),(\theta,x,\xi_2,\phi),z\bigr)=\tfrac{\sfi\,\sfd z}{z}+\theta^\a\,\sfd\bigl(\xi_{2\,\a}-\xi_{1\,\a}\bigr)\equiv\tfrac{\sfi\,\sfd z}{z}+\sfY^{[2]}\pi^*\txa\bigl((\theta,x,\xi_1,\phi),(\theta,x,\xi_2,\phi)\bigr)\,.
\qqq
Writing
\qq\nn
\widetilde\xcL\pi\equiv\pr_1\ :\ \xcL\x{\rm Spin}(d,1)\too\xcL\,,
\qqq
with
\qq\nn
\pi_\xcL\circ\widetilde\xcL\pi=\pi\circ\pi_{\widetilde\xcL}\,,
\qqq
we obtain
\qq\nn
\z\equiv\widetilde\xcL\pi^*\unl\z\,.
\qqq
Once again, we end up with the structure of a (trivial) principal $\bC^\x$-bundle 
\qq\nn
\pi_{\widetilde\xcL}\ :\ \widetilde\xcL\too\sfY^{[2]}{\rm sISO}(d,1|D_{d,1})
\qqq
with the LI principal connection super-1-form
\qq\nn
\underset{\tx{\ciut{(1)}}}{\cA}{}_{\widetilde\xcL}\equiv\z
\qqq
of curvature $\,\underset{\tx{\ciut{(2)}}}{\cF}$,
\qq\nn
\sfd\underset{\tx{\ciut{(1)}}}{\cA}{}_{\widetilde\xcL}=\pi_{\widetilde\xcL}^*\underset{\tx{\ciut{(2)}}}{\cF}\,.
\qqq
A reasoning fully analogous to the one presented in the case of $\,\xcL\,$ leads to the trivial groupoid structure
\qq\nn
\mu_{\widetilde\xcL}\equiv\bd1\ :\ \pr_{1,2}^*\widetilde\xcL\ox\pr_{2,3}^*\widetilde\xcL\xrightarrow{\ \cong\ }\pr_{1,3}^*\widetilde\xcL
\qqq
that isomorphically maps the Lie-supergroup structure on its domain to the one on its codomain.\ Also the latter admits a Lie-superalgebraic description,\ which we state hereunder for later reference.\ Its reconstruction starts with the self-explanatory definition of the pullback Lie superalgebras
\qq\nn
\alxydim{@C=2.5cm@R=2cm}{ \pr_{i,j}^*\widetilde\lgt\equiv\sfY^{[3]}\gt{siso}(d,1|D_{d,1})\hspace{2pt}{}_{\pr_{i,j}}\hspace{-3pt}\oplus_{\pi_{\widetilde\lgt}}\widetilde\lgt \ar[r]^{\quad\qquad\qquad\qquad\pr_2} \ar[d]_{\pr_1} & \widetilde\lgt \ar[d]^{\pi_{\widetilde\lgt}} \\ \sfY^{[3]}\gt{siso}(d,1|D_{d,1}) \ar[r]_{\pr_{i,j}} & \sfY^{[2]}\gt{siso}(d,1|D_{d,1}) }\,,\qquad(i,j)\in\{(1,2),(2,3),(1,3)\}\,,
\qqq
with the respective bases
\qq\nn
\pr_{1,2}^*\widetilde\lgt&=&\bigoplus_{\a=1}^{D_{d,1}}\,\corr{\bigl(\bigl(\sfY Q_\a,\sfY Q_\a,\sfY Q_\a\bigr),\widetilde\xcL Q_\a\bigr)\equiv\widetilde\xcL Q^{(1,2)}_\a}\oplus\bigoplus_{a=0}^d\,\corr{\bigl(\bigl(\sfY P_a,\sfY P_a,\sfY P_a\bigr),\widetilde\xcL P_a\bigr)\equiv\widetilde\xcL P^{(1,2)}_a}\cr\cr
&&\oplus\bigoplus_{\b=1}^{D_{d,1}}\,\corr{\bigl(\bigl(Z^\b,0,0\bigr),\widetilde\xcL Z_{(1)}^\b\bigr)\equiv\widetilde\xcL Z_{(1)}^{(1,2)\,\b}}\oplus\bigoplus_{\g=1}^{D_{d,1}}\,\corr{\bigl(\bigl(0,Z^\g,0\bigr),\widetilde\xcL Z_{(2)}^\g\bigr)\equiv\widetilde\xcL Z_{(2)}^{(1,2)\,\g}}\cr\cr
&&\oplus\bigoplus_{\d=1}^{D_{d,1}}\,\corr{\bigl(\bigl(0,0,Z^\d\bigr),0\bigr)\equiv\widetilde\xcL Z_{(3)}^{(1,2)\,\d}}\oplus\corr{\bigl((0,0,0),\cZ\bigr)\equiv\cZ^{(1,2)}}\cr\cr
&&\oplus\bigoplus_{b<c=0}^d\,\corr{\bigl(\bigl(\sfY J_{bc},\sfY J_{bc},\sfY J_{bc}\bigr),\widetilde\xcL J_{bc}\bigr)\equiv\widetilde\xcL J^{(1,2)}_{bc}}\,,\cr\cr\cr
\pr_{2,3}^*\widetilde\lgt&=&\bigoplus_{\a=1}^{D_{d,1}}\,\corr{\bigl(\bigl(\sfY Q_\a,\sfY Q_\a,\sfY Q_\a\bigr),\widetilde\xcL Q_\a\bigr)\equiv\widetilde\xcL Q^{(2,3)}_\a}\oplus\bigoplus_{a=0}^d\,\corr{\bigl(\bigl(\sfY P_a,\sfY P_a,\sfY P_a\bigr),\widetilde\xcL P_a\bigr)\equiv\widetilde\xcL P^{(2,3)}_a}\cr\cr
&&\oplus\bigoplus_{\b=1}^{D_{d,1}}\,\corr{\bigl(\bigl(0,Z^\b,0\bigr),\widetilde\xcL Z_{(1)}^\b\bigr)\equiv\widetilde\xcL Z_{(1)}^{(2,3)\,\b}}\oplus\bigoplus_{\g=1}^{D_{d,1}}\,\corr{\bigl(\bigl(0,0,Z^\g\bigr),\widetilde\xcL Z_{(2)}^\g\bigr)\equiv\widetilde\xcL Z_{(2)}^{(2,3)\,\g}}\cr\cr
&&\oplus\bigoplus_{\d=1}^{D_{d,1}}\,\corr{\bigl(\bigl(Z^\d,0,0\bigr),0\bigr)\equiv\widetilde\xcL Z_{(3)}^{(2,3)\,\d}}\oplus\corr{\bigl((0,0,0),\cZ\bigr)\equiv\cZ^{(2,3)}}\cr\cr
&&\oplus\bigoplus_{b<c=0}^d\,\corr{\bigl(\bigl(\sfY J_{bc},\sfY J_{bc},\sfY J_{bc}\bigr),\widetilde\xcL J_{bc}\bigr)\equiv\widetilde\xcL J^{(2,3)}_{bc}}\,,\cr\cr\cr
\pr_{1,3}^*\widetilde\lgt&=&\bigoplus_{\a=1}^{D_{d,1}}\,\corr{\bigl(\bigl(\sfY Q_\a,\sfY Q_\a,\sfY Q_\a\bigr),\widetilde\xcL Q_\a\bigr)\equiv\widetilde\xcL Q^{(1,3)}_\a}\oplus\bigoplus_{a=0}^d\,\corr{\bigl(\bigl(\sfY P_a,\sfY P_a,\sfY P_a\bigr),\widetilde\xcL P_a\bigr)\equiv\widetilde\xcL P^{(1,3)}_a}\cr\cr
&&\oplus\bigoplus_{\b=1}^{D_{d,1}}\,\corr{\bigl(\bigl(Z^\b,0,0\bigr),\widetilde\xcL Z_{(1)}^\b\bigr)\equiv\widetilde\xcL Z_{(1)}^{(1,3)\,\b}}\oplus\bigoplus_{\d=1}^{D_{d,1}}\,\corr{\bigl(\bigl(0,0,Z^\g\bigr),\widetilde\xcL Z_{(2)}^\g\bigr)\equiv\widetilde\xcL Z_{(2)}^{(1,3)\,\b}}\cr\cr
&&\oplus\bigoplus_{\d=1}^{D_{d,1}}\,\corr{\bigl(\bigl(0,Z^\d,0\bigr),0\bigr)\equiv\widetilde\xcL Z_{(3)}^{(1,3)\,\b}}\oplus\corr{\bigl((0,0,0),\cZ\bigr)\equiv\cZ^{(1,3)}}\cr\cr
&&\oplus\bigoplus_{b<c=0}^d\,\corr{\bigl(\bigl(\sfY J_{bc},\sfY J_{bc},\sfY J_{bc}\bigr),\widetilde\xcL J_{bc}\bigr)\equiv\widetilde\xcL J^{(1,3)}_{bc}}
\qqq
and the superbracket obtained from the direct-sum one on $\,\sfY^{[3]}\gt{siso}(d,1|D_{d,1})\oplus\widetilde\lgt\,$ through restriction.\ Finally,\ we form the `tensor product' of the first two,
\qq\nn
\pr_{1,2}^*\widetilde\lgt\ox\pr_{2,3}^*\widetilde\lgt\equiv\bigl(\pr_{1,2}^*\widetilde\lgt\hspace{2pt}{}_{\pr_1}\hspace{-2pt}\oplus_{\pr_1}\pr_{2,3}^*\widetilde\lgt\bigr)/_{\sim_\bR}\,,
\qqq
by identifying the generators
\qq\nn
\bigl(\cZ^{(1,2)},0\bigr)\sim_\bR\bigl(0,\cZ^{(2,3)}\bigr)\,,
\qqq
so that
\qq\nn
\pr_{1,2}^*\widetilde\lgt\ox\pr_{2,3}^*\widetilde\lgt&=&\bigoplus_{\a=1}^{D_{d,1}}\,\corr{\bigl(\widetilde\xcL Q^{(1,2)}_\a,\widetilde\xcL Q^{(2,3)}_\a\bigr)\equiv\widetilde\xcL Q^{(1,2;2,3)}_\a}\oplus\bigoplus_{a=0}^d\,\corr{\bigl(\widetilde\xcL P^{(1,2)}_a,\widetilde\xcL P^{(2,3)}_a\bigr)\equiv\widetilde\xcL P^{(1,2;2,3)}_a}\cr\cr
&&\oplus\bigoplus_{\b=1}^{D_{d,1}}\,\corr{\bigl(\widetilde\xcL Z_{(1)}^{(1,2)\,\b},\widetilde\xcL Z_{(3)}^{(2,3)\,\b}\bigr)\equiv\widetilde\xcL Z_{(1,3)}^{(1,2;2,3)\,\b}}\oplus\bigoplus_{\g=1}^{D_{d,1}}\,\corr{\bigl(\widetilde\xcL Z_{(3)}^{(1,2)\,\g},\widetilde\xcL Z_{(2)}^{(1,2)\,\g}\bigr)\equiv\widetilde\xcL Z_{(3,2)}^{(1,2;2,3)\,\g}}\cr\cr
&&\oplus\bigoplus_{\d=1}^{D_{d,1}}\,\corr{\bigl(\widetilde\xcL Z_{(2)}^{(1,2)\,\d},\widetilde\xcL Z_{(1)}^{(2,3)\,\d}\bigr)\equiv\widetilde\xcL Z_{(2,1)}^{(1,2;2,3)\,\d}}\oplus\corr{\bigl[\bigl(\cZ^{(1,2)},0\bigr)\bigr]_{\sim_\bR}\equiv\cZ^{1,2;2,3}}\cr\cr
&&\oplus\bigoplus_{b<c=0}^d\,\corr{\bigl(\bigl(\widetilde\xcL J^{(1,2)}_{bc},\widetilde\xcL J^{(2,3)}_{bc}\bigr)\equiv\widetilde\xcL J^{(1,2;2,3)}_{bc}}\,,
\qqq
with the superbracket (defined as the projection of the restricted direct-sum superbracket on $\,\pr_{1,2}^*\widetilde\lgt\oplus_{\sfY^{[3]}\gt{siso}(d,1|D_{d,1})}\pr_{2,3}^*\widetilde\lgt$,\ computed for arbitrary representatives of the equivalence classes of arguments,\ back to the quotient)
\qq\nn
&\{\widetilde\xcL Q^{(1,2;2,3)}_\a,\widetilde\xcL Q^{(1,2;2,3)}_\b\}=\ovl\G{}^a_{\a\b}\,\widetilde\xcL P^{(1,2;2,3)}_a\,,\qquad\qquad[\widetilde\xcL P^{(1,2;2,3)}_a,\widetilde\xcL P^{(1,2;2,3)}_b]=0\,,&\cr\cr
&[\widetilde\xcL Q^{(1,2;2,3)}_\a,\widetilde\xcL P^{(1,2;2,3)}_a]=\ovl\G_{a\,\a\b}\,\bigl(\widetilde\xcL Z_{(1,3)}^{(1,2;2,3)\,\b}+\widetilde\xcL Z_{(3,2)}^{(1,2;2,3)\,\b}+\widetilde\xcL Z_{(2,1)}^{(1,2;2,3)\,\b}\bigr)\,,&\cr\cr
&-\{\widetilde\xcL Q^{(1,2;2,3)}_\a,\widetilde\xcL Z_{(1,3)}^{(1,2;2,3)\,\b}\}=\d_\a^{\ \b}\,\cZ^{(1,2;2,3)}=\{\widetilde\xcL Q^{(1,2;2,3)}_\a,\widetilde\xcL Z_{(3,2)}^{(1,2;2,3)\,\b}\}\,,\qquad\qquad\{\widetilde\xcL Q^{(1,2;2,3)}_\a,\widetilde\xcL Z_{(2,1)}^{(1,2;2,3)\,\b}\}=0\,,&\cr\cr
&[\widetilde\xcL P^{(1,2;2,3)}_a,\widetilde\xcL Z_{(m,n)}^{(1,2;2,3)\,\a}]=0\,,\qquad\qquad\{\widetilde\xcL Z_{(m,n)}^{(1,2;2,3)\,\a},\widetilde\xcL Z_{(r,s)}^{(1,2;2,3)\,\b}\}=0\,,&\cr\cr
&[\widetilde\xcL Q^{(1,2;2,3)}_\a,\cZ^{(1,2;2,3)}]=0\,,\qquad\qquad[\widetilde\xcL P^{(1,2;2,3)}_a,\cZ^{(1,2;2,3)}]=0\,,\qquad\qquad[\widetilde\xcL Z_{(m,n)}^{(1,2;2,3)\,\a},\cZ^{(1,2;2,3)}]=0\,,&\cr\cr
&[\cZ^{(1,2;2,3)},\cZ^{(1,2;2,3)}]=0\,,&\cr\cr
&[\widetilde\xcL J^{(1,2;2,3)}_{ab},\widetilde\xcL Q^{(1,2;2,3)}_\a]=\tfrac{1}{2}\,\G_{ab}{}^\b_{\ \a}\,\widetilde\xcL Q^{(1,2;2,3)}_\b\,,\qquad\qquad[\widetilde\xcL J^{(1,2;2,3)}_{ab},\widetilde\xcL P^{(1,2;2,3)}_c]=\eta_{bc}\,\widetilde\xcL P^{(1,2;2,3)}_a-\eta_{ac}\,\widetilde\xcL P^{(1,2;2,3)}_b\,,&\cr\cr
&[\widetilde\xcL J^{(1,2;2,3)}_{ab},\widetilde\xcL Z_{(m,n)}^{(1,2;2,3)\,\a}]=-\tfrac{1}{2}\,\G_{ab}{}^\a_{\ \b}\,\widetilde\xcL Z_{(m,n)}^{(1,2;2,3)\,\b}\,,\qquad\qquad[\widetilde\xcL J^{(1,2;2,3)}_{ab},\cZ^{(1,2;2,3)}]=0\,,&\cr\cr
&[\widetilde\xcL J^{(1,2;2,3)}_{ab},\widetilde\xcL J^{(1,2;2,3)}_{cd}]=\eta_{ad}\,\widetilde\xcL J^{(1,2;2,3)}_{bc}-\eta_{ac}\,\widetilde\xcL J^{(1,2;2,3)}_{bd}+\eta_{bc}\,\widetilde\xcL J^{(1,2;2,3)}_{ad}-\eta_{bd}\,\widetilde\xcL J^{(1,2;2,3)}_{ac}\,.&
\qqq
Comparing the above with the superbracket of $\,\pr_{2,3}^*\widetilde\lgt$,\ we infer the existence of a Lie-superalgebra isomorphism 
\qq\nn
\mu_{\widetilde\lgt}\ :\ \pr_{1,2}^*\widetilde\lgt\ox\pr_{2,3}^*\widetilde\lgt\xrightarrow{\ \cong\ }\pr_{1,3}^*\widetilde\lgt
\qqq
given by the unique linear extension of the assignment
\qq\nn
&&\bigl(\widetilde\xcL Q^{(1,2;2,3)}_\a,\widetilde\xcL P^{(1,2;2,3)}_a,\widetilde\xcL Z_{(1,3)}^{(1,2;2,3)\,\b},\widetilde\xcL Z_{(3,2)}^{(1,2;2,3)\,\g},\widetilde\xcL Z_{(2,1)}^{(1,2;2,3)\,\d},\cZ^{1,2;2,3},\widetilde\xcL J^{(1,2;2,3)}_{bc}\bigr)\cr\cr
&\longmapsto&\bigl(\widetilde\xcL Q^{(1,3)}_\a,\widetilde\xcL P^{(1,3)}_a,\widetilde\xcL Z_{(1)}^{(1,3)\,\b},\widetilde\xcL Z_{(2)}^{(1,3)\,\g},\widetilde\xcL Z_{(3)}^{(1,3)\,\d},\cZ^{1,3},\widetilde\xcL J^{(1,3)}_{bc}\bigr)\,.
\qqq
This is the Lie-superalgebraic counterpart of the groupoid structure $\,\mu_{\widetilde\xcL}$,\ its triviality being reflected in the identity
\qq\nn
\mu_{\widetilde\lgt}\bigl(\cZ^{1,2;2,3}\bigr)=\cZ^{1,3}\,,
\qqq
which we encode in the same notation:
\qq\nn
\mu_{\widetilde\lgt}\equiv\bd1
\qqq
as the one used for the trivial $\,\mu_{\widetilde\xcL}$.

By the end of the long day,\ we conclude that

\bethe\label{thm:liftGSs1g}
The GS super-1-gerbe over $\,{\rm sMink}(d,1|D_{d,1})\,$ canonically lifts (${\rm Spin}(d,1)$-equivariantly) to a CaE super-1-gerbe over $\,{\rm sISO}(d,1|D_{d,1})$.
\ethe

\noindent The resulting CaE super-1-gerbe
\qq\nn
\widetilde\cG{}^{(1)}_{\rm GS}:=\bigl(\sfY{\rm sISO}(d,1|D_{d,1}),\pi_{\sfY{\rm sISO}(d,1|D_{d,1})},\underset{\tx{\ciut{(2)}}}{\sfY\b},\widetilde\xcL,\pi_{\widetilde\xcL},\underset{\tx{\ciut{(1)}}}{\cA}{}_{\widetilde\xcL},\mu_{\widetilde\xcL}\bigr)\equiv\pi^*\cG^{(1)}_{\rm GS}\,,
\qqq
a \emph{distinguished} (${\rm Spin}(d,1)$-equivariant) pullback of $\,\cG^{(1)}_{\rm GS}$,\ shall be called the \textbf{lifted Green--Schwarz super-1-gerbe over $\,{\rm sISO}(d,1|D_{d,1})$}.\ It constitutes the point of departure of a full-fledged `gerbification' of the GS super-$\si$-model in the purely topological HP formulation that we shall carry out in what follows.\ Its first step consists in extending $\,\widetilde\cG{}^{(1)}_{\rm GS}\,$ by the trivial CaE super-1-gerbe 
\qq\nn
\cI^{(1)}_{2\Vol(\tgt^{(0)}_{\rm vac})}\equiv\bigl({\rm sISO}(d,1|D_{d,1}),\id_{{\rm sISO}(d,1|D_{d,1})},2\Vol\bigl(\tgt^{(0)}_{\rm vac}\bigr),{\rm sISO}(d,1|D_{d,1})\x\bC^\x,\pr_1,\pr_2^*\vartheta_{\bC^\x},\bd1\bigr)
\qqq 
over the supersymmetry group $\,{\rm sISO}(d,1|D_{d,1})\,$ associated with the LI super-2-form $\,2\Vol(\tgt^{(0)}_{\rm vac})\,$ (featuring as its curving).\ Above,\ the total space $\,{\rm sISO}(d,1|D_{d,1})\x\bC^\x\ni(\theta^\a,x^a,z)\,$ of the trivial principal $\bC^\x$-bundle 
\qq\nn
\pr_1\ :\ {\rm sISO}(d,1|D_{d,1})\x\bC^\x\too{\rm sISO}(d,1|D_{d,1})
\qqq
carries the product Lie-supergroup structure and comes equipped with the trivial principal connection super-1-form $\,\pr_2^*\vartheta_{\bC^\x}\,$ with the coordinate presentation
\qq\nn
\pr_2^*\vartheta_{\bC^\x}\bigl((\theta,x),z\bigr)\equiv\vartheta_{\bC^\x}(z):=\tfrac{\sfi\,\sfd z}{z}\,,
\qqq
manifestly LI with respect to the said Lie-supergroup structure.\ The tensor product 
\qq\nn
\widetilde\cG{}^{(1)}_{\rm GS}\ox\cI^{(1)}_{2\Vol(\tgt^{(0)}_{\rm vac})}=\bigl(\sfY{\rm sISO}(d,1|D_{d,1}),\pi_{\sfY{\rm sISO}(d,1|D_{d,1})},\underset{\tx{\ciut{(2)}}}{\sfY\b}+2\sfY\pi^*\Vol\bigl(\tgt^{(0)}_{\rm vac}\bigr)\equiv\widehat{\underset{\tx{\ciut{(2)}}}{\sfY\b}},\widetilde\xcL,\pi_{\widetilde\xcL},\underset{\tx{\ciut{(1)}}}{\cA}{}_{\widetilde\xcL},\mu_{\widetilde\xcL}\bigr)\equiv\widehat\cG{}^{(1)}_{\rm HP}
\qqq
is also a CaE super-1-gerbe,\ to be referred to -- after \Rxcite{Def.\,6.5}{Suszek:2019cum}, but taking into account its supersymmetry established above -- the \textbf{extended Hughes--Polchinski super-1-gerbe over} $\,{\rm sISO}(d,1|D_{d,1})$.\ The 1-gerbe over $\,\Si^{\rm HP}\,$ with restrictions $\,\widehat\cG{}^{(1)}_{\rm HP}\rstr_{\cV_i}\,$ over the components $\,\cV_i\,$ of that supermanifold shall be denoted as
\qq\nn
\widehat\cG{}^{(1)}_{\Si^{\rm HP}}\equiv\bigsqcup_{i\in I}\,\widehat\cG{}^{(1)}_{\rm HP}\rstr_{\cV_i}\,.
\qqq 
In the remainder of the present paper,\ we investigate at great length structural properties of its \textbf{vacuum restriction}
\qq\nn
\cG{}^{(1)}_{\rm vac}\equiv\iota_{\rm vac}^*\widehat\cG{}^{(1)}_{\Si^{\rm HP}}\equiv\widehat\cG{}^{(1)}_{\Si^{\rm HP}}\rstr_{\Si^{\rm HP}_{\rm vac}}=:\bigl(\sfY\Si^{\rm HP}_{\rm vac},\pi_{\sfY\Si^{\rm HP}_{\rm vac}},\widehat{\underset{\tx{\ciut{(2)}}}{\sfY\b}}{}_{\rm vac},\widetilde\xcL{}_{\rm vac},\pi_{\widetilde\xcL{}_{\rm vac}},\underset{\tx{\ciut{(1)}}}{\cA}{}_{\widetilde\xcL{}_{\rm vac}},\mu_{\widetilde\xcL{}_{\rm vac}}\bigr)\,,
\qqq
with view to understanding the quantum-mechanical aspect of the vacuum of the GS super-$\si$-model and of its global and local supersymmetry,\ as encoded by the (super-)gerbe theory of the field theory of interest.

\section{The supersymmetry of the super-1-gerbe(s)}\label{sec:susyequiv}

Prequantisable symmetries of the (super-)$\si$-model have specific gerbe-theoretic manifestations that ensure the existence of their consistent lift to the Hilbert space of the (super)field theory.\ These have been known for quite some time from the extensive study of the subject carried out in the non-$\bZ/2\bZ$-graded geometric category.\ They fall into the two classes,\ mentioned previously,\ with a fundamentaly different ontological status and,\ accordingly,\ a different higher-geometric implementation,\ to wit,\ the global and the local symmetries that we discuss in sequence hereunder.

\subsection{Higher global supersymmetry}\label{sub:hgsusy} We begin with \emph{global} symmetries that set in correspondence \emph{inequivalent} field configurations.\ In the non-$\bZ/2\bZ$-graded setting,\ these are represented by families of 1-gerbe 1-isomorphisms indexed by elements of the symmetry group that identify the 1-gerbe $\,\cG^{(1)}\,$ of the $\si$-model as \emph{invariant} under the element-wise realisation of the group,\ a fact established firmly in Refs.\,\cite{Suszek:2011hg,Suszek:2011}.\ More specifically,\ given a Lie group $\,\txG\,$ of those isometries of the target $\,M\,$ whose action on fields of the $\si$-model induced from its action
\qq\nn
\la_\cdot\ :\ \txG\x M\too M\ :\ (g,m)\longmapsto\la_g(m)
\qqq
on the target preserves the DF amplitude,\ we demand the existence of 1-isomorphisms
\qq\nn
\Phi_g\ :\ \la_g^*\cG^{(1)}\xrightarrow{\ \cong\ }\cG^{(1)}\,,\qquad g\in\txG\,.
\qqq
In the case of a homogeneous space $\,\txG/\txK\,$ of a supersymmetry Lie supergroup $\,\txG\,$ (relative to its Lie subgroup $\,\txK$),\ this simple scenario requires,\ in general,\ a straightforward sheaf-theoretic adaptation that separately takes into account invariance under the element-wise action 
\qq\nn
[|\ell|]^\txK_\cdot\ :\ |\txG|\too{\rm Aut}_\sMan\bigl(\txG/\txK\bigr)\ :\ g\longmapsto[\ell]^\txK\circ\bigl(\widehat g\x\id_{\txG/\txK}\bigr)
\qqq
of the body Lie group $\,|G|$,
\qq\nn
|\Phi|_g\ :\ [|\ell|]^{\txK\,*}_g\cG^{(1)}\xrightarrow{\ \cong\ }\cG^{(1)}\,,\qquad g\in|\txG|\,,
\qqq
and that under the element-wise tangential action of the Lie superalgebra $\,\ggt\,$ of the Lie supergroup $\,\txG$,
\qq\nn
\sfd\Phi_X\ :\ \pLie{\cK_X}\cG^{(1)}\xrightarrow{\ \cong\ }\cI^{(1)}_0\,,\qquad X\in\ggt\,.
\qqq
Here,\ $\,\pLie{\cK_X}\cG^{(1)}\,$ is a super-1-gerbe obtained from $\,\cG^{(1)}\,$ by Lie-differentiating local data of the latter in the direction of the fundamental vector field $\,\cK_X\,$ for the induced action $\,[\ell]^\txK\,$ of $\,\txG\,$ on $\,\txG/\txK$,\ and $\,\cI^{(1)}_0\,$ is the flat trivial super-1-gerbe with a null curving.\ In the super-minkowskian setting,\ we may readily put the components of the structure sheaf of the supersymmetry supergroup $\,{\rm sISO}(d,1|D_{d,1})\,$ of both parities on the same footing by using the global generators $\,\theta^\a\,$ of the Gra\ss mann-odd component of that sheaf,\ alongside the remaining coordinates $\,(x^a,\phi^{bc})$,\ and demand the existence of 1-isomorphisms that we write, in a self-explanatory notation\footnote{That is,\ in the coordinate picture,\ in which $[\ell]^{{\rm Spin}(d,1)}_{(\vep,y,\psi)}\equiv[\ell]^{{\rm Spin}(d,1)}((\vep,y,\psi),\cdot)$.},\ as
\qq\nn
\Phi_{(\vep,y,\psi)}\ :\ [\ell]^{{\rm Spin}(d,1)\,*}_{(\vep,y,\psi)}\cG^{(1)}_{\rm GS}\xrightarrow{\ \cong\ }\cG^{(1)}_{\rm GS}\,,\qquad(\vep,y,\psi)\in{\rm sISO}(d,1|D_{d,1})\,.
\qqq
Actually,\ we shall go one step further and consider,\ instead,\ the corresponding 1-isomorphisms
\qq\nn
\widetilde\Phi{}_{(\vep,y,\psi)}\ :\ \ell^*_{(\vep,y,\psi)}\widetilde\cG{}^{(1)}_{\rm GS}\xrightarrow{\ \cong\ }\widetilde\cG{}^{(1)}_{\rm GS}\,,\qquad(\vep,y,\psi)\in{\rm sISO}(d,1|D_{d,1})\,.
\qqq
for its lift.\ In so doing,\ we get a chance to appreciate the structural merits of the geometrisation scheme adopted in which the implementation of the global supersymmetry is seen to essentially trivialise.\ Thus,\ we take as the surjective submersion of the pullback super-1-gerbe $\,\ell^*_{(\vep,y,\psi)}\widetilde\cG{}^{(1)}_{\rm GS}\,$ the very same one as for $\,\widetilde\cG{}^{(1)}_{\rm GS}\,$ -- that this makes sense follows from the identity
\qq\nn
\pi_{\sfY{\rm sISO}(d,1|D_{d,1})}\circ\widehat\ell{}_{(\vep,y,\psi)}=\ell_{(\vep,y,\psi)}\circ\pi_{\ell^*_{(\vep,y,\psi)}\sfY{\rm sISO}(d,1|D_{d,1})}\,,\qquad\qquad\pi_{\ell^*_{(\vep,y,\psi)}\sfY{\rm sISO}(d,1|D_{d,1})}\equiv\pi_{\sfY{\rm sISO}(d,1|D_{d,1})}\,,
\qqq
written for $\,\widehat\ell{}_{(\vep,y,\psi)}\equiv\sfY\ell_{(\vep,y,0,\psi)}\equiv\sfY\ell((\vep,y,0,\psi),\cdot)\,$ and ensured by the equivariance of $\,\pi_{\sfY{\rm sISO}(d,1|D_{d,1})}$.\ For this choice of the surjective submersion,\ we find 
\qq\nn
\widehat\ell{}^*_{(\vep,y,\psi)}\underset{\tx{\ciut{(2)}}}{\sfY\b}=\underset{\tx{\ciut{(2)}}}{\sfY\b}\,,
\qqq
and so we infer that $\,\underset{\tx{\ciut{(2)}}}{\sfY\b}\,$ is the curving of the pullback super-1-gerbe.\ Continuing along these lines,\ we take $\,\widetilde\xcL\,$ as the principal $\bC^\x$-bundle of the pullback super-1-gerbe,\ a choice legitimised by the identity
\qq\nn
\pi_{\widetilde\xcL}\circ\widehat\ell{}^{[2]}_{(\vep,y,\psi)}= \sfY^{[2]}\ell_{(\vep,y,0,\psi)}\circ\pi_{\widehat\ell{}^{\x 2\,*}_{(\vep,y,\psi)}\widetilde\xcL}\,,\qquad\qquad\pi_{\widehat\ell{}^{\x 2\,*}_{(\vep,y,\psi)}\widetilde\xcL}\equiv\pi_{\widetilde\xcL}
\qqq 
in which $\,\widehat\ell{}^{[2]}_{(\vep,y,\psi)}\equiv\xcL\ell_{(\vep,y,0,\psi,1)}\equiv\xcL\ell((\vep,y,0,\psi,1),\cdot)$.\ The left-invariance of $\,\underset{\tx{\ciut{(1)}}}{\cA}{}_{\widetilde\xcL}$, 
\qq\nn
\widehat\ell{}^{[2]\,*}_{(\vep,y,\psi)}\underset{\tx{\ciut{(1)}}}{\cA}{}_{\widetilde\xcL}=\underset{\tx{\ciut{(1)}}}{\cA}{}_{\widetilde\xcL}\,,
\qqq
now permits us to take $\,\underset{\tx{\ciut{(1)}}}{\cA}{}_{\widetilde\xcL}\,$ as the principal $\bC^\x$-connection on the pullback principal $\bC^\x$-bundle.\ The construction is consistently completed by taking
\qq\nn
\widehat\ell{}^{[3]\,*}_{(\vep,y,\psi)}\mu_{\widetilde\xcL}\equiv\mu_{\widetilde\xcL}
\qqq
as the groupoid structure of the pullback super-1-gerbe (for an obvious definition of $\,\widehat\ell{}^{[3]}_{(\vep,y,\psi)}$).\ Altogether, then, we obtain
\qq\nn
\ell^*_{(\vep,y,\psi)}\widetilde\cG{}^{(1)}_{\rm GS}\equiv\widetilde\cG{}^{(1)}_{\rm GS}\,,
\qqq
whence also 
\qq\nn
\widetilde\Phi{}_{(\vep,y,\psi)}\equiv\id_{\widetilde\cG{}^{(1)}_{\rm GS}}\,.
\qqq
Given the nature of the trivial correction $\,\cI^{(1)}_{2\Vol(\tgt^{(0)}_{\rm vac})}$,\ we ultimately obtain 1-isomorphisms
\qq\nn
\widehat\Phi{}_{(\vep,y,\psi)}\equiv\id_{\widehat\cG{}^{(1)}_{\rm GS}}\ :\ \ell^*_{(\vep,y,\psi)}\widehat\cG{}^{(1)}_{\rm GS}\xrightarrow{\ \cong\ }\widehat\cG{}^{(1)}_{\rm GS}\,,\qquad(\vep,y,\psi)\in{\rm sISO}(d,1|D_{d,1})\,.
\qqq
This is the anticipated higher-geometric realisation of the global supersymmetry of the GS super-$\si$-model in the HP formulation.\medskip

\subsection{Higher $\k$-symmetry \& the ${\bf sLieAlg}$-skeleton of the vacuum}\label{sub:hkappa} Next, we pass to \emph{local} symmetries that relate (\emph{gauge}-)\emph{equivalent} field configurations,\ or -- according to the passive interpretation of symmetry -- different (and equivalent) coordinate \emph{descriptions} of a given field configuration,\ and signal reducibility of the set of degrees of freedom of the field theory to those charting the space of orbits of the action of the gauge group.\ Whenever an action $\,\la\,$ of a group $\,\txG\,$ of global symmetries of a $\si$-model with the target $\,M\,$ is rendered local,\ or gauged,\ the ensuing (gauged) field theory effectively describes a $\si$-model on the orbispace $\,M/\txG$,\ or actually descends to the quotient manifold if the latter exists,\ {\it cp} \Rxcite{Sec.\,8}{Suszek:2012ddg} and \Rxcite{Sec.\,9}{Gawedzki:2012fu},\ which happens,\ {\it e.g.},\ when $\,\la\,$ is free and proper.\ A quantum-mechanical consistency of the descent of the $\si$-model to the orbispace calls for the existence of a $\txG$-equivariant structure on the associated 1-gerbe $\,\cG^{(1)}\,$ (of,\ say,\ curvature $\,{\rm curv}\,(\cG^{(1)})\in Z_{\rm dR}^3(M)$),\ to arise over the nerve $\,\sfN^\bullet(\txG\lx M)\equiv\txG^\bullet\x M$  
\qq\label{eq:actgrpdsimpl}
\alxydim{@C=1.5cm@R=1.5cm}{ \cdots \ar@<.75ex>[r]^{d_\bullet^{(3)}\quad} \ar@<.25ex>[r] \ar@<-.25ex>[r]
\ar@<-.75ex>[r] & \txG^{\x 2}\x M \ar@<.5ex>[r]^{\ d_\bullet^{(2)}} \ar@<0.ex>[r]
\ar@<-.5ex>[r] & \txG\x M \ar@<.5ex>[r]^{\quad d_\bullet^{(1)}} \ar@<-.5ex>[r] & M}
\qqq
of the action (Lie) groupoid $\,\txG\lx M$,\ {\it i.e.},\ a simplicial manifold with face maps (written for $\,x\in M,\ g,g_k\in\txG,\ k\in\ovl{1,m}\,$ with $\,m\in\bN^\x$)
\qq\nn
&&d_0^{(1)}(g,x)=x\equiv\pr_2(g,x)\,,\qquad\qquad d_1^{(1)}(g,x)=\la_g(x)\,,\cr\cr\cr
&&d_0^{(m)}(g_{m},g_{m-1},\ldots,g_1,x)=(g_{m-1},g_{m-2},
\ldots,g_1,x)\,,\cr\cr
&&d_{m}^{(m)}(g_{m},g_{m-1},\ldots,g_1
,x)=\bigl(g_{m},g_{m-1},\ldots,g_2 ,\ell_{g_1}(x)\bigr)\,,\cr\cr
&&d_i^{(m)}(g_{m},g_{m-1},\ldots,g_1,x)=(g_{m},g_{m-1},\ldots,g_{m+2-i}
,g_{m+1-i}\cdot g_{m-i},g_{m-1-i},\ldots,g_1,x)\,,\quad i\in\ovl{1,m-1}\,,
\qqq
{\it cp} \Rcite{Gawedzki:2010rn}.\ The first component of the structure is a 1-isomorphism
\qq\nn
\Upsilon\ :\ d_1^{(1)\,*}\cG^{(1)}\xrightarrow{\ \cong\ }d_0^{(1)\,*}\cG^{(1)}\ox\cI^{(1)}_{\varrho_{\theta_{\rm L}}}
\qqq
of 1-gerbes over the arrow manifold $\,\txG\x M\,$ of $\,\txG\lx M$,\ written in terms of the 2-form
\qq\nn
\varrho_{\theta_{\rm L}}=\pr_2^*\k_A\wedge\pr_1^*\theta_{\rm L}^A-\tfrac{1}{2}\,\pr_2^*(\cK_A\con\k_B)\,\pr_1^*\bigr(\theta_{\rm L}^A\wedge\theta_{\rm L}^B\bigl)\in\Om^2(\txG\x M)
\qqq 
in whose definition $\,\cK_A\equiv\cK_{t_A}$,\ the $\,\theta^A_{\rm L}\,$ are components of the LI $\ggt$-valued Maurer--Cartan 1-form
\qq\nn
\theta_{\rm L}=\theta^A_{\rm L}\ox t_A\in\Om^1(\txG)\ox\ggt
\qqq
associated with the generators $\,\{t_A\}_{A\in\ovl{1,\dim\,\ggt}}\,$ of the Lie algebra $\,\ggt$,\ and the $\,\k_A\,$ are 1-forms on $\,M\,$ satisfying the identities
\qq\nn
\cK_A\con{\rm curv}\,\bigl(\cG^{(1)}\bigr)=-\sfd\k_A\,,
\qqq
as required for $\,\txG\,$ to be a symmetry of the $\si$-model in the first place.\ The second component is a 2-isomorphism 
\qq\nn
\qquad\alxydim{@C=1.5cm@R=2cm}{ \bigl(d^{(1)}_1\circ d^{(2)}_1
\bigr)^*\cG^{(1)} \ar[r]^{d^{(2)\,*}_2\Upsilon\hspace{1cm}}
\ar[d]_{d^{(2)\,*}_1\Upsilon} & \bigl(d^{(1)}_1\circ
d^{(2)}_0\bigr)^*\cG^{(1)}\ox\cI^{(1)}_{d^{(2)\,*}_2\varrho_{\theta_{\rm L}}}
\ar[d]^{d^{(2)\,*}_0\Upsilon \ox\id_{\cI^{(1)}_{d^{(2)\,*}_2\varrho_{\theta_{\rm L}}}}} \ar@{=>}[dl]|{\ \g\ } \\
\bigl(d^{(1)}_0\circ d^{(2)}_1\bigr)^*\cG^{(1)}\ox\cI^{(1)}_{d^{(2)\,*}_1\varrho_{\theta_{\rm L}}}
\ar@{=}[r] & \bigl(d^{(1)}_0\circ
d^{(2)}_0\bigr)^*\cG^{(1)}\ox\cI^{(1)}_{d^{(2) *}_0\varrho_{\theta_{\rm L}}+d^{(2)\,
*}_2\varrho_{\theta_{\rm L}}}}
\qqq
between the 1-isomorphisms over $\,\txG^{\x 2}\x M$,\ satisfying, over $\,\txG^{\x 3}\x M$,\ the coherence condition
\qq\nn
d_1^{(3)\,*}\g\bullet\bigl(\id_{(d_2^{(2)}\circ d_1^{(3)})^*\Upsilon}\circ d_3^{(3)\,*}\g
\bigr)=d_2^{(3)\,*}\g\bullet\bigl(\bigl(d_0^{(3)\,*}\g
\ox\id_{\id_{\cI^{(1)}_{(d_2^{(2)}\circ d_1^{(3)})^*\varrho_{\theta_{\rm L}}}}}\bigr)\circ\id_{(d_2^{(2)}\circ d_3^{(3)})^*\Upsilon}\bigr)
\qqq 
in which $\,\circ\,$ and $\,\bullet\,$ are the horizontal and vertical compositions of 1-gerbe 2-isomorphisms,\ respectively.\ Thus,\ altogether,\ a 1-gerbe with a $\txG$-equivariant structure relative to the 2-form $\,\varrho_{\theta_{\rm L}}\,$ is the triple
\qq\nn
\bigl(\cG^{(1)},\Upsilon,\g\bigr)
\qqq
as defined and constrained by the conditions of coherence above.\ The principle of descent for $\,\la\,$ such that $\,M/\txG\,$ \emph{is} a smooth manifold is now encoded in the equivalence
\qq\label{equivcorresp}
\gt{Grb}^\nabla(M/\txG)\cong\gt{Grb}^\nabla(M)^\txG_{\varrho_{\theta_{\rm L}}\equiv 0}
\qqq
between the bicategory $\,\gt{Grb}^\nabla(M/\txG)\,$ of 1-gerbes (with a connective structure) over $\,M/\txG\,$ and the bicategory $\,\gt{Grb}^\nabla(M)^\txG_{\varrho_{\theta_{\rm L}}\equiv 0}\,$ of 1-gerbes (with a connective structure) over $\,M\,$ with a $\txG$-equivariant structure \emph{relative to} $\,\varrho_{\theta_{\rm L}}\equiv 0$,\
{\it cp} \Rxcite{Thm.\,5.3}{Gawedzki:2010rn}.\ Generically,\ the 2-form $\,\varrho_{\theta_{\rm L}}\,$ does \emph{not} vanish,\ and then a fairly complex construction of Refs.\,\cite{Gawedzki:2010rn,Suszek:2012ddg,Gawedzki:2012fu,Suszek:2013} has to be carried out to formulate loop dynamics with the global symmetry $\,\txG\,$ gauged.\ The construction employs a bundle $\,\sfP_\txG\x_\la M\,$ associated with a principal $\txG$-bundle $\,\sfP_\txG\,$ over $\,\Si\,$ and endowed with the Crittenden connection induced from that on $\,\sfP_\txG\,$ and with an action of the gauge group $\,\G(\Ad\,\sfP_\txG)\,$ of global sections of the adjoint bundle $\,\Ad\,\sfP_\txG\equiv\sfP_\txG\x_\Ad\txG$.\ It goes well beyond the classic minimal-coupling scheme.\ The loop dynamics descends to $\,M/\txG\,$ if the latter exists as a smooth manifold,\ or is taken to \emph{model} it otherwise.\ It may also happen that the 1-gerbe of the $\si$-model carries a $\txG$-equivariant structure relative to the vanishing 2-form $\,\varrho_{\theta_{\rm L}}\equiv 0$,\ in which case the 1-gerbe,\ and with it the $\si$-model,\ directly descends to resp.\ models loop dynamics on the orbispace $\,M/\txG$.\ This is the very special situation that we encouter below.

Bearing in mind that the existence of a gerbe-theoretic realisation of a symmetry is requisite for its quantum-mechanical consistency,\ we shall,\ now,\ put the $\k$-symmetry of Sec.\,\ref{sub:kappasym} in the above framework.\ In trying to do that,\ though,\ we stumble upon a peculiarity\footnote{For a general supertarget $\,\txG/\txH_{\rm vac}\,$ realised patchwise within $\,\txG\,$ by means of local sections of the principal $\txH_{\rm vac}$-bundle $\,\txG\too\txG/\txH_{\rm vac}$,\ there is yet another problematic peculiarity that we encounter,\ to wit,\ the symmetry seems to be well-defined (\emph{on} $\,\txG$) only in its infinitesimal (tangential) form due to the intrinsic ambiguities of the patchwise realisation over intersections of elements of the trivialising cover of $\,\txG/\txH_{\rm vac}$.} of the symmetry that takes us out of the standard scheme:\ The symmetry is realised in its full and integrated form only after imposition of the Euler--Lagrange equations of the super-$\si$-model,\ {\it i.e.},\ it is a gauge symmetry of the vacuum.\ Therefore,\ when looking for a higher-geometric signature of $\k$-symmetry,\ we should investigate the vacuum restriction $\,\cG{}^{(1)}_{\rm vac}\,$ of the extended HP super-1-gerbe.\ Taking into account the higher-geometric interpretation of gauge symmetries,\ we seek to establish a ${\rm sISO}(d,1|D_{d,1})_{\rm vac}$-equivariant structure on the latter.\ The very definition of the $\k$-symmetry superdistribution (and of the limit of its weak derived flag) makes it obvious that the structure,\ if present,\ is descendable -- indeed, $\,\cG{}^{(1)}_{\rm vac}\,$ is a flat 1-gerbe.\ But then the correspondence \eqref{equivcorresp} in conjunction with our earlier description of the leaves of $\,\Si^{\rm HP}_{\rm vac}\,$ as full orbits of $\,{\rm sISO}(d,1|D_{d,1})_{\rm vac}\,$ leads us to posit,\ as a hypothesis to be verified,\ the nullity of $\,\cG{}^{(1)}_{\rm vac}$,\ that is the existence of a 1-isomorphism
\qq\nn
\t\ :\ \cG{}^{(1)}_{\rm vac}\xrightarrow{\ \cong\ }\cI^{(1)}_0\,,
\qqq
in which $\,\cI^{(1)}_0\,$ is to be understood as (the pullback of) the unique 1-gerbe over the 0-dimensional orbispace of a leaf of the lifted vacuum foliation with respect to the action of the $\k$-symmetry group.\ This hypothesis was first formulated (without a mention of the lift) in \Rxcite{Rem.\,7.15}{Suszek:2020xcu},\ {\it cp} also \Rxcite{Sec.\,7}{Suszek:2020rev}.\ Below,\ we prove it directly (essentially in the de Rham cohomology) and in a more manifestly supersymmetric procedure,\ suggested by the track of thought delineated in \Rcite{Suszek:2020rev},\ in which we stay in the tangent sheaf of a leaf of $\,\Si^{\rm HP}_{\rm vac}\,$ and exploit the Lie-superalgebra structure on its model $\,\gt{vac}(\gt{sB}^{{\rm (HP)}}_{1,2})$.

We begin our investigation on the leaf $\,\cD_{i,\upsilon_i}\subset\Si^{\rm HP}_{\rm vac}\cap\cV_i\,$ of the vacuum superdistribution \eqref{eq:vacfol},\ embedded in the superdomain with the previously introduced local coordinates $\,(\theta^\a,x^a,\phi^{\unl b\widehat c}_i)$.\ We have (keeping the pullbacks by $\,\iota_{\rm vac}\,$ implicit to unburden the notation)
\qq\nn
\underset{\tx{\ciut{(3)}}}{\widehat\chi}\rstr_{\cD_{i,\upsilon_i}}=0\,,
\qqq
and so we pass to consider the curving of the extended HP super-1-gerbe restricted to
\qq\nn
\sfY\cD_{i,\upsilon_i}\equiv\sfY{\rm sISO}(d,1|D_{d,1})\rstr_{\cD_{i,\upsilon_i}}\ni\bigl(\theta^\a,x^a,\xi_\b,\phi^{\unl b\widehat c}_i\bigr)\,,
\qqq
whereby we find
\qq\nn
\widehat{\underset{\tx{\ciut{(2)}}}{\sfY\b}}(\theta,x,\xi,\phi_i)=\sfd\xi_\a\wedge\sfd\theta^\a+2L(\phi_i)^{-1\,0}_{\ \ \ \ a}\,L(\phi_i)^{-1\,1}_{\ \ \ \ b}\,\sfd x^a\wedge\sfd x^b+\widetilde\D(\theta,x,\phi_i)\,,
\qqq
with
\qq\nn
\widetilde\D(\theta,x,\phi_i)&=&L(\phi_i)^{-1\,0}_{\ \ \ \ a}\,L(\phi_i)^{-1\,1}_{\ \ \ \ b}\,\bigl(\sfd x^a\wedge\theta\,\ovl\G{}^b\,\sfd\theta-\sfd x^b\wedge\theta\,\ovl\G{}^a\,\sfd\theta+\tfrac{1}{2}\,\theta\,\ovl\G{}^a\,\sfd\theta\wedge\theta\,\ovl\G{}^b\,\sfd\theta\bigr)\cr\cr
&&-p^{\unl a}(\theta,x,\phi_i)\wedge S(\phi_i)^{-1}\,\theta\,\ovl\G{}_{\unl a}\,q(\theta,x,\phi_i)\,.
\qqq
However,\ on $\,\cD_{i,\upsilon_i}$,\ where
\qq\nn
\ovl\G{}^1\,q\rstr_{\cD_{i,\upsilon_i}}=-\ovl\G{}^0\,q\rstr_{\cD_{i,\upsilon_i}}\,,
\qqq
we obtain the identity
\qq\nn
L(\phi_i)^{-1\,0}_{\ \ \ \ a}\,L(\phi_i)^{-1\,1}_{\ \ \ \ b}\,\theta\,\ovl\G{}^a\,\sfd\theta\wedge\theta\,\ovl\G{}^b\,\sfd\theta=0\,,
\qqq
and so also
\qq\nn
\widetilde\D(\theta,x,\phi_i)=0\,,
\qqq
whence also (for the totally skew tensor $\,\ep_{\unl a\unl b}=-\ep_{\unl b\unl a}\,$ with $\,\ep_{01}=1$)
\qq\label{eq:Ybetonvac}
\widehat{\underset{\tx{\ciut{(2)}}}{\sfY\b}}(\theta,x,\xi,\phi_i)=\sfd\bigl(\theta^\a\,\sfd\xi_\a+\ep_{\unl a\unl b}\,L(\phi_i)^{-1\,\unl a}_{\ \ \ \ c}\,L(\phi_i)^{-1\,\unl b}_{\ \ \ \ d}\,x^c\,\sfd x^d\bigr)\,.
\qqq
Indeed, the identities
\qq\nn
\sfd L(\phi_i)^{-1\,\unl a}_{\ \ \ \ b}=-j^{\unl a\unl c}(\theta,x,\phi_i)\,\eta_{\unl c\unl d}\,L(\phi_i)^{-1\,\unl d}_{\ \ \ \ b}
\qqq
that obtain on $\,\cD_{i,\upsilon_i}\,$ yield the desired result
\qq\nn
\sfd\bigl(\ep_{\unl a\unl b}\,L(\phi_i)^{-1\,\unl a}_{\ \ \ \ c}\,L(\phi_i)^{-1\,\unl b}_{\ \ \ \ d}\,x^c\,\sfd x^d\bigr)-2L(\phi_i)^{-1\,0}_{\ \ \ \ a}\,L(\phi_i)^{-1\,1}_{\ \ \ \ b}\,\sfd x^a\wedge\sfd x^b=0\,.
\qqq
Following the standard gerbe-theoretic procedure,\ we erect a trivial principal $\bC^\x$-bundle over $\,\cD_{i,\upsilon_i}$,
\qq\nn
\pi_{\xcE_{i,\upsilon_i}}\equiv\pr_1\ :\ \xcE_{i,\upsilon_i}\equiv\sfY\cD_{i,\upsilon_i}\x\bC^\x\too\sfY\cD_{i,\upsilon_i}\,,
\qqq
and endow it with the principal connection super-1-form ($(\theta^\a,x^a,\phi^{\unl b\widehat c}_i,\xi_\b,z)\in\sfY\cD_{i,\upsilon_i}\x\bC^\x$)
\qq\nn
\underset{\tx{\ciut{(1)}}}{\cA}{}_{\xcE_{i,\upsilon_i}}(\theta,x,\xi,\phi_i,z)=\tfrac{\sfi\,\sfd z}{z}-\theta^\a\,\sfd\xi_\a-\ep_{\unl a\unl b}\,L(\phi_i)^{-1\,\unl a}_{\ \ \ \ c}\,L(\phi_i)^{-1\,\unl b}_{\ \ \ \ d}\,x^c\,\sfd x^d=:\tfrac{\sfi\,\sfd z}{z}+\txA_{i,\upsilon_i}(\theta,x,\xi,\phi_i)\,.
\qqq
The bundle may subsequently be pulled back to $\,\sfY^{[2]}\cD_{i,\upsilon_i}\equiv\sfY\cD_{i,\upsilon_i}\x_{\cD_{i,\upsilon_i}}\sfY\cD_{i,\upsilon_i}\,$ along the canonical projections $\,\pr_n\ :\ \sfY^{[2]}\cD_{i,\upsilon_i}\too\sfY\cD_{i,\upsilon_i},\ n\in\{1,2\}$,\ whereupon we obtain the two (trivial) principal $\bC^\x$-bundles
\qq\nn
\pi_{\pr_n^*\xcE_{i,\upsilon_i}}\equiv\pr_1\ :\ \pr_n^*\xcE_{i,\upsilon_i}\equiv\sfY^{[2]}\cD_{i,\upsilon_i}\hspace{2pt}{}_{\pr_n}\hspace{-3pt}\x_{\pi_{\xcE_{i,\upsilon_i}}}\xcE_{i,\upsilon_i}\too\sfY^{[2]}\cD_{i,\upsilon_i}\,,\qquad n\in\{1,2\}
\qqq
with
\qq\nn
\widehat\pr{}_n\equiv\pr_2\ :\ \pr_n^*\xcE_{i,\upsilon_i}\too\xcE_{i,\upsilon_i}
\qqq
and with the respective principal connection super-1-forms
\qq\nn
\widehat\pr{}^*_n\underset{\tx{\ciut{(1)}}}{\cA}{}_{\xcE_{i,\upsilon_i}}\equiv\pr_2^*\underset{\tx{\ciut{(1)}}}{\cA}{}_{\xcE_{i,\upsilon_i}}\,.
\qqq
Next,\ we tensor the second of these bundles,\ $\,\pr_2^*\xcE_{i,\upsilon_i}$,\ with the restriction of the principal $\bC^\x$-bundle $\,\widetilde\xcL\,$ of the extended HP super-1-gerbe to $\,\sfY^{[2]}\cD_{i,\upsilon_i}\,$ and look for a connection-preserving principal $\bC^\x$-bundle isomorphism
\qq\nn
\a_{\xcE_{i,\upsilon_i}}\ :\ \widetilde\xcL\rstr_{\sfY^{[2]}\cD_{i,\upsilon_i}}\ox\pr_2^*\xcE_{i,\upsilon_i}\xrightarrow{\ \cong\ }\pr_1^*\xcE_{i,\upsilon_i}\,.
\qqq
Direct comparison of the base components of the respective connection super-1-forms,
\qq\nn
\bigl(\sfY^{[2]}\pi^*\txa+\pr_2^*\txA_{i,\upsilon_i}\bigr)\bigl((\theta,x,\xi_1,\phi_i),(\theta,x,\xi_2,\phi_i)\bigr)=\pr_2^*\txA_{i,\upsilon_i}\bigl((\theta,x,\xi_1,\phi_i),(\theta,x,\xi_2,\phi_i)\bigr)\,,
\qqq
indicates that we may take the isomorphism in the trivial form,\ with the coordinate presentation
\qq\nn
&&\a_{\xcE_{i,\upsilon_i}}\bigl(\bigl((\theta,x,\xi_1,\phi_i),(\theta,x,\xi_2,\phi_i),1\bigr)\ox\bigl((\theta,x,\xi_1,\phi_i),(\theta,x,\xi_2,\phi_i),(\theta,x,\xi_2,\phi_i,z)\bigr)\bigr)\cr\cr 
&=&\bigl((\theta,x,\xi_1,\phi_i),(\theta,x,\xi_2,\phi_i),(\theta,x,\xi_1,\phi_i,z)\bigr)\,,
\qqq
or, symbolically,
\qq\nn
\a_{\xcE_{i,\upsilon_i}}\equiv\bd1\,.
\qqq
This is automatically compatible with the (trivial) groupoid structure on (the fibres of) $\,\widetilde\xcL\rstr_{\sfY^{[2]}\cD_{i,\upsilon_i}}$,\ and so we conclude that the quadruple
\qq\nn
\t_{i,\upsilon_i}=\bigl(\xcE_{i,\upsilon_i},\pi_{\xcE_{i,\upsilon_i}},\underset{\tx{\ciut{(1)}}}{\cA}{}_{\xcE_{i,\upsilon_i}},\a_{\xcE_{i,\upsilon_i}}\bigr)
\qqq
defines a trivialisation
\qq\nn
\t_{i,\upsilon_i}\ :\ \cG{}^{(1)}_{\rm vac}\rstr_{\cD_{i,\upsilon_i}}\xrightarrow{\ \cong\ }\cI^{(1)}_0\rstr_{\cD_{i,\upsilon_i}}\,.
\qqq
Combining the local trivialisations over the entire vacuum foliation gives us the sought-after global trivialisation
\qq\label{eq:exts1gtrivac}\qquad\qquad
\t\equiv(\xcE,\pi_\xcE,\underset{\tx{\ciut{(1)}}}{\cA}{}_\xcE,\a_\xcE\bigr)\ :\ \cG{}^{(1)}_{\rm vac}\xrightarrow{\ \cong\ }\cI^{(1)}_0\,,\qquad\qquad\t\equiv\bigsqcup_{i\in I}\,\bigsqcup_{\upsilon_i\in\Upsilon_i}\,\t_{i,\upsilon_i}\,.
\qqq
Of course, ultimately, we want to make statements about the extended HP super-1-gerbe \emph{descended} to the \emph{physical} vacuum foliation $\,\Si^{\rm HP}_{\rm phys\ vac}$.\ For the results of the above analysis to descend to the homogeneous space $\,{\rm sISO}(d,1|D_{d,1})/{\rm Spin}(d,1)_{\rm vac}\supset\Si^{\rm HP}_{\rm phys\ vac}$,\ we need essentially to equip $\,\t\,$ with a \emph{descendable} ${\rm Spin}(d,1)_{\rm vac}$-equivariant structure.\ Luckily,\ our construction provides us with such a structure,\ namely -- the trivial one.\ Indeed,\ upon invoking the relation between the local sections $\,\si^{\rm vac}_i\,$ of \Reqref{eq:sivac} over nonempty intersections $\,\cU^{\rm vac}_{ij}\equiv\cU^{\rm vac}_i\cap\cU^{\rm vac}_j\,$ of superdomains $\,\cU^{\rm vac}_i\,$ (with global coordinates $\,(\theta^\a,x^a,\xi_\b,\phi^{\unl b\widehat c}_i)$) and $\,\cU^{\rm vac}_j\,$ (with global coordinates $\,(\theta^\a,x^a,\xi_\b,\phi^{\unl b\widehat c}_j)$),
\qq\nn
\si^{\rm vac}_j\rstr_{\cU^{\rm vac}_{ij}}=|\wp|_{h_{ij}\circ\pr_2}\bigl(\si^{\rm vac}_i\rstr_{\cU^{\rm vac}_{ij}}\bigr)\,,
\qqq
expressed in terms of the transition maps $\,h_{ij}\ :\ \cO_{ij}\too{\rm Spin}(d,1)_{\rm vac}\,$ of the principal ${\rm Spin}(d,1)_{\rm vac}$-bundle $\,{\rm Spin}(d,1)\too{\rm Spin}(d,1)/{\rm Spin}(d,1)_{\rm vac}\,$ (inherited by that of \Reqref{diag:princsbndlvac}),\ we readily verify the desired gluing property for the base components of the relevant restrictions $\,\underset{\tx{\ciut{(1)}}}{\cA}{}_{\xcE_{i,\upsilon_i}}\,$ and $\,\underset{\tx{\ciut{(1)}}}{\cA}{}_{\xcE_{j,\upsilon_j}}\,$ of the principal $\bC^\x$-connection $\,\underset{\tx{\ciut{(1)}}}{\cA}{}_\xcE\,$ on $\,\xcE$,
\qq\nn
\txA_{j,\upsilon_j}(\theta,x,\xi,\phi_j)=\txA_{i,\upsilon_i}(\theta,x,\xi,\phi_i)\,,
\qqq 
which follows from the block-diagonal structure of the matrix $\,L(h)\,$ for $\,h\in{\rm Spin}(d,1)_{\rm vac}\,$ with respect to the decomposition $\,\gt{mink}(d,1)=\corr{P_0,P_1}\oplus(\bigoplus_{\widehat a=2}^d\,\corr{P_{\widehat a}})$.\ We conclude that

\bethe\label{thm:vactriv}
The super-1-gerbe of the GS super-$\si$-model in the HP formulation descended from the extended HP super-1-gerbe to the supertarget $\,{\rm sISO}(d,1|D_{d,1})/{\rm Spin}(d,1)_{\rm vac}\,$ trivialises upon restriction to the vacuum of the (super)field theory.
\ethe

\noindent The main principle underlying the scheme of geometrisation proposed in \Rcite{Suszek:2017xlw} (and recalled in Sec.\,\ref{sec:geometrise}) is the \emph{invariance} of all structures under consideration with respect to the global supersymmetry \emph{enforced through restriction} of the standard constructions (of 1-gerbes,\ their 1- and 2-isomorphisms) due to Murray {\it et al.}\ to the category of Lie supergroups.\ There seems to be no obvious way of implementing this principle in the last construction that leads up to Theorem 2. -- quite simply because there is no natural Lie-supergroup structure on the vacuum of the super-$\si$-model.\ Rather than trying to save the day,\ at least partially,\ by imposing invariance with respect to a residual global supersymmetry of the vacuum\footnote{{\it Cp} \Rxcite{Prop.\,5.5}{Suszek:2020xcu}.} (an idea that we leave for a future study),\ we rectify the present situation by passing to the tangent sheaves of the various supermanifolds entering the definition of the vacuum restriction $\,\cG{}^{(1)}_{\rm vac}\,$ of the extended HP super-1-gerbe of the superstring and read off the Lie-superalgebraic trace of the trivialisation \eqref{eq:exts1gtrivac}.\medskip

The point of departure of our analysis is the concise restatement of a faithful Lie-superalgebraic model of the composite diagram 
{\small\qq\nn
\alxydim{@C=1cm@R=1cm}{ & & & \bd1 \ar[d] \ar[dll] \\ \mu_{\widetilde\xcL}\equiv\bd1\ :\ \pr_{1,2}^*\widetilde\xcL\ox\pr_{2,3}^*\widetilde\xcL\xrightarrow{\ \cong\ }\pr_{1,3}^*\widetilde\xcL \ar[d] & \bC^\x \ar[r] & \widetilde\xcL,\underset{\tx{\ciut{(1)}}}{\cA}{}_{\widetilde\xcL} \ar[d]^{\pi_{\widetilde\xcL}} & \bR^{0|D_{d,1}} \ar[d] \\ \sfY^{[3]}{\rm sISO}(d,1|D_{d,1}) \ar@/^.75pc/[rr]^{\pr_{1,2}} \ar@<0ex>[rr]|-{\pr_{2,3}} \ar@/^-.75pc/[rr]_{\pr_{1,3}} & & \sfY^{[2]}{\rm sISO}(d,1|D_{d,1}) \ar@<-.75ex>[r]_{\pr_2} \ar@<.75ex>[r]^{\pr_1} \ar[d] & \sfY{\rm sISO}(d,1|D_{d,1}),\widehat{\underset{\tx{\ciut{(2)}}}{\sfY\b}} \ar[d]^{\pi_{\sfY{\rm sISO}(d,1|D_{d,1})}} \\ & & \bd1 & {\rm sISO}(d,1|D_{d,1}),\underset{\tx{\ciut{(3)}}}{\widehat\chi} \ar[l] }
\qqq}
\hspace{-.1cm}in $\,{\bf sMan}\,$ (actually,\ in $\,{\bf sLieGrp}$),\ decorated with the relevant CaE data (as well as indicators of the various supercentral extensions),\ in the form
{\small\qq\nn
\alxydim{@C=1cm@R=1cm}{ & & & \brd0 \ar[d] \ar[dll] \\ \mu_{\widetilde\lgt}\equiv\bd1\ :\ \pr_{1,2}^*\widetilde\lgt\ox\pr_{2,3}^*\widetilde\lgt\xrightarrow{\ \cong\ }\pr_{1,3}^*\widetilde\lgt \ar[d] & \bR \ar[r] & \widetilde\lgt,\z \ar[d]^{\pi_{\widetilde\lgt}} & \bR^{0|D_{d,1}} \ar[d] \\ \sfY^{[3]}\gt{siso}(d,1|D_{d,1}) \ar@/^.75pc/[rr]^{\pr_{1,2}} \ar@<0ex>[rr]|-{\pr_{2,3}} \ar@/^-.75pc/[rr]_{\pr_{1,3}} & & \sfY^{[2]}\gt{siso}(d,1|D_{d,1}) \ar@<-.75ex>[r]_{\pr_2} \ar@<.75ex>[r]^{\pr_1} \ar[d] & \sfY\gt{siso}(d,1|D_{d,1}),\widehat{\underset{\tx{\ciut{(2)}}}{\sfY\b}} \ar[d]^{\pi_{\sfY\gt{siso}(d,1|D_{d,1})}} \\ & & \brd0 & \gt{siso}(d,1|D_{d,1}),\underset{\tx{\ciut{(3)}}}{\widehat\chi} \ar[l] }\,,
\qqq}
\hspace{-.1cm}to be referred to as the ${\bf sLieAlg}${\bf -skeleton} of the supermanifold diagram.\ Now,\ given the supermanifold diagram describing the trivialisation of the extended HP super-1-gerbe over the vacuum foliation,
\qq\label{diag:trivac}\hspace{1cm}
\alxydim{@C=1cm@R=1cm}{ & \bd1 \ar[d] & & \\ \a_\xcE\equiv\bd1\ :\ \widetilde\xcL{}_{\rm vac}\ox\pr_2^*\xcE\xrightarrow{\ \cong\ }\pr_1^*\xcE \ar[d] & \bC^\x \ar[r] & \xcE,\underset{\tx{\ciut{(1)}}}{\cA}{}_\xcE \ar[d]^{\pi_\xcE} & \\ \sfY^{[2]}\Si^{\rm HP}_{\rm vac} \ar@<-.75ex>[rr]_{\pr_2} \ar@<.75ex>[rr]^{\pr_1} & & \sfY\Si^{\rm HP}_{\rm vac},\widehat{\underset{\tx{\ciut{(2)}}}{\sfY\b}}{}_{\rm vac} \ar[d]^{\pi_{\sfY\Si^{\rm HP}_{\rm vac}}} \ar[r] & \bd1 \\ & & \Si^{\rm HP}_{\rm vac},\underset{\tx{\ciut{(3)}}}{\widehat\chi}{}_{\rm vac}\equiv 0 & }\,,
\qqq
and the Lie-superalgebra monomorphism
\qq\nn
\brd0\too\gt{vac}\bigl(\gt{sB}^{{\rm (HP)}}_{1,2}\bigr)\xrightarrow{\ \jmath_{\rm vac}\ }\gt{siso}(d,1|D_{d,1})
\qqq
at its root,\ regarded as a ${\bf sLieAlg}$-skeleton of the embedding
\qq\nn
\Si^{\rm HP}_{\rm vac}\supset\cD_{i,\upsilon_i}\emb{\rm sISO}(d,1|D_{d,1})\,,
\qqq
we may enquire as to the existence of a consistent extension  
\qq\nn
\alxydim{@C=1cm@R=1cm}{ & \brd0 \ar[dr] \ar[dl] & \\ \bR^{0|\D_{d,1}} \ar[d] \ar@{^{(}->}[rr]^{} & & \bR^{0|D_{d,1}} \ar[d] \\ \sfY\gt{vac}\bigl(\gt{sB}^{{\rm (HP)}}_{1,2}\bigr) \ar@{^{(}->}[rr]^{\sfY\jmath_{\rm vac}} \ar[d]_{\pi_{\sfY\gt{vac}(\gt{sB}^{{\rm (HP)}}_{1,2})}} & & \sfY\gt{siso}(d,1|D_{d,1}) \ar[d]^{\pi_{\sfY\gt{siso}(d,1|D_{d,1})}} \\ \gt{vac}\bigl(\gt{sB}^{{\rm (HP)}}_{1,2}\bigr) \ar[dr] \ar@{^{(}->}[rr]_{\jmath_{\rm vac}} & & \gt{siso}(d,1|D_{d,1}) \ar[dl] \\ & \brd0 & }
\qqq
of the latter,\ written for some sub-superspace $\,\bR^{0|\D_{d,1}}\,$ of $\,\bR^{0|D_{d,1}}\,$ (with $\,\D_{d,1}\leq D_{d,1}$) and for a Lie superalgebra $\,\sfY\gt{vac}(\gt{sB}^{{\rm (HP)}}_{1,2})\,$ to be established together with a Lie-superalgebra monomorphism $\,\sfY\jmath_{\rm vac}\,$ and the extension $\,\pi_{\sfY\gt{vac}(\gt{sB}^{{\rm (HP)}}_{1,2})}$,\ and such that there exists a ${\bf sLieAlg}$-skeleton of Diag.\,\eqref{diag:trivac} of the form
\qq\label{diag:sLieAlgshad}\hspace{2cm}
\alxydim{@C=1cm@R=1cm}{ & \brd0 \ar[d] & & \\ \a_\egt\equiv\bd1\ :\ \sfY^{[2]}\jmath_{\rm vac}^*\widetilde\lgt\ox\pr_2^*\egt\xrightarrow{\ \cong\ }\pr_1^*\egt \ar[d] & \bR \ar[r] & \egt,\breve\z{}_\xcE \ar[d]^{\pi_\egt} \\ \sfY^{[2]}\gt{vac}\bigl(\gt{sB}^{{\rm (HP)}}_{1,2}\bigr) \ar@<-.75ex>[rr]_{\pr_2} \ar@<.75ex>[rr]^{\pr_1} & & \sfY\gt{vac}\bigl(\gt{sB}^{{\rm (HP)}}_{1,2}\bigr),\widehat{\underset{\tx{\ciut{(2)}}}{\sfY\b}}{}_{\rm vac} \ar[d]^{\pi_{\sfY\gt{vac}(\gt{sB}^{{\rm (HP)}}_{1,2})}} \ar[r] & \brd0 \\ & & \gt{vac}\bigl(\gt{sB}^{{\rm (HP)}}_{1,2}\bigr),\underset{\tx{\ciut{(3)}}}{\widehat\chi}{}_{\rm vac}\equiv 0 & }
\qqq
in which 
\qq\label{eq:egtext}
\brd0\too\bR\too\egt\xrightarrow{\ \pi_\egt\ }\sfY\gt{vac}\bigl(\gt{sB}^{{\rm (HP)}}_{1,2}\bigr)\too\brd0
\qqq
is a central extension determined by the super-2-cocycle 
\qq\nn
\widehat{\underset{\tx{\ciut{(2)}}}{\sfY\b}}{}_{\rm vac}=\sfY\jmath_{\rm vac}^*\widehat{\underset{\tx{\ciut{(2)}}}{\sfY\b}}
\qqq 
and such that the super-1-form $\,\breve\z{}_\xcE\,$ on $\,\egt\,$ dual to the central generator given as the image of $\,1\in\bR\,$ in $\,\sfY\gt{vac}(\gt{sB}^{{\rm (HP)}}_{1,2})\,$ trivialises the pullback of $\,\widehat{\underset{\tx{\ciut{(2)}}}{\sfY\b}}{}_{\rm vac}\,$ along $\,\pi_\egt$,
\qq\label{eq:trivtrivB}
\widehat\d\breve\z{}_\xcE=-\pi_\egt^*\widehat{\underset{\tx{\ciut{(2)}}}{\sfY\b}}{}_{\rm vac}\,,
\qqq
and in which $\,\sfY^{[2]}\jmath_{\rm vac}^*\widetilde\lgt\,$ is a central extension
\qq\nn
\brd0\too\bR\too\sfY^{[2]}\jmath_{\rm vac}^*\widetilde\lgt\xrightarrow{\ \pi_{\sfY^{[2]}\jmath_{\rm vac}^*\widetilde\lgt}\equiv\pr_1\ }\sfY^{[2]}\gt{vac}\bigl(\gt{sB}^{{\rm (HP)}}_{1,2}\bigr)\too\brd0
\qqq
consistent with that of \Reqref{eq:extbunext} in the sense expressed by the diagram
\qq\nn
\alxydim{@C=1cm@R=1cm}{ & \brd0 \ar[d] & \\ & \bR \ar[dl] \ar[dr] & \\ \sfY^{[2]}\jmath_{\rm vac}^*\widetilde\lgt \ar@{^{(}->}[rr]^{\widetilde\xcL\jmath_{\rm vac}} \ar[d]_{\pi_{\sfY^{[2]}\jmath_{\rm vac}^*\widetilde\lgt}} & & \widetilde\lgt \ar[d]^{\pi_{\widetilde\lgt}} \\ \sfY^{[2]}\gt{vac}\bigl(\gt{sB}^{{\rm (HP)}}_{1,2}\bigr) \ar[dr] \ar@{^{(}->}[rr]_{\sfY^{[2]}\jmath_{\rm vac}} & & \sfY^{[2]}\gt{siso}(d,1|D_{d,1}) \ar[dl] \\ & \brd0 & }\,,
\qqq
with the Lie-superalgebra monomorphism $\,\widetilde\xcL\jmath_{\rm vac}$.\ A constructive positive answer to the question thus posed is laid out below.

The structure of the first extension, $\,\sfY\gt{vac}(\gt{sB}^{{\rm (HP)}}_{1,2})$,\ is readily read off from the commutator $\,[\sfY Q_\a,\sfY P_a]\,$ of $\,\sfY\gt{siso}(d,1|D_{d,1})$:\ Upon restricting to the generators $\,\sfY\breve Q{}_{\unl\a}\,$ spanning $\,{\rm im}\,\sfP^{(1)}{}^{\rm T}\,$ within the sub-superspace $\,\bigoplus_{\a=1}^{D_{d,1}}\,\corr{\sfY Q_\a}\,$ and to the $\,\sfY P_{\unl a}\,$ spanning the lift of $\,\tgt_{\rm vac}^{(0)}$,\ we are naturally restricted to the subspace 
\qq\nn
{\rm im}\,\sfP^{(1)}\equiv\corr{\ \sfP^{(1)}{}^\a_{\ \b}\,Z^\b\ \vert\ \a\in\ovl{1,D_{d,1}}\ }\subset\bigoplus_{\a=1}^{D_{d,1}}\,\corr{Z^\a}\,.
\qqq 
Denote its basis as $\,\{\breve Z{}^{\unl\a}\}_{\unl\a\in\ovl{1,\frac{D_{d,1}}{2}}}\,$ to postulate
\qq
\sfY\gt{vac}(\gt{sB}^{{\rm (HP)}}_{1,2})&:=&\bigl(\bigoplus_{\unl\a=1}^{\frac{D_{d,1}}{2}}\,\corr{\sfY\breve Q{}_{\unl\a}}\oplus\corr{\sfY P_0,\sfY P_1}\oplus\bigoplus_{\unl\b=1}^{\frac{D_{d,1}}{2}}\,\corr{\breve Z{}^{\unl\b}}\bigr)\oplus\bigl(\corr{\sfY J_{01}}\oplus\bigoplus_{\widehat a<\widehat b=2}^d\,\corr{\sfY J_{\widehat a\widehat b}}\bigr)\cr\cr
&\equiv&\sfY\gt{smink}(d,1|D_{d,1})_{\rm vac}\oplus\gt{spin}(d,1)_{\rm vac} \label{eq:Yvacbas}
\qqq
with the Lie-superalgebra structure induced by the restriction of the superbracket of $\,\sfY\gt{siso}(d,1|D_{d,1})$.\ We confirm the self-consistency of the postulate by inspecting the brackets
\qq\nn
\bigl[\sfY J_{01},\bigl(\sfP^{(1)}\,Z\bigr)^\a]=-\tfrac{1}{2}\,\G_{01}{}^\a_{\ \b}\,\bigl(\sfP^{(1)}\,Z\bigr)^\b
\qqq
and
\qq\nn
\bigl[\sfY J_{\widehat a\widehat b},\bigl(\sfP^{(1)}\,Z\bigr)^\a]=-\tfrac{1}{2}\,\G_{\widehat a\widehat b}{}^\a_{\ \b}\,\bigl(\sfP^{(1)}\,Z\bigr)^\b\,.
\qqq
Thus,\ we have 
\qq\nn
\D_{d,1}\equiv\tfrac{D_{d,1}}{2}
\qqq
and
\qq\nn
\pi_{\sfY\gt{vac}(\gt{sB}^{{\rm (HP)}}_{1,2})}\equiv\pi_{\sfY\gt{siso}(d,1|D_{d,1})}\rstr_{\sfY\gt{vac}(\gt{sB}^{{\rm (HP)}}_{1,2})}\,.
\qqq
On the new Lie superalgebra,\ we find the super-2-form
\qq\nn
\widehat{\underset{\tx{\ciut{(2)}}}{\sfY\b}}{}_{\rm vac}=\breve z{}_{\unl\a}\wedge\pi_{\sfY\gt{vac}(\gt{sB}^{{\rm (HP)}}_{1,2})}^*\breve q{}^{\unl\a}+2 \pi_{\sfY\gt{vac}(\gt{sB}^{{\rm (HP)}}_{1,2})}^*\bigl(p^0\wedge p^1\bigr)\,,
\qqq
written in terms of the duals $\,\breve z{}_{\unl\a}\,$ of the $\,\breve Z{}^{\unl\a}\,$ and the duals $\,\breve q{}^{\unl\a}\,$ of the $\,\breve Q{}_{\unl\a}$.\ The super-2-form satifies the identity
\qq\nn
\widehat\d\widehat{\underset{\tx{\ciut{(2)}}}{\sfY\b}}{}_{\rm vac}=\jmath_{\rm vac}^*\widehat{\underset{\tx{\ciut{(3)}}}{\chi}}\equiv 0\,,
\qqq
and so it determines a central extension \eqref{eq:egtext} with the supervector-space structure
\qq\nn
\egt&=&\bigoplus_{\unl\a=1}^{\frac{D_{d,1}}{2}}\,\corr{\xcE\breve Q{}_{\unl\a}}\oplus\corr{\xcE P_0,\xcE P_1}\oplus\bigoplus_{\unl\b=1}^{\frac{D_{d,1}}{2}}\,\corr{\xcE\breve Z{}^{\unl\b}}\oplus\corr{\breve\cZ}\oplus\corr{\xcE J_{01}}\oplus\bigoplus_{\widehat a<\widehat b=2}^d\,\corr{\xcE J_{\widehat a\widehat b}}\cr\cr
&\cong&\bigl(\sfY\gt{smink}(d,1|D_{d,1})_{\rm vac}\oplus\bR\bigr)\oplus\gt{spin}(d,1)_{\rm vac}
\qqq
and the associated projection
\qq\nn
\pi_\egt\equiv\pr_1\oplus\id_{\gt{spin}(d,1)_{\rm vac}}\ :\ \bigl(\sfY\gt{smink}(d,1|D_{d,1})_{\rm vac}\oplus\bR\bigr)\oplus\gt{spin}(d,1)_{\rm vac}\too\sfY\gt{smink}(d,1|D_{d,1})_{\rm vac}\oplus\gt{spin}(d,1)_{\rm vac}\,,
\qqq
and with the superbrackets
\qq\nn
&\{\xcE\breve Q{}_{\unl\a},\xcE\breve Q{}_{\unl\b}\}=\ovl\g{}^{\unl a}_{\unl\a\unl\b}\,\xcE P_{\unl a}\,,\qquad\qquad[\xcE P_0,\xcE P_1]=2\breve\cZ\,,\qquad\qquad[\xcE\breve Q{}_{\unl\a},\xcE P_{\unl a}]=\ovl\g_{\unl a}{}_{\unl\a\unl\b}\,\xcE\breve Z{}^{\unl\b}\,,&\cr\cr
&\{\xcE\breve Q{}_{\unl\a},\xcE\breve Z{}^{\unl\b}\}=-\d_{\unl\a}^{\ \unl\b}\,\breve\cZ\,,\qquad\qquad[\xcE P_{\unl a},\xcE\breve Z{}^{\unl\a}]=0\,,\qquad\qquad\{\xcE\breve Z{}^{\unl\a},\xcE\breve Z{}^{\unl\b}\}=0\,,&\cr\cr
&[\xcE\breve Q{}_{\unl\a},\breve\cZ]=0\,,\qquad\qquad[\xcE P_{\unl a},\breve\cZ]=0\,,\qquad\qquad[\breve\cZ,\breve\cZ]=0\,,&\cr\cr
&[\xcE J_{01},\xcE\breve Q{}_{\unl\a}]=\tfrac{1}{2}\,\g_{01}{}^{\unl\b}_{\ \unl\a}\,\xcE\breve Q{}_{\unl\b}\,,\qquad\qquad[\xcE J_{\widehat a\widehat b},\xcE\breve Q{}_{\unl\a}]=\tfrac{1}{2}\,\g_{\widehat a\widehat b}{}^{\unl\b}_{\ \unl\a}\,\xcE\breve Q{}_{\unl\b}\,,&\cr\cr
&[\xcE J_{01},\xcE P_{\unl a}]=\d_{1\unl a}\,\xcE P_0+\d_{0\unl a}\,P_1\,,\qquad\qquad[\xcE J_{\widehat a\widehat b},\xcE P_{\unl c}]=0\,,&\cr\cr
&[\xcE J_{01},\xcE\breve Z{}^{\unl\a}]=-\tfrac{1}{2}\,\g_{01}{}^{\unl\a}_{\ \unl\b}\,\xcE\breve Z{}^{\unl\b}\,,\qquad\qquad[\xcE J_{\widehat a\widehat b},\xcE\breve Z{}^{\unl\a}]=-\tfrac{1}{2}\,\g_{\widehat a\widehat b}{}^{\unl\a}_{\ \unl\b}\,\xcE\breve Z{}^{\unl\b}\,,&\cr\cr
&[\xcE J_{01},\breve\cZ]=0\,,\qquad\qquad[\xcE J_{\widehat a\widehat b},\breve\cZ]=0\,,&\cr\cr
&[\xcE J_{01},\xcE J_{\widehat a\widehat b}]=0\,,\qquad\qquad[\xcE J_{\widehat a\widehat b},\xcE J_{\widehat c\widehat d}]=\d_{\widehat a\widehat d}\,\xcE J_{\widehat b\widehat c}-\d_{\widehat a\widehat c}\,\xcE J_{\widehat b\widehat d}+\d_{\widehat b\widehat c}\,\xcE J_{\widehat a\widehat d}-\d_{\widehat b\widehat d}\,\xcE J_{\widehat a\widehat c}\,.&
\qqq
Let $\,\breve\z{}_\xcE\,$ be the dual of $\,\breve\cZ$.\ Clearly,\ it satisfies the desired relation \eqref{eq:trivtrivB},\ and so the first stage of the construction is complete.

In the next step,\ we form the pullback Lie superalgebra
\qq\nn
\sfY^{[2]}\jmath_{\rm vac}^*\widetilde\lgt\equiv\sfY^{[2]}\gt{vac}\bigl(\gt{sB}^{{\rm (HP)}}_{1,2}\bigr)\hspace{2pt}{}_{\sfY^{[2]}\jmath_{\rm vac}}\hspace{-3pt}\oplus_{\pi_{\widetilde\lgt}}\widetilde\lgt 
\qqq
with
\qq\nn
\pi_{\sfY^{[2]}\jmath_{\rm vac}^*\widetilde\lgt}\equiv\pr_1\ :\ \sfY^{[2]}\jmath_{\rm vac}^*\widetilde\lgt\too\sfY^{[2]}\gt{vac}\bigl(\gt{sB}^{{\rm (HP)}}_{1,2}\bigr)\,,\qquad\qquad
\widetilde\xcL\jmath_{\rm vac}\equiv\pr_2\ :\ \sfY^{[2]}\jmath_{\rm vac}^*\widetilde\lgt\too\widetilde\lgt
\qqq
and with the basis
\qq\nn
\sfY^{[2]}\jmath_{\rm vac}^*\widetilde\lgt&=&\bigoplus_{\unl\a=1}^{\frac{D_{d,1}}{2}}\,\corr{\bigl(\bigl(\sfY\breve Q{}_{\unl\a},\sfY\breve Q{}_{\unl\a}\bigr),\widetilde\xcL\breve Q{}_{\unl\a}\bigr)\equiv\widetilde\xcL_{\rm vac}\breve Q{}_{\unl\a}}\oplus\bigoplus_{\unl a\in\{0,1\}}\,\corr{\bigl(\bigl(\sfY P_{\unl a},\sfY P_{\unl a}\bigr),\widetilde\xcL P_{\unl a}\bigr)\equiv\widetilde\xcL_{\rm vac} P_{\unl a}}\cr\cr
&&\oplus\bigoplus_{\unl\b=1}^{\frac{D_{d,1}}{2}}\,\corr{\bigl(\bigl(\breve Z{}^{\unl\b},0\bigr),\widetilde\xcL\breve Z{}_{(1)}^{\unl\b}\bigr)\equiv\widetilde\xcL_{\rm vac}\breve Z{}_{(1)}^{\unl\b}}\oplus\bigoplus_{\unl\g=1}^{\frac{D_{d,1}}{2}}\,\corr{\bigl(\bigl(0,\breve Z{}^{\unl\g}\bigr),\widetilde\xcL\breve Z{}_{(2)}^{\unl\g}\bigr)\equiv\widetilde\xcL_{\rm vac}\breve Z{}_{(2)}^{\unl\g}}\oplus\corr{\bigl((0,0),\cZ\bigr)\equiv\cZ_{\rm vac}}\cr\cr
&&\oplus\corr{\bigl(\bigl(\sfY J_{01},\sfY J_{01}\bigr),\widetilde\xcL J_{01}\bigr)\equiv\widetilde\xcL_{\rm vac} J_{01}}\oplus\bigoplus_{\widehat a<\widehat b=2}^d\,\corr{\bigl(\bigl(\sfY J_{\widehat a\widehat b},\sfY J_{\widehat a\widehat b}\bigr),\widetilde\xcL J_{\widehat a\widehat b}\bigr)\equiv\widetilde\xcL_{\rm vac} J_{\widehat a\widehat b}}
\qqq
and the superbracket obtained from the direct-sum one on $\,\sfY^{[2]}\gt{vac}\bigl(\gt{sB}^{{\rm (HP)}}_{1,2}\bigr)\oplus\widetilde\lgt\,$ through restriction.\ With this Lie superalgebra in hand,\ we may,\ at long last,\ finish the construction of the ${\bf sLieAlg}$-skeleton of Diag.\,\eqref{diag:trivac}.\ To this end,\ consider the pullback Lie superalgebras
\qq\nn
\pr_n^*\egt\equiv\sfY^{[2]}\gt{vac}\bigl(\gt{sB}^{{\rm (HP)}}_{1,2}\bigr)\hspace{2pt}{}_{\pr_n}\hspace{-3pt}\oplus_{\pi_\egt}\egt\,,\qquad n\in\{1,2\}
\qqq
with the respective bases
\qq\nn
\pr_1^*\egt&=&\bigoplus_{\unl\a=1}^{\frac{D_{d,1}}{2}}\,\corr{\bigl(\bigl(\sfY\breve Q{}_{\unl\a},\sfY\breve Q{}_{\unl\a}\bigr),\xcE\breve Q{}_{\unl\a}\bigr)\equiv\xcE\breve Q{}^{(1)}_{\unl\a}}\oplus\bigoplus_{\unl a\in\{0,1\}}\,\corr{\bigl(\bigl(\sfY P_{\unl a},\sfY P_{\unl a}\bigr),\xcE P_{\unl a}\bigr)\equiv\xcE P^{(1)}_{\unl a}}\cr\cr
&&\oplus\bigoplus_{\unl\b=1}^{\frac{D_{d,1}}{2}}\,\corr{\bigl(\bigl(\breve Z{}^{\unl\b},0\bigr),\xcE\breve Z{}^{\unl\b}\bigr)\equiv\xcE\breve Z{}^{(1)\,\unl\b}_{(1)}}\oplus\bigoplus_{\unl\g=1}^{\frac{D_{d,1}}{2}}\,\corr{\bigl(\bigl(0,\breve Z{}^{\unl\g}\bigr),0\bigr)\equiv\xcE\breve Z{}^{(1)\,\unl\g}_{(2)}}\oplus\corr{\bigl((0,0),\breve\cZ\bigr)\equiv\xcE\breve\cZ{}^{(1)}}\cr\cr
&&\oplus\corr{\bigl(\bigl(\sfY J_{01},\sfY J_{01}\bigr),\xcE J_{01}\bigr)\equiv\xcE J^{(1)}_{01}}\oplus\bigoplus_{\widehat b<\widehat c=2}^d\,\corr{\bigl(\bigl(\sfY J_{\widehat b\widehat c},\sfY J_{\widehat b\widehat c}\bigr),\xcE J_{\widehat b\widehat c}\bigr)\equiv\xcE J^{(1)}_{\widehat b\widehat c}}\,,\cr\cr\cr
\pr_2^*\egt&=&\bigoplus_{\unl\a=1}^{\frac{D_{d,1}}{2}}\,\corr{\bigl(\bigl(\sfY\breve Q{}_{\unl\a},\sfY\breve Q{}_{\unl\a}\bigr),\xcE\breve Q{}_{\unl\a}\bigr)\equiv\xcE\breve Q{}^{(2)}_{\unl\a}}\oplus\bigoplus_{\unl a\in\{0,1\}}\,\corr{\bigl(\bigl(\sfY P_{\unl a},\sfY P_{\unl a}\bigr),\xcE P_{\unl a}\bigr)\equiv\xcE P^{(2)}_{\unl a}}\cr\cr
&&\oplus\bigoplus_{\unl\b=1}^{\frac{D_{d,1}}{2}}\,\corr{\bigl(\bigl(\breve Z{}^{\unl\g},0\bigr),0\bigr)\equiv\xcE\breve Z{}^{(2)\,\unl\b}_{(1)}}\oplus\bigoplus_{\unl\g=1}^{\frac{D_{d,1}}{2}}\,\corr{\bigl(\bigl(0,\breve Z{}^{\unl\b}\bigr),\xcE\breve Z{}^{\unl\b}\bigr)\equiv\xcE\breve Z{}^{(2)\,\unl\g}_{(2)}}\oplus\corr{\bigl((0,0),\breve\cZ\bigr)\equiv\xcE\breve\cZ{}^{(2)}}\cr\cr
&&\oplus\corr{\bigl(\bigl(\sfY J_{01},\sfY J_{01}\bigr),\xcE J_{01}\bigr)\equiv\xcE J^{(2)}_{01}}\oplus\bigoplus_{\widehat b<\widehat c=2}^d\,\corr{\bigl(\bigl(\sfY J_{\widehat b\widehat c},\sfY J_{\widehat b\widehat c}\bigr),\xcE J_{\widehat b\widehat c}\bigr)\equiv\xcE J^{(2)}_{\widehat b\widehat c}}\,.
\qqq
and superbrackets induced from the direct-sum ones,\ and form the `tensor-product' Lie superalgebra
\qq\nn
\sfY^{[2]}\jmath_{\rm vac}^*\widetilde\lgt\ox\pr_2^*\egt\equiv\bigl(\sfY^{[2]}\jmath_{\rm vac}^*\widetilde\lgt\hspace{2pt}{}_{\pi_{\sfY^{[2]}\jmath_{\rm vac}^*\widetilde\lgt}}\hspace{-3pt}\oplus_{\pr_1}\pr_2^*\egt\bigr)/_{\sim_\bR}\,,
\qqq
based on the identification
\qq\nn
\bigl(\cZ_{\rm vac},0\bigr)\sim_\bR\bigl(0,\xcE\breve\cZ{}^{(2)}\bigr)\,.
\qqq
The latter is the supervector space
\qq\nn
\sfY^{[2]}\jmath_{\rm vac}^*\widetilde\lgt\ox\pr_2^*\egt&=&\bigoplus_{\unl\a=1}^{\frac{D_{d,1}}{2}}\,\corr{\bigl(\widetilde\xcL_{\rm vac}\breve Q{}_{\unl\a},\xcE\breve Q{}^{(2)}_{\unl\a}\bigr)\equiv\breve Q{}^\ox_{\unl\a}}\oplus\bigoplus_{\unl a\in\{0,1\}}\,\corr{\bigl(\widetilde\xcL_{\rm vac} P_{\unl a},\xcE P^{(2)}_{\unl a}\bigr)\equiv P^\ox_{\unl a}}\cr\cr
&&\oplus\bigoplus_{\unl\b=1}^{\frac{D_{d,1}}{2}}\,\corr{\bigl(\widetilde\xcL_{\rm vac}\breve Z{}_{(1)}^{\unl\b},\xcE\breve Z{}^{(2)\,\unl\b}_{(1)}\bigr)\equiv\breve Z{}_{(1)}^{\ox\,\unl\b}}\oplus\bigoplus_{\unl\g=1}^{\frac{D_{d,1}}{2}}\,\corr{\bigl(\widetilde\xcL_{\rm vac}\breve Z{}_{(2)}^{\unl\g},\xcE\breve Z{}^{(2)\,\unl\g}_{(2)}\bigr)\equiv\breve Z{}_{(2)}^{\ox\,\unl\g}}\cr\cr
&&\oplus\corr{[(\cZ_{\rm vac},0)]_{\sim_\bR}\equiv\cZ^\ox}\oplus\corr{\bigl(\widetilde\xcL_{\rm vac} J_{01},\xcE J^{(2)}_{01}\bigr)\equiv J^\ox_{01}}\oplus\bigoplus_{\widehat b<\widehat c=2}^d\,\corr{\bigl(\widetilde\xcL_{\rm vac} J_{\widehat b\widehat c},\xcE J^{(2)}_{\widehat b\widehat c}\bigr)\equiv J^\ox_{\widehat b\widehat c}}
\qqq
endowed with the superbracket
\qq\nn
&\{\breve Q{}^\ox_{\unl\a},\breve Q{}^\ox_{\unl\b}\}=\ovl\g{}^{\unl a}_{\unl\a\unl\b}\,P^\ox_{\unl a}\,,\qquad\qquad[P^\ox_0,P^\ox_1]=2\cZ^\ox\,,\qquad\qquad[\breve Q{}^\ox_{\unl\a},P^\ox_{\unl a}]=\ovl\g{}_{\unl a\,\unl\a\unl\b}\,\bigl(\breve Z{}_{(1)}^{\ox\,\unl\b}+\breve Z{}_{(2)}^{\ox\,\unl\b}\bigr)\,,&\cr\cr
&\{\breve Q{}^\ox_{\unl\a},\breve Z{}_{(1)}^{\ox\,\unl\b}\}=-\d_{\unl\a}^{\ \unl\b}\,\breve\cZ{}^\ox\,,\qquad\qquad\{\breve Q{}^\ox_{\unl\a},\breve Z{}_{(2)}^{\ox\,\unl\b}\}=0\,,&\cr\cr
&[P^\ox_{\unl a},\breve Z{}_{(m)}^{\ox\,\unl\a}]=0\,,\qquad\qquad\{\breve Z{}_{(m)}^{\ox\,\unl\a},\breve Z{}_{(n)}^{\ox\,\unl\b}\}=0\,,&\cr\cr
&[\breve Q{}^\ox_{\unl\a},\breve\cZ{}^\ox]=0\,,\qquad\qquad[P^\ox_{\unl a},\breve\cZ{}^\ox]=0\,,\qquad\qquad[\breve Z{}_{(m)}^{\ox\,\unl\a},\breve\cZ{}^\ox]=0\,,\qquad\qquad[\breve\cZ{}^\ox,\breve\cZ{}^\ox]=0\,,&\cr\cr
&[J^\ox_{01},\breve Q{}^\ox_{\unl\a}]=\tfrac{1}{2}\,\g_{01}{}^{\unl\b}_{\ \unl\a}\,\breve Q{}^\ox_{\unl\b}\,,\qquad\qquad[J^\ox_{\widehat a\widehat b},\breve Q{}^\ox_{\unl\a}]=\tfrac{1}{2}\,\g_{\widehat a\widehat b}{}^{\unl\b}_{\ \unl\a}\,\breve Q{}^\ox_{\unl\b}\,,&\cr\cr
&[J^\ox_{01},P^\ox_{\unl c}]=\d_{1\unl c}\,P^\ox_0+\d_{0\unl c}\,P^\ox_1\,,\qquad\qquad[J^\ox_{\widehat a\widehat b},P^\ox_{\unl c}]=0\,,&\cr\cr
&[J^\ox_{01},\breve Z{}_{(m)}^{\ox\,\unl\a}]=-\tfrac{1}{2}\,\g_{01}{}^{\unl\a}_{\ \unl\b}\,\breve Z{}_{(m)}^{\ox\,\unl\b}\,,\qquad\qquad[J^\ox_{\widehat a\widehat b},\breve Z{}_{(m)}^{\ox\,\unl\a}]=-\tfrac{1}{2}\,\g_{\widehat a\widehat b}{}^{\unl\a}_{\ \unl\b}\,\breve Z{}_{(m)}^{\ox\,\unl\b}\,,&\cr\cr
&[J^\ox_{01},\breve\cZ{}^\ox]=0\,,\qquad\qquad[J^\ox_{\widehat a\widehat b},\breve\cZ{}^\ox]=0\,,&\cr\cr
&[J^\ox_{01},J^\ox_{\widehat a\widehat b}]=0\,,\qquad\qquad[J^\ox_{\widehat a\widehat b},J^\ox_{\widehat c\widehat d}]=\d_{\widehat a\widehat d}\,J^\ox_{\widehat b\widehat c}-\d_{\widehat a\widehat c}\,J^\ox_{\widehat b\widehat d}+\d_{\widehat b\widehat c}\,J^\ox_{\widehat a\widehat d}-\d_{\widehat b\widehat d}\,J^\ox_{\widehat a\widehat c}\,.&
\qqq
Comparison of the above with the superbracket of $\,\pr_1^*\egt\,$ reveals the existence of the sought-after Lie-superalgebra isomorphism
\qq\nn
\a_{\widetilde\egt}\equiv\bd1\ :\ \sfY^{[2]}\jmath_{\rm vac}^*\widetilde\lgt\ox\pr_2^*\egt\xrightarrow{\ \cong\ }\pr_1^*\egt
\qqq
given by the unique linear extension of the assignment
\qq\nn
\bigl(\breve Q{}^\ox_{\unl\a},P^\ox_{\unl a},\breve Z{}_{(1)}^{\ox\,\unl\b},\breve Z{}_{(2)}^{\ox\,\unl\g},\cZ^\ox,J^\ox_{01},J^\ox_{\widehat b\widehat c}\bigr)\longmapsto\bigl(\xcE\breve Q{}^{(1)}_{\unl\a},\xcE P^{(1)}_{\unl a},\xcE\breve Z{}^{(1)\,\unl\b}_{(1)},\xcE\breve Z{}^{(1)\,\unl\g}_{(2)},\xcE\breve\cZ{}^{(1)},\xcE J^{(1)}_{01},\xcE J^{(1)}_{\widehat b\widehat c}\bigr)\,.
\qqq
We summarise our findings in

\bethe\label{thm:salgskel} 
The null trivialisation $\,\t\,$ of the descended super-1-gerbe of the GS super-$\si$-model in the HP formulation stated in Theorem 2.\ admits a ${\bf sLieAlg}$-skeleton.
\ethe

The ${\bf sLieAlg}$-skeleton determines a formal setting in which we may quite naturally address the question of existence of a ${\bf sLieGrp}${\bf -model} of the vacuum,\ by which we mean a diagram 
\qq\label{diag:sLieGrpmod}\hspace{2cm}
\alxydim{@C=1cm@R=1cm}{ & \bd1 \ar[d] & & \\ \a_\txE\equiv\bd1\ :\ \sfY^{[2]}J_{\rm vac}^*\widetilde\xcL\ox\pr_2^*\txE\xrightarrow{\ \cong\ }\pr_1^*\txE \ar[d] & \bC^\x \ar[r] & \txE,\underset{\tx{\ciut{(1)}}}{\breve\cA}{}_\txE \ar[d]^{\pi_\txE} \\ \sfY^{[2]}{\rm sISO}(d,1|D_{d,1})_{\rm vac} \ar@<-.75ex>[rr]_{\pr_2} \ar@<.75ex>[rr]^{\pr_1} & & \sfY{\rm sISO}(d,1|D_{d,1})_{\rm vac},\underset{\tx{\ciut{(2)}}}{\breve\b} \ar[d]^{\pi_{\sfY{\rm sISO}(d,1|D_{d,1})_{\rm vac}}} \ar[r] & \bd1 \\ & & {\rm sISO}(d,1|D_{d,1})_{\rm vac},\underset{\tx{\ciut{(3)}}}{\breve\chi}\equiv 0 & }
\qqq
in $\,{\bf sLieGrp}\,$ (decorated by constitutive CaE data) that integrates Diag.\,\eqref{diag:sLieAlgshad} in a self-explanatory manner,\ elaborated below.\ This is the anticipated Lie-supergroup structure that we were unable to associate \emph{directly} with the vacuum foliation in Diag.\,\eqref{diag:trivac} -- it seems fitting to call it the {\bf extended $\k$-symmetry group of the superstring in} $\,{\rm sMink}(d,1|D_{d,1})$.\ Here,\ we start by passing to the (sub-)supermanifold description of the Lie sub-supergroup $\,{\rm sISO}(d,1|D_{d,1})_{\rm vac}\subset{\rm sISO}(d,1|D_{d,1})\,$ as the locus of the coordinate equations
\qq\nn
&\bigl(\bd1-\sfP^{(1)}\bigr)^\a_{\ \b}\,\theta^\b=0\,,\qquad\a\in\ovl{1,D_{d,1}}\,,\qquad\qquad x^{\widehat a}=0\,,\qquad\widehat a\in\ovl{2,d}\,,&\cr\cr
&\phi^{\unl b\widehat c}=0\,,\qquad(\unl b,\widehat c)\in\{0,1\}\x\ovl{2,d}\,.&
\qqq
Using an adapted basis in $\,\gt{siso}(d,1|D_{d,1})\,$ obtained through completion of the one for $\,\gt{vac}(\gt{sB}^{{\rm (HP)}}_{1,2})\,$ given in \Reqref{eq:vacbas},\ we thus obtain an embedding
\qq\nn
J_{\rm vac}\ :\ {\rm sISO}(d,1|D_{d,1})_{\rm vac}\equiv{\rm sMink}(d,1|D_{d,1})_{\rm vac}\rx_{L,S}{\rm Spin}(d,1)_{\rm vac}\emb{\rm sISO}(d,1|D_{d,1})\,,
\qqq
with $\,{\rm sMink}(d,1|D_{d,1})_{\rm vac}\subset{\rm sMink}(d,1|D_{d,1})\,$ defined by the top-line equations above.\ The embedding admits the explicit coordinate description
\qq\nn
J_{\rm vac}\bigl(\breve\theta{}^{\unl\a},\breve x{}^{\unl a},\breve\phi{}^{\unl S}\bigr)=\bigl(\breve\theta{}^{\unl\a},0,\breve x{}^{\unl a},0,\breve\phi{}^{\unl S},0\bigr)\,,
\qqq
where the zeros correspond to the nullified coordinates on $\,{\rm ker}\,\sfP^{(1)}$,\ the $\,x^{\widehat a}\,$ and the $\phi^{\unl b\widehat c}$,\ respectively.\ The binary operation $\,\txm_{\rm vac}\,$ on the embedded Lie supergroup is inherited,\ through restriction,\ from that on $\,{\rm sISO}(d,1|D_{d,1})$,
\qq\nn
\txm_{\rm vac}\equiv J_{\rm vac}^{-1}\circ\txm\circ\bigl(J_{\rm vac}\x J_{\rm vac}\bigr)\ :\ {\rm sISO}(d,1|D_{d,1})_{\rm vac}\x{\rm sISO}(d,1|D_{d,1})_{\rm vac}\too{\rm sISO}(d,1|D_{d,1})_{\rm vac}\,,
\qqq
and reads,\ in coordinates,
\qq\nn
\txm_{\rm vac}\bigl(\bigl(\breve\theta{}^{\unl\a}_1,\breve x{}^{\unl a}_1,\breve\phi{}^{\unl S}_1\bigr),\bigl(\breve\theta{}^{\unl\a}_2,\breve x{}^{\unl a}_2,\breve\phi{}^{\unl S}_2\bigr)\bigr)=\bigl(\breve\theta{}^{\unl\a}_1+\breve S\bigl(\breve\phi_1\bigr)^{\unl\a}_{\ \unl\b}\,\breve\theta{}^{\unl\b}_2,\breve x{}^{\unl a}_1+\breve L\bigl(\breve\phi_1\bigr)^{\unl a}_{\ \unl b}\,\breve x{}^{\unl b}_2-\tfrac{1}{2}\,\breve\theta{}_1\,\ovl\g{}^{\unl a}\,\breve S\bigl(\breve\phi_1\bigr)\,\breve\theta{}_2,\bigl(\breve\phi{}_1\star\breve\phi{}_2\bigr)^{\unl S}\bigr)\,,
\qqq
where we have used the notation $\,\breve S(\breve\phi_1)\,$ and $\,\breve L(\breve\phi_1)\,$ for the `vacuum' blocks of the block-diagonal (in the adapted basis) matrices $\,S(\breve\phi_1,0)\,$ and $\,L(\breve\phi_1,0)$,\ respectively.\ In the next step,\ we take the sub-supermanifold
\qq\nn
\sfY{\rm sMink}(d,1|D_{d,1})_{\rm vac}\equiv{\rm sMink}(d,1|D_{d,1})_{\rm vac}\x\bR^{0|\frac{D_{d,1}}{2}}
\qqq
of $\,\sfY{\rm sMink}(d,1|D_{d,1})\,$ with the second cartesian factor defined by the coordinate equations
\qq\nn
\xi_\b\,\bigl(\bd1-\sfP^{(1)}\bigr)^\b_{\ \a}=0\,,\qquad\a\in\ovl{1,D_{d,1}}\,,
\qqq
and use it to write the desired embedding
\qq\nn
\sfY J_{\rm vac}\ :\ \sfY{\rm sISO}(d,1|D_{d,1})_{\rm vac}\equiv\sfY{\rm sMink}(d,1|D_{d,1})_{\rm vac}\rx_{L,S,S^{-{\rm T}}}{\rm Spin}(d,1)_{\rm vac}\emb\sfY{\rm sISO}(d,1|D_{d,1})
\qqq
in adapted coordinates ({\it cp} \Reqref{eq:Yvacbas}) as
\qq\nn
\sfY J_{\rm vac}\bigl(\breve\theta{}^{\unl\a},\breve x{}^{\unl a},\breve\xi{}_{\unl\b},\breve\phi{}^{\unl S}\bigr)=\bigl(\breve\theta{}^{\unl\a},0,\breve x{}^{\unl a},0,\breve\xi{}_{\unl\b},0,\breve\phi{}^{\unl S},0\bigr)\,.
\qqq
The sub-supermanifold submerses surjectively onto $\,{\rm sISO}(d,1|D_{d,1})_{\rm vac}$,
\qq\nn
\pi_{\sfY{\rm sISO}(d,1|D_{d,1})_{\rm vac}}\ :\ \sfY{\rm sISO}(d,1|D_{d,1})_{\rm vac}\too{\rm sISO}(d,1|D_{d,1})_{\rm vac}\,,
\qqq
as
\qq\nn
\pi_{\sfY{\rm sISO}(d,1|D_{d,1})_{\rm vac}}\bigl(\breve\theta{}^{\unl\a},\breve x{}^{\unl a},\breve\xi{}_{\unl\b},\breve\phi{}^{\unl S}\bigr)=\bigl(\breve\theta{}^{\unl\a},\breve x{}^{\unl a},\breve\phi{}^{\unl S}\bigr)\,.
\qqq
The inherited binary operation 
\qq\nn
\sfY\txm_{\rm vac}\equiv\sfY J_{\rm vac}^{-1}\circ\sfY\txm\circ\bigl(\sfY J_{\rm vac}\x\sfY J_{\rm vac}\bigr)\ :\ \sfY{\rm sISO}(d,1|D_{d,1})_{\rm vac}\x\sfY{\rm sISO}(d,1|D_{d,1})_{\rm vac}\too\sfY{\rm sISO}(d,1|D_{d,1})_{\rm vac}
\qqq
uses the same objects $\,\breve S\,$ and $\,\breve L\,$ as $\,\txm_{\rm vac}$,
\qq
\sfY\txm_{\rm vac}\bigl(\bigl(\breve\theta{}^{\unl\a}_1,\breve x{}^{\unl a}_1,\breve\xi{}_{1\,\unl\b},\breve\phi{}^{\unl S}_1\bigr),\bigl(\breve\theta{}^{\unl\a}_2,\breve x{}^{\unl a}_2,\breve\xi{}_{2\,\unl\b},\breve\phi{}^{\unl S}_2\bigr)\bigr)&=&\bigl(\breve\theta{}^{\unl\a}_1+\breve S\bigl(\breve\phi_1\bigr)^{\unl\a}_{\ \unl\b}\,\breve\theta{}^{\unl\b}_2,\breve x{}^{\unl a}_1+\breve L\bigl(\breve\phi_1\bigr)^{\unl a}_{\ \unl b}\,\breve x{}^{\unl b}_2-\tfrac{1}{2}\,\breve\theta{}_1\,\ovl\g{}^{\unl a}\,\breve S\bigl(\breve\phi_1\bigr)\,\breve\theta{}_2,\cr\cr
&&\breve\xi{}_{1\,\unl\b}+\breve\xi{}_{2\,\unl\g}\,\breve S\bigl(\breve\phi_1\bigr)^{-1\,\unl\g}_{\ \ \ \ \unl\b}+\ovl\g{}_{\unl d\,\unl\b\unl\g}\,\breve\theta{}_1^{\unl\g}\,\breve L\bigl(\unl\phi_1\bigr)^{\unl d}_{\ \unl e}\,\breve x{}_2^{\unl e},\bigl(\breve\phi{}_1\star\breve\phi{}_2\bigr)^{\unl S}\bigr)\,,
\qqq
and gives rise,\ as usual,\ to the left regular action 
\qq\nn
\sfY\breve\ell\equiv\sfY\txm_{\rm vac}\,.
\qqq
of the Lie supergroup $\,\sfY{\rm sISO}(d,1|D_{d,1})_{\rm vac}\,$ on itself.\ In the coordinate presentation of $\,\sfY\txm_{\rm vac}$,\ we have taken into account the identity
\qq\nn
\ovl\g{}_{\unl a}\ox\ovl\g{}^{\unl a}=0
\qqq
and dropped the term in the coordinate expression for $\,\sfY\txm\,$ trilinear in the Gra\ss mann-odd coordinates accordingly. 

On the Lie supergroup $\,\sfY{\rm sISO}(d,1|D_{d,1})_{\rm vac}$,\ we find the standard $\sfY\gt{vac}(\gt{sB}^{{\rm (HP)}}_{1,2})$-valued LI Maurer--Cartan super-1-form
\qq\nn
\breve\theta{}_{\rm L}=\pi_{\sfY{\rm sISO}(d,1|D_{d,1})_{\rm vac}}^*\breve q{}^{\unl\a}\ox\sfY\breve Q{}_{\unl\a}+\pi_{\sfY{\rm sISO}(d,1|D_{d,1})_{\rm vac}}^*\breve p{}^{\unl a}\ox\sfY P_{\unl a}+\breve z{}_{\unl\b}\ox\breve Z{}^{\unl\b}+\pi_{\sfY{\rm sISO}(d,1|D_{d,1})_{\rm vac}}^*\breve j^{\unl S}\ox\sfY J_{\unl S}
\qqq
whose components along $\,\sfY\gt{smink}(d,1|D_{d,1})$,\ with the coordinate presentations (derived in a procedure analogous to the one leading to the formul\ae ~for their counterparts on $\,\sfY{\rm sISO}(d,1|D_{d,1})$)
\qq\nn
\breve q{}^{\unl\a}\bigl(\breve\theta,\breve x,\breve\phi\bigr)&=&\breve S\bigl(\breve\phi\bigr)^{-1\,\unl\a}_{\ \ \ \unl\b}\,\sfd\breve\theta{}^{\unl\b}=:\breve S\bigl(\breve\phi\bigr)^{-1\,\unl\a}_{\ \ \ \unl\b}\,\breve{\unl q}{}^{\unl\a}\bigl(\breve\theta,\breve x\bigr)\,,\cr\cr
\breve p{}^{\unl a}\bigl(\breve\theta,\breve x,\breve\phi\bigr)&=&\breve L\bigl(\breve\phi\bigr)^{-1\,\unl a}_{\ \ \ \unl b}\,\bigl(\sfd\breve x{}^{\unl b}+\tfrac{1}{2}\,\breve\theta\,\ovl\g{}^{\unl b}\,\sfd\breve\theta\bigr)=:\breve L\bigl(\breve\phi\bigr)^{-1\,\unl a}_{\ \ \ \unl b}\,\breve{\unl p}{}^{\unl b}\bigl(\breve\theta,\breve x\bigr)\,,\cr\cr
\breve z{}_{\unl\b}\bigl(\breve\theta,\breve x,\breve\xi,\breve\phi\bigr)&=&\bigl(\sfd\breve\xi{}_{\unl\b}-\ovl\g{}_{\unl a\,\unl\b\unl\g}\,\breve\theta{}^{\unl\g}\,\sfd\breve x^{\unl a}\bigr)\,\breve S\bigl(\breve\phi\bigr)^{\unl\b}_{\ \unl\a}=:\breve{\unl z}{}_{\unl\b}\bigl(\breve\theta,\breve x,\breve\xi\bigr)\,\breve S\bigl(\breve\phi\bigr)^{\unl\b}_{\ \unl\a}\,,
\qqq
enter the definition of the LI super-2-cocycle
\qq\nn
\underset{\tx{\ciut{(2)}}}{\breve\b}&=&\pi_{\sfY{\rm sISO}(d,1|D_{d,1})_{\rm vac}}^*\breve q{}^{\unl\a}\wedge\breve z{}_{\unl\a}+2\pi_{\sfY{\rm sISO}(d,1|D_{d,1})_{\rm vac}}^*\bigl(\breve p{}^0\wedge\breve p{}^1\bigr)\cr\cr
&=&\pi_{\sfY{\rm sISO}(d,1|D_{d,1})_{\rm vac}}^*\pr_2^*\breve{\unl q}{}^{\unl\a}\wedge\pr_2^*\breve{\unl z}{}_{\unl\a}+2\pi_{\sfY{\rm sISO}(d,1|D_{d,1})_{\rm vac}}^*\pr_2^*\bigl(\breve{\unl p}{}^0\wedge\breve{\unl p}{}^1\bigr)\,,
\qqq
whose final (descended) form follows from the unimodularity of $\,\rho_{{\rm Spin}(d,1)_{\rm vac}}\rstr_{\tgt^{(0)}_{\rm vac}}$,
\qq\nn
\bigl(\breve p{}^0\wedge\breve p{}^1\bigr)\bigl(\breve\theta,\breve x,\breve\phi\bigr)=\bigl(\breve{\unl p}{}^0\wedge\breve{\unl p}{}^1\bigr)\bigl(\breve\theta,\breve x\bigr)\,.
\qqq
Following the logic of the geometrisation programme,\ we seek to associate with the latter a central extension
\qq\label{eq:Eshes}\hspace{1cm}
\bd1\too\bC^\x\too\txE\equiv\sfY{\rm sISO}(d,1|D_{d,1})_{\rm vac}\x\bC^\x\xrightarrow{\ \pi_\txE\equiv\pr_1\ }\sfY{\rm sISO}(d,1|D_{d,1})_{\rm vac}\too\bd1
\qqq
integrating the formerly obtained Lie-superalgebra extension \eqref{eq:egtext}.\ To this end,\ we compute the non-LI primitive of $\,\underset{\tx{\ciut{(2)}}}{\breve\b}$,
\qq\nn
\underset{\tx{\ciut{(2)}}}{\breve\b}\bigl(\breve\theta,\breve x,\breve\xi,\breve\phi\bigr)=\sfd\bigl(\breve\theta{}^{\unl\a}\,\sfd\breve\xi{}_{\unl\a}+\ep_{\unl a\unl b}\,\breve x{}^{\unl a}\,\sfd\breve x{}^{\unl b}\bigr)\,,
\qqq
and define the principal $\bC^\x$-connection super-1-form 
\qq\nn
\underset{\tx{\ciut{(1)}}}{\breve\cA}{}_\txE\in\Om^1(\txE)
\qqq
on the total space $\,\txE\ni(\breve\theta{}^{\unl\a},\breve x{}^{\unl a},\breve\xi{}_{\unl\b},\breve\phi{}^{\unl S},\breve z)\,$ of the principal $\bC^\x$-bundle
\qq\nn
\pi_\txE\ :\ \txE\too\sfY{\rm sISO}(d,1|D_{d,1})_{\rm vac}
\qqq
explicitly as
\qq\nn
\underset{\tx{\ciut{(1)}}}{\breve\cA}{}_\txE\bigl(\breve\theta,\breve x,\breve\xi,\breve\phi,\breve z\bigr)=\tfrac{\sfi\,\sfd\breve z}{\breve z}-\breve\theta{}^{\unl\a}\,\sfd\breve\xi{}_{\unl\a}-\ep_{\unl a\unl b}\,\breve x{}^{\unl a}\,\sfd\breve x{}^{\unl b}=:\tfrac{\sfi\,\sfd\breve z}{\breve z}+\breve\txA\bigl(\breve\theta,\breve x,\breve\xi,\breve\phi\bigr)\,,
\qqq
determining the Lie-supergroup structure on $\,\txE\,$ through imposition of the usual demand that the primitive $\,-\underset{\tx{\ciut{(1)}}}{\breve\cA}{}_\txE\,$ of the pullback $\,\pi_\txE^*\underset{\tx{\ciut{(2)}}}{\breve\b}\,$ of the super-2-cocycle $\,\underset{\tx{\ciut{(2)}}}{\breve\b}\,$ be LI.\ From the direct computation (carried out for $\,(\breve\vep{}^{\unl\a},\breve y{}^{\unl a},\breve\z{}_{\unl\b},\breve\psi{}^{\unl S})\in\sfY{\rm sISO}(d,1|D_{d,1})_{\rm vac}$) of
\qq\nn
\sfY\breve\ell_{(\breve\vep,\breve y,\breve\z,\breve\psi)}^*\breve\txA\bigl(\breve\theta,\breve x,\breve\xi,\breve\phi\bigr)&=&\breve\txA\bigl(\breve\theta,\breve x,\breve\xi,\breve\phi\bigr)+\sfd\bigl(\breve\xi\,\breve S\bigl(\breve\psi\bigr)^{-1}\,\breve\vep-\ep_{\unl a\unl b}\,\breve y{}^{\unl a}\,\bigl(\breve L\bigl(\breve\psi\bigr)^{\unl b}_{\ \unl c}\,\breve x{}^{\unl c}-\tfrac{1}{2}\,\breve\vep\,\ovl\g{}^{\unl b}\,\breve S\bigl(\breve\psi\bigr)\,\breve\theta\bigr)\cr\cr
&&-\tfrac{1}{2}\,\ep_{\unl a\unl b}\,\breve\vep\,\ovl\g{}^{\unl a}\,\breve S\bigl(\breve\psi\bigr)\,\breve\theta\,\breve L\bigl(\breve\psi\bigr)^{\unl b}_{\ \unl c}\,\breve x{}^{\unl c}\bigr)\,,
\qqq
using the identities
\qq\nn
\ep_{\unl a\unl b}\,\ovl\g{}^{\unl a}\ox\ovl\g{}^{\unl b}=0\,,\qquad\qquad\ep_{\unl a\unl b}\,\ovl\g{}^{\unl a}=-\ovl\g{}_{\unl b}\,,
\qqq
and hence leading to
\qq\nn
&&\bigl(\sfY\breve\ell_{(\breve\vep,\breve y,\breve\z,\breve\psi)}^*-\id_{\sfY{\rm sISO}(d,1|D_{d,1})_{\rm vac}}^*\bigr)\breve\txA\bigl(\breve\theta,\breve x,\breve\xi,\breve\phi\bigr)\cr\cr
&=&\sfd\bigl(\breve\xi\,\breve S\bigl(\breve\psi\bigr)^{-1}\,\breve\vep-\ep_{\unl a\unl b}\,\breve y{}^{\unl a}\,\breve L\bigl(\breve\psi\bigr)^{\unl b}_{\ \unl c}\,\breve x{}^{\unl c}+\tfrac{1}{2}\,\bigl(\breve y{}^{\unl a}+\breve L\bigl(\breve\psi\bigr)^{\unl a}_{\ \unl b}\,\breve x{}^{\unl b}\bigr)\,\breve\vep{}\,\ovl\g{}_{\unl a}\,\breve S\bigl(\breve\psi\bigr)\,\breve\theta\bigr)\bigr)\,,
\qqq
we read off the candidate for the binary operation:
\qq\nn
\txE\txm_{\rm vac}\ :\ \txE\x\txE\too\txE
\qqq
in the coordinate form
\qq\nn
\txE\txm_{\rm vac}\bigl(\bigl(\breve\theta{}^{\unl\a}_1,\breve x{}^{\unl a}_1,\breve\xi{}_{1\,\unl\b},\breve\phi{}^{\unl S}_1,\breve z{}_1\bigr),\bigl(\breve\theta{}^{\unl\a}_2,\breve x{}^{\unl a}_2,\breve\xi{}_{2\,\unl\b},\breve\phi{}^{\unl S}_2,\breve z{}_2\bigr)\bigr)=\bigl(\sfY\txm_{\rm vac}\bigl(\bigl(\breve\theta{}^{\unl\a}_1,\breve x{}^{\unl a}_1,\breve\xi{}_{1\,\unl\b},\breve\phi{}^{\unl S}_1\bigr),\bigl(\breve\theta{}^{\unl\a}_2,\breve x{}^{\unl a}_2,\breve\xi{}_{2\,\unl\b},\breve\phi{}^{\unl S}_2\bigr)\bigr),\cr\cr
\ee^{\sfi\,(\breve\xi{}_2\,\breve S(\breve\phi{}_1)^{-1}\,\breve\theta{}_1-\ep_{\unl a\unl b}\,\breve x{}_1^{\unl a}\,\breve L(\breve\phi{}_1)^{\unl b}_{\ \unl c}\,\breve x{}_2^{\unl c}+\frac{1}{2}\,(\breve x{}_1^{\unl a}+\breve L(\breve\phi{}_1)^{\unl a}_{\ \unl b}\,\breve x{}_2^{\unl b})\,\breve\theta{}_1\,\ovl\g{}_{\unl a}\,\breve S(\breve\phi{}_1)\,\breve\theta{}_2)}\cdot\breve z{}_1\cdot\breve z{}_2\bigr)\,.
\qqq 
Through inspection,\ we readily prove
\berop
The supermanifold $\,\txE$,\ together with the supermanifold morphism $\,\txE\txm_{\rm vac}\,$ defined above  as the binary operation and the pair of supermanifold morphisms
\qq\nn
\txE{\rm Inv}_{\rm vac}\ :\ \txE\too\txE\,,\qquad\qquad\txE\vep_{\rm vac}\ :\ \bR^{0|0}\too\txE
\qqq
with the coordinate presentations
\qq\nn
\txE{\rm Inv}_{\rm vac}\bigl(\breve\theta{}^{\unl\a},\breve x{}^{\unl a},\breve\xi{}_{\unl\b},\breve\phi{}^{\unl S},\breve z\bigr)&=&\bigl(-\breve S\bigl(\breve\phi\bigr)^{-1\,\unl\a}_{\ \ \ \unl\g}\,\breve\theta{}^{\unl\g},-\breve L\bigl(\breve\phi\bigr)^{-1\,\unl a}_{\ \ \ \unl b}\,\breve x{}^{\unl b},\bigl(-\breve\xi{}_{\unl\g}+\breve x{}^{\unl a}\,\ovl\g{}_{\unl a\,\unl\g\unl\d}\,\breve\theta{}^{\unl\d}\bigr)\,\,\breve S\bigl(\breve\phi\bigr)^{\unl\g}_{\ \unl\b},-\breve\phi{}^{\unl S},\ee^{\sfi\,\breve\xi\,\breve\theta}\cdot\breve z{}^{-1}\bigr)\,,\cr\cr
\txE\vep_{\rm vac}(\bullet)=(0,0,0,0,1)
\qqq
as the inverse and the unit,\ respectively,\ is a Lie supergroup that centrally extends $\,\sfY{\rm sISO}(d,1|D_{d,1})_{\rm vac}\,$ as in \Reqref{eq:Eshes}.
\eerop

At this stage,\ it remains to establish the existence of the \emph{trivial} Lie-supergroup isomorphism 
\qq\nn
\a_\txE\equiv\bd1\ :\ \sfY^{[2]}J_{\rm vac}^*\widetilde\xcL\ox\pr_2^*\txE\xrightarrow{\ \cong\ }\pr_1^*\txE
\qqq
over the fibred square 
\qq\nn
\sfY^{[2]}{\rm sISO}(d,1|D_{d,1})_{\rm vac}\equiv\sfY{\rm sISO}(d,1|D_{d,1})_{\rm vac}\x_{{\rm sISO}(d,1|D_{d,1})_{\rm vac}}\sfY{\rm sISO}(d,1|D_{d,1})_{\rm vac}
\qqq
of the surjective submersion $\,\sfY{\rm sISO}(d,1|D_{d,1})_{\rm vac}\,$ (endowed with the Lie-supergroup structure induced from the product one on $\,\sfY{\rm sISO}(d,1|D_{d,1})_{\rm vac}\x\sfY{\rm sISO}(d,1|D_{d,1})_{\rm vac}\,$ through restriction) and check that it is simultaneously a connection-preserving principal $\bC^\x$-bundle isomorphism.\  Here,\ $\,\sfY^{[2]}J_{\rm vac}^*\widetilde\xcL\,$ is the Lie sub-supergroup of the product Lie supergroup $\,\sfY^{[2]}{\rm sISO}(d,1|D_{d,1})_{\rm vac}\x\widetilde\xcL\,$ whose support is the principal $\bC^\x$-bundle
\qq\nn
\pi_{\sfY^{[2]}J_{\rm vac}^*\widetilde\xcL}\equiv\pr_1\ :\ \sfY^{[2]}J_{\rm vac}^*\widetilde\xcL\equiv\sfY^{[2]}{\rm sISO}(d,1|D_{d,1})_{\rm vac}\hspace{2pt}{}_{\sfY^{[2]}J_{\rm vac}}\hspace{-3pt}\x_{\pi_{\widetilde\xcL}}\widetilde\xcL \too\sfY^{[2]}{\rm sISO}(d,1|D_{d,1})_{\rm vac}
\qqq
with coordinates
\qq\nn
\bigl(\bigl(\bigl(\breve\theta{}^{\unl\a},\breve x{}^{\unl a},\breve\xi{}_{1\,\unl\b},\breve\phi{}^{\unl S}\bigr),\bigl(\breve\theta{}^{\unl\a},\breve x{}^{\unl a},\breve\xi{}_{2\,\unl\b},\breve\phi{}^{\unl S}\bigr)\bigr),\bigl(\bigl(\breve\theta{}^{\unl\a},0,\breve x{}^{\unl a},0,\breve\xi{}_{1\,\unl\b},0,\breve\phi{}^{\unl S},0\bigr),\bigl(\breve\theta{}^{\unl\a},0,\breve x{}^{\unl a},0,\breve\xi{}_{2\,\unl\b},0,\breve\phi{}^{\unl S},0\bigr),z\bigr)\bigr)\in\sfY^{[2]}J_{\rm vac}^*\widetilde\xcL\,,
\qqq
and the tensor product in $\,\sfY^{[2]}J_{\rm vac}^*\widetilde\xcL\ox\pr_2^*\txE\,$ is defined analogously to the one on p.\,\pageref{pref:tensLie}.\ Comparing the base components of the principal $\bC^\x$-connection super-1-forms of the two principal $\bC^\x$-bundles to be related by $\,\a_\txE\equiv\bd1$,
\qq\nn
\sfY^{[2]}\pi^*\txa\bigl(\bigl(\breve\theta{}^{\unl\a},0,\breve x{}^{\unl a},0,\breve\xi{}_{1\,\unl\b},0,\breve\phi{}^{\unl S},0\bigr),\bigl(\breve\theta{}^{\unl\a},0,\breve x{}^{\unl a},0,\breve\xi{}_{2\,\unl\b},0,\breve\phi{}^{\unl S},0\bigr)\bigr)+\breve\txA\bigl(\breve\theta{}^{\unl\a},\breve x{}^{\unl a},\breve\xi{}_{2\,\unl\b},\breve\phi{}^{\unl S}\bigr)-\breve\txA\bigl(\breve\theta{}^{\unl\a},\breve x{}^{\unl a},\breve\xi{}_{1\,\unl\b},\breve\phi{}^{\unl S}\bigr)=0\,,
\qqq
we conclude that the bundles are related by the connection-preserving isomorphism indicated.\ We merely need to check if the latter is a Lie-supergroup homomorphism.\ That this is,\ indeed,\ the case follows from the equality
\qq\nn
&&\ee^{\sfi\,\breve\theta{}_1^{\unl\a}\,(\breve\xi{}_{2,2\,\unl\b}-\breve\xi{}_{2,1\,\unl\b})\,\breve S(\breve\phi{}_1)^{-1\,\unl\b}_{\ \ \ \unl\a}}\cdot\ee^{\sfi\,(\breve\xi{}_{2,2}\,\breve S(\breve\phi{}_1)^{-1}\,\breve\theta{}_1-\ep_{\unl a\unl b}\,\breve x{}_1^{\unl a}\,\breve L(\breve\phi{}_1)^{\unl b}_{\ \unl c}\,\breve x{}_2^{\unl c}+\frac{1}{2}\,(\breve x{}_1^{\unl a}+\breve L(\breve\phi{}_1)^{\unl a}_{\ \unl c}\,\breve x{}_2^{\unl c})\,\breve\theta{}_1\,\ovl\g{}_{\unl a}\,\breve S(\breve\phi{}_1)\,\breve\theta{}_2)}\cr\cr
&=&\ee^{\sfi\,(\breve\xi{}_{2,1}\,\breve S(\breve\phi{}_1)^{-1}\,\breve\theta{}_1-\ep_{\unl a\unl b}\,\breve x{}_1^{\unl a}\,\breve L(\breve\phi{}_1)^{\unl b}_{\ \unl c}\,\breve x{}_2^{\unl c}+\frac{1}{2}\,(\breve x{}_1^{\unl a}+\breve L(\breve\phi{}_1)^{\unl a}_{\ \unl c}\,\breve x{}_2^{\unl c})\,\breve\theta{}_1\,\ovl\g{}_{\unl a}\,\breve S(\breve\phi{}_1)\,\breve\theta{}_2)}
\qqq
of the `phase' factors in -- on the one hand (the left-hand side of the equality sign) -- the $\a_\txE$-image of the product of the
\qq\nn
\bigl(\bigl(\bigl(\breve\theta{}_n^{\unl\a},\breve x{}_n^{\unl a},\breve\xi{}_{n,1\,\unl\b},\breve\phi{}_n^{\unl S}\bigr),\bigl(\breve\theta{}_n^{\unl\a},\breve x{}_n^{\unl a},\breve\xi{}_{n,2\,\unl\b},\breve\phi{}_n^{\unl S}\bigr)\bigr),\bigl(\bigl(\breve\theta{}_n^{\unl\a},0,\breve x{}_n^{\unl a},0,\breve\xi{}_{n,1\,\unl\b},0,\breve\phi{}_n^{\unl S},0\bigr),\bigl(\breve\theta{}_n^{\unl\a},0,\breve x{}_n^{\unl a},0,\breve\xi{}_{n,2\,\unl\b},0,\breve\phi{}_n^{\unl S},0\bigr),1\bigr)\bigr)\cr\cr
\ox\bigl(\bigl(\bigl(\breve\theta{}_n^{\unl\a},\breve x{}_n^{\unl a},\breve\xi{}_{n,1\,\unl\b},\breve\phi{}_n^{\unl S}\bigr),\bigl(\breve\theta{}_n^{\unl\a},\breve x{}_n^{\unl a},\breve\xi{}_{n,2\,\unl\b},\breve\phi{}_n^{\unl S}\bigr)\bigr),\bigl(\breve\theta{}_n^{\unl\a},\breve x{}_n^{\unl a},\breve\xi{}_{n,2\,\unl\b},\breve\phi{}_n^{\unl S},\breve z{}_n\bigr)\bigr)\in\sfY^{[2]}J_{\rm vac}^*\widetilde\xcL\ox\pr_2^*\txE
\qqq
with $\,n\in\{1,2\}$,\ and -- on the other hand (the right-hand side of the equality sign) -- the product of their $\a_\txE$-images
\qq\nn
\bigl(\bigl(\bigl(\breve\theta{}_n^{\unl\a},\breve x{}_n^{\unl a},\breve\xi{}_{n,1\,\unl\b},\breve\phi{}_n^{\unl S}\bigr),\bigl(\breve\theta{}_n^{\unl\a},\breve x{}_n^{\unl a},\breve\xi{}_{n,2\,\unl\b},\breve\phi{}_n^{\unl S}\bigr)\bigr),\bigl(\breve\theta{}_n^{\unl\a},\breve x{}_n^{\unl a},\breve\xi{}_{n,1\,\unl\b},\breve\phi{}_n^{\unl S},\breve z{}_n\bigr)\bigr)\in\pr_1^*\txE\,.
\qqq
Thus,\ altogether,\ we arrive at

\bethe\label{thm:sgpmod}
The ${\bf sLieAlg}$-skeleton of the null trivialisation $\,\t\,$ of the super-1-gerbe of the GS super-$\si$-model in the HP formulation from theorem 3.\ integrates to a ${\bf sLieGrp}$-model.
\ethe

Theorems \ref{thm:vactriv}-\ref{thm:sgpmod} attest to the veracity of our expectation with regard to (the triviality of) the vacuum restriction of the extended HP super-1-gerbe over $\,{\rm sISO}(d,1|D_{d,1})$.\ We conclude our study by demonstrating how this fact \emph{implies} the existence of a descendable ${\rm sISO}(d,1|D_{d,1})_{\rm vac}$-equivariant structure on the restriction.\ In so doing,\ we localise our analysis over a single leaf $\,\cD_{i,\upsilon_i}\,$ of the vacuum foliation within $\,\Si^{\rm HP}_{\rm vac}$,\ with the understanding that the mechanisms discovered in its course descend to the physical vacuum in $\,\Si^{\rm HP}_{\rm phys\ vac}$.\ Moreover,\ we exclude the `hidden' gauge-symmetry subgroup $\,{\rm Spin}(d,1)_{\rm vac}\subset{\rm sISO}(d,1|D_{d,1})_{\rm vac}\,$ from our discussion as the latter is an artifact of the realisation of the physical supertarget $\,{\rm sISO}(d,1|D_{d,1})/{\rm Spin}(d,1)_{\rm vac}\,$ in the mother Lie supergroup $\,{\rm sISO}(d,1|D_{d,1})$,\ with the corresponding equivariance explicitly built into the construction of the (extended) super-1-gerbe.\ This leaves us with the {\bf visible $\k$-symmetry group} 
\qq\nn
\k_{\rm vis}\equiv{\rm sMink}(d,1|D_{d,1})_{\rm vac}\xrightarrow{\ \jmath_{\rm vis}\ }{\rm sISO}(d,1|D_{d,1})
\qqq 
{\bf of the superstring in} $\,{\rm Mink}(d,1|D_{d,1})\,$ (embedded in an obvious manner in the mother supersymmetry group) as the Lie supergroup for which we are to establish an equivariant structure on $\,\cG{}^{(1)}_{\rm vac}\rstr_{\cD_{i,\upsilon_i}}$.

The point of departure of our considerations is the action (super)groupoid
\qq\nn
\alxydim{@C=4.5cm@R=2cm}{\k^1_{{\rm vis}\,i,\upsilon_i}\equiv\k_{\rm vis}\x\cD_{i,\upsilon_i} \ar@<-.75ex>[r]_{\pr_2} \ar@<.75ex>[r]^{\la_{\rm vis}\equiv\wp\circ(\Inv\circ\jmath_{\rm vis}\x\id_{\cD_{i,\upsilon_i}})} & \cD_{i,\upsilon_i}\equiv\k^0_{{\rm vis}\,i,\upsilon_i} }\,,
\qqq
with the target map $\,\la_{\rm vis}\,$ purposefully turned into a \emph{left} action,\ so that we may directly employ the construction \eqref{eq:actgrpdsimpl} introduced previously.\ Over the arrow supermanifold of this category,\ we set up a pair of surjective submersions
\qq\nn
\pi_{f^*\sfY\cD_{i,\upsilon_i}}\equiv\pr_1\ :\ f^*\sfY\cD_{i,\upsilon_i}\equiv\k^1_{{\rm vis}\,i,\upsilon_i}\hspace{2pt}{}_f\hspace{-2pt}\x_{\pi_{\sfY{\rm sISO}(d,1|D_{d,1})}}\sfY\cD_{i,\upsilon_i}\too\k^1_{{\rm vis}\,i,\upsilon_i}\,,\qquad\qquad f\in\{\la_{\rm vis},\pr_2\}
\qqq
with
\qq\nn
\widehat f\equiv\pr_2\ :\ f^*\sfY\cD_{i,\upsilon_i}\too\sfY\cD_{i,\upsilon_i}\,,
\qqq
and demand the existence of a principal $\bC^\x$-bundle
\qq\nn
\pi_\cF\ :\ \cF\too\la_{\rm vis}^*\sfY\cD_{i,\upsilon_i}\hspace{2pt}{}_{\pr_1}\hspace{-3pt}\x_{\pr_1}\pr_2^*\sfY\cD_{i,\upsilon_i}\equiv\sfY_{\la_{\rm vis}2}\cD_{i,\upsilon_i}
\qqq
over the $\k^1_{{\rm vis}\,i,\upsilon_i}$-fibred product of the two,\ with a principal $\bC^\x$-connection super-1-form
\qq\nn
\underset{\tx{\ciut{(1)}}}{\cA}{}_\cF\in\Om^1(\cF)
\qqq
satisfying the identity
\qq\nn
\sfd\underset{\tx{\ciut{(1)}}}{\cA}{}_\cF=\pi_\cF^*\bigl(\pr_2^*\circ\widehat\pr{}_2^*-\pr_1^*\circ\widehat\la{}_{\rm vis}^*\bigr)\widehat{\underset{\tx{\ciut{(2)}}}{\sfY\b}}\,,
\qqq
and such that there exists,\ over  
\qq\nn
\la_{\rm vis}^*\sfY\cD_{i,\upsilon_i}\hspace{2pt}{}_{\pr_1}\hspace{-3pt}\x_{\pr_1}\pr_2^*\sfY\cD_{i,\upsilon_i} \hspace{2pt}{}_{\pr_1}\hspace{-3pt}\x_{\pr_1}\la_{\rm vis}^*\sfY\cD_{i,\upsilon_i}\hspace{2pt}{}_{\pr_1}\hspace{-3pt}\x_{\pr_1}\pr_2^*\sfY\cD_{i,\upsilon_i}\equiv\sfY_{\la_{\rm vis}2\la_{\rm vis}2}\cD_{i,\upsilon_i}\,,
\qqq
a connection-preserving isomorphism
\qq\nn
\a_\cF\ :\ \pr_{1,3}^*\widehat\la{}^{\x 2\,*}_{\rm vis}\widetilde\xcL\ox\pr_{3,4}^*\cF\xrightarrow{\ \cong\ }\pr_{1,2}^*\cF\ox\pr_{2,4}^*\widehat\pr{}^{\x 2\,*}_2\widetilde\xcL
\qqq
of the principal $\bC^\x$-bundles obtained by tensoring,\ in the manner discussed earlier,\ the pullbacks of the bundles 
\qq\nn
\pi_{\widehat f{}^{\x 2\,*}\widetilde\xcL}\equiv\pr_1\ :\ \widehat f{}^{\x 2\,*}\widetilde\xcL\equiv\sfY_{ff}\cD_{i,\upsilon_i}\hspace{2pt}{}_{{\widehat f{}^{\x 2}}}\hspace{-3pt}\x_{\pi_{\widetilde\xcL}}\widetilde\xcL\too\sfY_{ff}\cD_{i,\upsilon_i}\equiv f^*\sfY\cD_{i,\upsilon_i}\x_{\k^1_{{\rm vis}\,i,\upsilon_i}}f^*\sfY\cD_{i,\upsilon_i}
\qqq
given by
\qq
\pi_{\pr_{1,3}^*\widehat\la{}_{\rm vis}^{\x 2\,*}\widetilde\xcL}\equiv\pr_1\ :\ \pr_{1,3}^*\widehat\la{}_{\rm vis}^{\x 2\,*}\widetilde\xcL\equiv\sfY_{\la_{\rm vis}2\la_{\rm vis}2}\cD_{i,\upsilon_i}\hspace{2pt}{}_{\pr_{1,3}}\hspace{-3pt}\x_{\pr_1}\widehat\la{}_{\rm vis}^{\x 2\,*}\widetilde\xcL\too\sfY_{\la_{\rm vis}2\la_{\rm vis}2}\cD_{i,\upsilon_i}
\qqq
(for $\,f=\la{}_{\rm vis}$) and
\qq\nn
\pi_{\pr_{2,4}^*\widehat\pr_2^{\x 2\,*}\widetilde\xcL}\equiv\pr_1\ :\ \pr_{2,4}^*\widehat\pr_2^{\x 2\,*}\widetilde\xcL\equiv\sfY_{\la_{\rm vis}2\la_{\rm vis}2}\cD_{i,\upsilon_i}\hspace{2pt}{}_{\pr_{2,4}}\hspace{-3pt}\x_{\pr_1}\widehat\pr_2^{\x 2\,*}\widetilde\xcL \too\sfY_{\la_{\rm vis}2\la_{\rm vis}2}\cD_{i,\upsilon_i}
\qqq
(for $\,f=\pr_2$),\ respectively,\ with the pullback bundles
\qq\nn
\pi_{\pr_{i,j}^*\cF}\equiv\pr_1\ :\ \pr_{i,j}^*\cF\equiv\sfY_{\la_{\rm vis}2\la_{\rm vis}2}\cD_{i,\upsilon_i}\hspace{2pt}{}_{\pr_{i,j}}\hspace{-3pt}\x_{\pi_\cF}\cF\too\sfY_{\la_{\rm vis}2\la_{\rm vis}2}\cD_{i,\upsilon_i}\,,\qquad (i,j)\in\{(1,2),(3,4)\}\,.
\qqq
Using \Reqref{eq:Ybetonvac},\ we readily find
\qq\nn
\bigl(\pr_2^*\circ\widehat\pr{}_2^*-\pr_1^*\circ\widehat\la{}_{\rm vis}^*\bigr)\widehat{\underset{\tx{\ciut{(2)}}}{\sfY\b}}=\sfd\bigl[\bigl(\pr_1^*\circ\widehat\la{}_{\rm vis}^*-\pr_2^*\circ\widehat\pr{}_2^*\bigr)\txA_{i,\upsilon_i}\bigr]\,,
\qqq
and so we postulate
\qq\nn
&&\cF\equiv\sfY_{\la_{\rm vis}2\la_{\rm vis}2}\cD_{i,\upsilon_i}\x\bC^\x\ni\bigl(\bigl(\bigl(\bigl(\breve\vep,\breve y\bigr),\bigl(\theta,x,\phi\bigr)\bigr),\bigl(\theta-S(\phi)\,\breve\vep,x-L(\phi)\,\breve y+\tfrac{1}{2}\,\theta\,\ovl\G\,S(\phi)\,\breve\vep,\xi_1,\phi\bigr)\bigr),\cr\cr
&&\hspace{4cm}\bigl(\bigl(\bigl(\breve\vep,\breve y\bigr),\bigl(\theta,x,\phi\bigr)\bigr),\bigl(\theta,x,\xi_2,\phi\bigr)\bigr),z\bigr)\equiv\varphi
\qqq
with $\,\pi_\cF\equiv\pr_1\,$ and
\qq\nn
\underset{\tx{\ciut{(1)}}}{\cA}{}_\cF(\varphi)&=&\tfrac{\sfi\,\sfd z}{z}+\txA_{i,\upsilon_i}\bigl(\theta-S(\phi)\,\breve\vep,x-L(\phi)\,\breve y+\tfrac{1}{2}\,\theta\,\ovl\G\,S(\phi)\,\breve\vep,\xi_1,\phi\bigr)-\txA_{i,\upsilon_i}\bigl(\theta,x,\xi_2,\phi\bigr)\cr\cr
&=:&\tfrac{\sfi\,\sfd z}{z}+\pi_\cF^*\sfa_{i,\upsilon_i}(\varphi)\,.
\qqq
Inspection of the base components of the relevant principal $\bC^\x$-connection super-1-forms,
\qq\nn
&&\txa\bigl(\bigl(\theta-S(\phi)\,\breve\vep,x-L(\phi)\,\breve y+\tfrac{1}{2}\,\theta\,\ovl\G\,S(\phi)\,\breve\vep,\xi_1\bigr),\bigl(\theta-S(\phi)\,\breve\vep,x-L(\phi)\,\breve y+\tfrac{1}{2}\,\theta\,\ovl\G\,S(\phi)\,\breve\vep,\xi_3\bigr)\bigr)+\pi_\cF^*\sfa_{i,\upsilon_i}\bigl(\varphi_2\bigr)\cr\cr
&=&\pi_\cF^*\sfa_{i,\upsilon_i}\bigl(\varphi_1\bigr)+\txa\bigl(\bigl(\theta,x,\xi_2\bigr),\bigl(\theta,x,\xi_4\bigr)\bigr)\,,
\qqq
written for
\qq\nn
\varphi_A&=&\bigl(\bigl(\bigl(\bigl(\breve\vep,\breve y\bigr),\bigl(\theta,x,\phi\bigr)\bigr),\bigl(\theta-S(\phi)\,\breve\vep,x-L(\phi)\,\breve y+\tfrac{1}{2}\,\theta\,\ovl\G\,S(\phi)\,\breve\vep,\xi_{2A-1},\phi\bigr)\bigr),\cr\cr
&&\bigl(\bigl(\bigl(\breve\vep,\breve y\bigr),\bigl(\theta,x,\phi\bigr)\bigr),\bigl(\theta,x,\xi_{2A},\phi\bigr)\bigr),z\bigr)\,,\qquad A\in\{1,2\}\,,
\qqq
reveals that we may take $\,\a_\cF\,$ in the trivial form
\qq\nn
\a_\cF\equiv\bd1\,,
\qqq
automatically compatible with $\,\mu_{\widetilde\xcL}\equiv\bd1$.\ Thus,\ we obtain data 
\qq\nn
\Upsilon_{\rm vis}\equiv\bigl(\sfY_{\la_{\rm vis}2}\cD_{i,\upsilon_i},\id_{\sfY_{\la_{\rm vis}2}\cD_{i,\upsilon_i}},\cF,\pi_\cF,\underset{\tx{\ciut{(1)}}}{\cA}{}_\cF,\a_\cF\bigr)
\qqq
of the 1-isomorphism 
\qq\nn
\Upsilon_{\rm vis}\ :\ \la_{\rm vis}^*\cG{}^{(1)}_{\rm vac}\xrightarrow{\ \cong\ }\pr_2^*\cG{}^{(1)}_{\rm vac}
\qqq
of the (descendable) $\k_{\rm vis}$-equivariant structure sought after.

In order to complete the construction,\ we move to 
\qq\nn
\k^2_{{\rm vis}\,i,\upsilon_i}\equiv\k_{\rm vis}^{\x 2}\x\cD_{i,\upsilon_i}\,,
\qqq 
with its face maps
\qq\nn
d^{(2)}_0=\pr_{2,3}\,,\qquad\qquad d^{(2)}_1=\id_{\k_{\rm vis}}\x\la_{\rm vis}\,,\qquad\qquad d^{(2)}_2=\unl\txm\x\id_{\cD_{i,\upsilon_i}}
\qqq
to $\,\k^1_{{\rm vis}\,i,\upsilon_i}$,\ and define,\ over the fibred products
\qq\nn
\pi_{\sfY^2_{\la_{\rm vis}2;i}\cD_{i,\upsilon_i}}\equiv\pr_1\ &:&\ \sfY^2_{\la_{\rm vis}2;i}\cD_{i,\upsilon_i}\equiv\bigl(\k^2_{{\rm vis}\,i,\upsilon_i}\hspace{2pt}{}_{d_i^{(2)}}\hspace{-3pt}\x_{\pr_1}\la_{\rm vis}^*\sfY\cD_{i,\upsilon_i}\bigr)\hspace{2pt}{}_{\pr_1}\hspace{-3pt}\x_{\pr_1}\bigl(\k^2_{{\rm vis}\,i,\upsilon_i}\hspace{2pt}{}_{d_i^{(2)}}\hspace{-3pt}\x_{\pr_1}\pr_2^*\sfY\cD_{i,\upsilon_i}\bigr)\cr\cr
&&\too\k^2_{{\rm vis}\,i,\upsilon_i}\hspace{2pt}{}_{d_i^{(2)}}\hspace{-3pt}\x_{\pr_1}\la_{\rm vis}^*\sfY\cD_{i,\upsilon_i}\,,
\qqq
the respective principal $\bC^\x$-bundles
\qq\nn
\pi_{\sfY^2_{\la_{\rm vis}2;i}\cD_{i,\upsilon_i}\hspace{2pt}{}_{\pr_2^{\x 2}}\hspace{-1pt}\x_{\pi_\cF}\cF}\equiv\pr_1\ :\ \sfY^2_{\la_{\rm vis}2;i}\cD_{i,\upsilon_i}\hspace{2pt}{}_{\pr_2^{\x 2}}\hspace{-3pt}\x_{\pi_\cF}\cF\too\sfY^2_{\la_{\rm vis}2;i}\cD_{i,\upsilon_i}\,.
\qqq
We identify\footnote{Coordinate expressions for the identification mappings can be read off directly from \Rxcite{Sec.\,4.2}{Suszek:2019cum}.}
\qq\nn
\iota_1\equiv\bigl(\pr_1,\bigl(\pr_{2,3}\circ\pr_1,\pr_2\circ\pr_2\bigr)\bigr)\ &:&\ d_2^{(2)\,*}\pr_2^*\sfY\cD_{i,\upsilon_i}\xrightarrow{\ \cong\ }d_0^{(2)\,*}\la_{\rm vis}^*\sfY\cD_{i,\upsilon_i}\,,\cr\cr
\iota_2\equiv\bigl(\pr_1,\bigl((\unl\txm\x\id_{\cD_{i,\upsilon_i}})\circ\pr_1,\pr_2\circ\pr_2\bigr)\bigr)\ &:&\ d_2^{(2)\,*}\la_{\rm vis}^*\sfY\cD_{i,\upsilon_i}\xrightarrow{\ \cong\ }d_1^{(2)\,*}\la_{\rm vis}^*\sfY\cD_{i,\upsilon_i}\,,\cr\cr
\iota_3\equiv\bigl(\pr_1,\bigl((\unl\txm\x\id_{\cD_{i,\upsilon_i}})\circ\pr_1,\pr_2\circ\pr_2\bigr)\bigr)\ &:&\ d_0^{(2)\,*}\pr_2^*\sfY\cD_{i,\upsilon_i}\xrightarrow{\ \cong\ }d_1^{(2)\,*}\pr_2^*\sfY\cD_{i,\upsilon_i}\,,
\qqq
and define (in an obvious shorthand notation)
\qq\nn
&&\hspace{-2.1cm}\sfY_{\la_{\rm vis}22}^3\cD_{i,\upsilon_i}\equiv d_2^{(2)\,*}\la_{\rm vis}^*\sfY\cD_{i,\upsilon_i}\x_{\k^2_{{\rm vis}\,i,\upsilon_i}}d_2^{(2)\,*}\pr_2^*\sfY\cD_{i,\upsilon_i}\x_{\k^2_{{\rm vis}\,i,\upsilon_i}}d_0^{(2)\,*}\pr_2^*\sfY\cD_{i,\upsilon_i}\cr\cr
&\xrightarrow{\ \cong_1\equiv\id\x\iota_1\x\id\ }&d_2^{(2)\,*}\la_{\rm vis}^*\sfY\cD_{i,\upsilon_i}\x_{\k^2_{{\rm vis}\,i,\upsilon_i}}d_0^{(2)\,*}\la_{\rm vis}^*\sfY\cD_{i,\upsilon_i}\x_{\k^2_{{\rm vis}\,i,\upsilon_i}}d_0^{(2)\,*}\pr_2^*\sfY\cD_{i,\upsilon_i}\cr\cr
&\xrightarrow{\ \cong_2\equiv\iota_2\x\id\x\iota_3\ }&d_1^{(2)\,*}\la_{\rm vis}^*\sfY\cD_{i,\upsilon_i}\x_{\k^2_{{\rm vis}\,i,\upsilon_i}}d_0^{(2)\,*}\la_{\rm vis}^*\sfY\cD_{i,\upsilon_i}\x_{\k^2_{{\rm vis}\,i,\upsilon_i}}d_1^{(2)\,*}\pr_2^*\sfY\cD_{i,\upsilon_i}\,,
\qqq
in terms of which we may write
\qq\nn
\pr_{1,2}^*\bigl(\sfY^2_{\la_{\rm vis}2;2}\cD_{i,\upsilon_i}\hspace{2pt}{}_{\pr_2^{\x 2}}\hspace{-3pt}\x_{\pi_\cF}\cF\bigr)&\equiv&\sfY^3_{\la_{\rm vis}22}\cD_{i,\upsilon_i}\hspace{2pt}{}_{\pr_{1,2}}\hspace{-3pt}\x_{\pr_1}\bigl(\sfY^2_{\la_{\rm vis}2;2}\cD_{i,\upsilon_i}\hspace{2pt}{}_{\pr_2^{\x 2}}\hspace{-3pt}\x_{\pi_\cF}\cF\bigr)\xrightarrow{\ \pr_1\ }\sfY^3_{\la_{\rm vis}22}\cD_{i,\upsilon_i}\,,\cr\cr
\bigl(\pr_{2,3}(\circ\cong_1)\bigr)^*\bigl(\sfY^2_{\la_{\rm vis}2;0}\cD_{i,\upsilon_i}\hspace{2pt}{}_{\pr_2^{\x 2}}\hspace{-3pt}\x_{\pi_\cF}\cF\bigr)&\equiv&\sfY^3_{\la_{\rm vis}22}\cD_{i,\upsilon_i}\hspace{2pt}{}_{\pr_{2,3}(\circ\cong_1)}\hspace{-3pt}\x_{\pr_1}\bigl(\sfY^2_{\la_{\rm vis}2;0}\cD_{i,\upsilon_i}\hspace{2pt}{}_{\pr_2^{\x 2}}\hspace{-3pt}\x_{\pi_\cF}\cF\bigr)\xrightarrow{\ \pr_1\ }\sfY^3_{\la_{\rm vis}22}\cD_{i,\upsilon_i}
\qqq
and
\qq\nn
\bigl(\pr_{1,3}(\circ\cong_2\circ\cong_1)\bigr)^*\bigl(\sfY^2_{\la_{\rm vis}2;1}\cD_{i,\upsilon_i}\hspace{2pt}{}_{\pr_2^{\x 2}}\hspace{-3pt}\x_{\pi_\cF}\cF\bigr)\equiv\sfY^3_{\la_{\rm vis}22}\cD_{i,\upsilon_i}\hspace{2pt}{}_{\pr_{1,3}(\circ\cong_2\circ\cong_1)}\hspace{-3pt}\x_{\pr_1}\bigl(\sfY^2_{\la_{\rm vis}2;1}\cD_{i,\upsilon_i}\hspace{2pt}{}_{\pr_2^{\x 2}}\hspace{-3pt}\x_{\pi_\cF}\cF\bigr)\xrightarrow{\ \pr_1\ }\sfY^3_{\la_{\rm vis}22}\cD_{i,\upsilon_i}\,.
\qqq
Subsequently,\ we form the tensor product of the former two,
\qq\nn
\pr_1\ : \ \pr_{1,2}^*\bigl(\sfY^2_{\la_{\rm vis}2;2}\cD_{i,\upsilon_i}\x_{\pr_2^{\x 2}}\cF\bigr)\ox\pr_{2,3}^*\bigl(\sfY^2_{\la_{\rm vis}2;0}\cD_{i,\upsilon_i}\x_{\pr_2^{\x 2}}\cF\bigr)\too\sfY^3_{\la_{\rm vis}22}\cD_{i,\upsilon_i}\,,
\qqq
and identify it as the principal $\bC^\x$-bundle of the product 1-isomorphism $\,d^{(2)\,*}_0\Upsilon_{\rm vis}\circ d^{(2)\,*}_2\Upsilon_{\rm vis}$,\ to be related to the latter one,\ in which we recognise the principal $\bC^\x$-bundle of $\,d^{(2)\,*}_1\Upsilon_{\rm vis}$.\ The relation is readily established through comparison of the base components of the respective principal $\bC^\x$-connection super-1-forms.\ For that purpose,\ write 
\qq\nn
k_{1,2}:=\bigl(\bigl(\breve\vep{}_1,\breve y{}_1\bigr),\bigl(\breve\vep{}_2,\breve y{}_2\bigr),\bigl(\theta,x,\phi\bigr)\bigr)\in\k^2_{{\rm vis}\,i,\upsilon_i}
\qqq
and consider
\qq\nn
\bigl(\bigl(k_{1,2},\bigl(\bigl(\bigl(\breve\vep{}_1,\breve y{}_1\bigr),\bigl(\theta-S(\phi)\,\breve\vep{}_2,x-L(\phi)\,\breve y{}_2+\tfrac{1}{2}\,\theta\,\ovl\G\,S(\phi)\,\breve\vep{}_2,\phi\bigr)\bigr),\cr\cr
\bigl(\theta-S(\phi)\,\bigl(\breve\vep{}_1+\breve\vep{}_2\bigr),x-L(\phi)\,\bigl(\breve y{}_1+\breve y{}_2+\tfrac{1}{2}\,\breve\vep{}_1\,\ovl\G\,\breve\vep{}_2\bigr)+\tfrac{1}{2}\,\theta\,\ovl\G\,S(\phi)\,\bigl(\breve\vep{}_1+\breve\vep{}_2\bigr),\xi_1,\phi\bigr)\bigr)\bigr),\cr\cr
\bigl(k_{1,2},\bigl(\bigl(\bigl(\breve\vep{}_1,\breve y{}_1\bigr),\bigl(\theta-S(\phi)\,\breve\vep{}_2,x-L(\phi)\,\breve y{}_2+\tfrac{1}{2}\,\theta\,\ovl\G\,S(\phi)\,\breve\vep{}_2,\phi\bigr)\bigr),\cr\cr
\bigl(\theta-S(\phi)\,\breve\vep{}_2,x-L(\phi)\,\breve y{}_2+\tfrac{1}{2}\,\theta\,\ovl\G\,S(\phi)\,\breve\vep{}_2,\phi,\xi_2,\phi\bigr)\bigr)\bigr),\bigl(k_{1,2},\bigl(\bigl(\bigl(\breve\vep{}_1,\breve y{}_1\bigr),\bigl(\theta,x,\phi\bigr)\bigr),\bigl(\theta,x,\xi_3,\phi\bigr)\bigr)\bigr)\bigr)\cr\cr
\equiv\bigl(\bigl(k_{1,2},\unl\varphi{}_1\bigr),\bigl(k_{1,2},\unl\varphi{}_2\bigr),\bigl(k_{1,2},\unl\varphi{}_3\bigr)\bigr)\in\sfY^3_{\la_{\rm vis}22}\cD_{i,\upsilon_i}\,.
\qqq
The connection super-1-forms now compare as
\qq\nn
&&\bigl(\pr_{1,2}^*\pr_2^{\x 2\,*}+\cong_1^*\pr_{2,3}^*\pr_2^{\x 2\,*}\bigr)\sfa_{i,\upsilon_i}\bigl(\bigl(k_{1,2},\unl\varphi{}_1\bigr),\bigl(k_{1,2},\unl\varphi{}_2\bigr),\bigl(k_{1,2},\unl\varphi{}_3\bigr)\bigr)\cr\cr
&=&\cong_1^*\cong_2^*\pr_{1,3}^*\pr_2^{\x 2\,*}\sfa_{i,\upsilon_i}\bigl(\bigl(k_{1,2},\unl\varphi{}_1\bigr),\bigl(k_{1,2},\unl\varphi{}_2\bigr),\bigl(k_{1,2},\unl\varphi{}_3\bigr)\bigr)\,,
\qqq
whence also the choice 
\qq\nn
\b\equiv\bd1\ :\ \pr_{1,2}^*\bigl(\sfY^2_{\la_{\rm vis}2;2}\cD_{i,\upsilon_i}\x_{\pr_2^{\x 2}}\cF\bigr)\ox\pr_{2,3}^*\bigl(\sfY^2_{\la_{\rm vis}2;0}\cD_{i,\upsilon_i}\x_{\pr_2^{\x 2}}\cF\bigr)\xrightarrow{\ \cong\ }\pr_{1,3}^*\bigl(\sfY^2_{\la_{\rm vis}2;1}\cD_{i,\upsilon_i}\x_{\pr_2^{\x 2}}\cF\bigr)\,,
\qqq
trivially compatible with $\,\a_\cF\equiv\bd1$.\ The ensuing 2-isomorphism 
\qq\nn
\g_{\rm vis}\equiv\bigl(\sfY_{\la_{\rm vis}22}^3\cD_{i,\upsilon_i},\id_{\sfY_{\la_{\rm vis}22}^3\cD_{i,\upsilon_i}},\b\bigr)\ :\ d^{(2)\,*}_0\Upsilon_{\rm vis}\circ d^{(2)\,*}_2\Upsilon_{\rm vis}\xLongrightarrow{\ \cong\ }d^{(2)\,*}_1\Upsilon_{\rm vis}
\qqq
of the $\k_{\rm vis}$-equivariant structure under reconstruction is manifestly (and trivially) coherent.\ We summarise the results of our check in\bigskip

\noindent {\bf Theorem 5.} {\it The vacuum restriction of the extended super-1-gerbe of the GS super-$\si$-model in the HP formulation from theorem 3.\ carries a canonical descendable $\k_{\rm vis}$-structure.}\bigskip

The last result -- a consequence of the trivialisation mechanism discussed earlier that we dissected above for the sake of illustration that may prove useful in geometrically more involved circumstances --  completes our systematic investigation of the higher-geometric and -algebraic content of the super-$\si$-model of the superstring in $\,{\rm sMink}(d,1|D_{d,1})\,$ in the purely topological HP formulation in which that content becomes particularly manifest and structured.\ It leaves us with a fairly complete picture of the (classical) vacuum of the theory and its global and local supersymmetries,\ alongside their very natural (super)gerbe-theoretic realisations with Lie-superalgebraic and -supergroup `skeleta'.\ We hope to return to the line of research drawn hereabove in the future.

\section{Conclusions \& Outlook}\label{ref:CandO}

In the present paper,\ we have associated with the classical vacuum of the super-$\si$-model for the superstring in the super-Minkowski spacetime ({\it i.e.},\ with the embedded superstring worldsheet) a higher Lie-superalgebraic object -- the ${\bf sLieAlg}$-skeleton of Theorem 3. -- that models,\ in the category of Lie superalgebras,\ the tangent sheaf of the null trivialisation of the vacuum restriction of an extended super-1-gerbe geometrising,\ through an adaptation of the general scheme of \Rcite{Murray:1994db} to the supergeometric setting proposed in \Rcite{Suszek:2017xlw},\ the topological action functional of the super-$\si$-model in the Hughes--Polchinski formulation consistently with the supersymmetries present.\ The geometrisation has been obtained,\ in Theorem 1.,\ as an equivariant lift of the one originally constructed in \Rcite{Suszek:2017xlw} over the physical supertarget $\,{\rm sMink}(d,1|D_{d,1})\,$ to the full supersymmetry group $\,{\rm sISO}(d,1|D_{d,1})$,\ and the said trivialisation,\ postulated in \Rcite{Suszek:2020xcu,Suszek:2020rev} and proven as Theorem 2.,\ can be viewed as a higher-geometric manifestation of the nature of the vacuum,\ which is that of an integral leaf of an involutive superdistribution (over the extended supertarget $\,{\rm sISO}(d,1|D_{d,1})$) of its tangential gauge supersymmetries,\ the latter being bracket-generated by the $\k$-symmetry superdistribution of Sec.\,\ref{sub:kappasym} and modelled on the Lie sub-superalgebra $\,\gt{vac}(\gt{sB}^{{\rm (HP)}}_{1,2})\,$ of the supersymmetry algebra $\,\gt{siso}(d,1|D_{d,1})\,$ of the super-$\si$-model given in \Reqref{eq:vacbas}.\ The ${\bf sLieAlg}$-skeleton has been demonstrated to integrate to a higher-geometric object -- the ${\bf sLieGrp}$-model of Theorem 4. -- in the category of Lie supergroups that acquires the interpretation of the higher gauge supersymmetry group of the vacuum.\ The statement of null trivialisation of the vacuum-restricted super-1-gerbe has been shown,\ in Theorem 5.,\ to strengthen the statement of descendable equivariance of that super-1-gerbe with respect to the extended $\k$-symmetry group of the superstring $\,{\rm sISO}(d,1|D_{d,1})_{\rm vac}\subset{\rm sISO}(d,1|D_{d,1})\,$ at the root of the ${\bf sLieGrp}$-model,\ anticipated,\ already in \Rcite{Suszek:2019cum} and in a more structured form in \Rcite{Suszek:2020xcu},\ on the basis of the interpretation of the latter supergroup as the gauge supersymmetry group of the vacuum,\ {\it cp} Refs.\,\cite{Gawedzki:2010rn,Gawedzki:2012fu}.\ In the light of the long-established interpretation of the higher-geometric objects associated with the cohomological content of the (super-)$\si$-model as structures encoding,\ through the transgression mechanism of Refs.\,\cite{Gawedzki:1987ak,Suszek:2011hg},\ the pre-quantisation of the (super)field theory under consideration,\ the findings of the present paper are to be understood as novel markers of quantum-mechanical coherence of the super-$\si$-model and,\ simultaneously,\ as strong and nontrivial evidence for the internal consistency of the gerbe-theoretic approach to Green--Schwarz-type super-$\si$-models advanced in \Rcite{Suszek:2017xlw} and developed in the series of papers \cite{Suszek:2019cum,Suszek:2018bvx,Suszek:2018ugf,Suszek:2020xcu,Suszek:2020rev} that followed.\ First and foremost,\ though,\ they provide us with a realisation of the goal set up in the Introduction,\ which consists in extracting a higher (super-)algebraic representation of the fundamental object of the super-field theory under consideration -- the superstring (trajectory/current) -- from the geometrisation of the background gauge field obtained through an extension of the supersymmetry algebra.\smallskip

The geometrisation of the \emph{supersymmetrically invariant} cohomological content of the super-mink\-ow\-skian super-$\si$-model reviewed and elaborated in the present paper can be understood,\ in the spirit of Refs.\,\cite{Rabin:1984rm,Rabin:1985tv},\ as standard geometrisation,\ \`a la Murray,\ of the cohomological content of a super-$\si$-model with an orbifold of the supermanifold $\,{\rm sMink}(d,1|D_{d,1})\,$ with respect to a natural action of the discrete Kosteleck\'y--Rabin group $\,\G_{\rm KR}\subset{\rm sMink}(d,1|D_{d,1})\,$ as the supertarget,\ the orbifold having a highly nontrivial topology,\ also in the Gra\ss mann-odd fibre.\ This remark disperses the illusion of triviality of the constructions considered which may arise as a consequence of the topological triviality of (the body of) the apparent supertarget $\,{\rm sMink}(d,1|D_{d,1})$.\ It legitimises the present choice of the superstring supergeometry as the one in which the novel phenomena entailed by the $\bZ/2\bZ$-grading of the target geometry and the supersymmetry of the dynamics that takes place in it and captured by the discrepancy between the de Rham cohomology and its physically favoured supersymmetric refinement are most neatly and tractably separated from the standard ones known from the study of $\si$-models with topologically nontrivial targets.\ That said,\ it is only natural,\ and very well justified from the physical point of view,\ to look for analogons of the structures and mechanisms reported herein in superstring superbackgrounds with topologically nontrivial curved supertargets.\ The results for the family of superbackgrounds over the $\,{\rm AdS}_p\x\bS^q\,$ obtained in Refs.\,\cite{Suszek:2018bvx,Suszek:2018ugf} and \Rcite{Suszek:2020xcu} provide a firm basis for such developments.

The key idea of the paper,\ which boils down to associating a particular diagram in the category of Lie superalgebras decorated with (and determined by) Chevalley--Eilenberg cohomological data and integrable to the corresponding diagram in the category of Lie supergroups,\ to the vacuum of the super-$\si$-model through a supersymmetrically invariant trivialisation of the super-1-gerbe that geometrises the relevant Cartan--Eilenberg super-3-cocycle upon restriction to the embedded (vacuum) superstring worldsheet,\ also admits an obvious generalisation to other species of BPS states encountered in superstring theory.\ Indeed,\ there are two independent sources of such structures,\ of a different physical status,\ that we may derive from the known super-$(p+2)$-cocycles that define Green--Schwarz-type super-$\si$-models for super-$p$-branes,\ to wit,
\bit
\item trivialisations of the super-$p$-gerbes geometrising the super-$(p+2)$-cocycles over the embedded super-$p$-brane vacua in the Hughes--Polchinski formulation,\ with curvatures given by the volume super-$(p+1)$-forms of the vacua (as in the present paper,\ in which $\,p=1$);
\item arbitrary modules of the same super-$p$-gerbes arising over sub-supermanifolds within the respective supertargets endowed with actions of subgroups of the supersymmetry groups and defining (Dirichlet) boundary conditions in the super-$p$-brane super-$\si$-models.
\eit
The latter class naturally extends to that of supersymmetric super-$p$-gerbe bimodules associated with worldvolume defects in these super-$\si$-models.\ We shall study their higher Lie-superalgebraic incarnation at length in an upcoming paper.\medskip

\noindent{\bf Acknowledgements:} The Author gratefully acknowledges the hospitality extended to him,\ in September 2020,\ by the Erwin Schr\"odinger International Institute for Mathematics and Physics where this work was brought to completion.\ He is also thankful to the Organisers of the Thematic Programme ``Higher Structures and Field Theory'',\ and to Thomas Strobl in particular,\ for the inspiring atmosphere of the meeting,\ for their interest in his research reported herein,\ and for financial support during the period of his sojourn in Vienna.

\appendix

\section{A convention}\label{app:conv}

Let $\,\cM_1,\cM_2\,$ and $\,\cN\,$ be supermanifolds,\ and let $\,\varphi_n\ :\ \cM_n\too\cN,\ n\in\{1,2\}\,$ be supermanifold morphisms of which (at least) one is a surjective submersion.\ We then define the fibred product of $\,\cM_1\,$ and $\,\cM_2\,$ over $\,\cN\,$ as the supermanifold 
\qq\nn
\cM_1\x_\cN\cM_2\equiv\cM_1\hspace{2pt}{}_{\varphi_1}\hspace{-2pt}\x_{\varphi_2}\cM_2
\qqq
embedded in the cartesian product $\,\cM_1\x\cM_2\,$ ({\it cp} \Rxcite{Sec.\,2.4.9}{Voronov:2014}) and described by the commutative diagram
\qq\nn
\alxydim{@C=1.5cm@R=1cm}{ \cM_1\x_\cN\cM_2 \ar[r]^{\qquad\pr_2} \ar[d]_{\pr_1} & \cM_2 \ar[d]^{\varphi_2} \\ \cM_1 \ar[r]_{\varphi_1} & \cN}\,.
\qqq
Its existence is ensured by \Rxcite{Prop.\,3.2.11}{Kessler:2019bwp}.

\end{document}